\documentclass[twocolumn,aps,prc,qshowpacs,floatfix,nofootinbib,superscriptaddress]{revtex4-1}

\usepackage{amsmath}
\usepackage{graphicx}
\usepackage{url}
\usepackage[table]{xcolor}

\newcommand{\gton}{\mathrel{\lower.9ex \hbox{$\stackrel{\displaystyle >}{\sim}$}}} 
\newcommand{\lton}{\mathrel{\lower.9ex \hbox{$\stackrel{\displaystyle <}{\sim}$}}}  
\newcommand{\dif}{\mathrm{d}}
\newcommand{\ch}{\mathrm{ch}}
\newcommand{\NN}{\mathrm{NN}}
\newcommand{\Tdec}{T_{\mathrm{dec}}}
\newcommand{\Tchem}{T_{\mathrm{chem}}}
\newcommand{\Th}{T_{\mathrm{high}}}
\newcommand{\x}{\times}
\newcommand{\Ksat}{K_{\mathrm{sat}}}
\newcommand{\etasmin}{(\eta/s)_{\mathrm{min}}}
\newcommand{\Tmin}{T_{\mathrm{H}}}
\newcommand{\Wmin}{W_{\mathrm{min}}}
\newcommand{\Shg}{S_{\mathrm{HG}}}
\newcommand{\Sqgp}{S_{\mathrm{QGP}}}

\begin{document}

\begin{flushright}
CERN-TH-2020-099
\end{flushright}

\title{Temperature dependence of $\eta/s$ of strongly interacting matter:
       effects of the equation of state and the parametric form of $(\eta/s)(T)$}

\author{Jussi Auvinen}
\email{auvinen@ipb.ac.rs}
\affiliation{Institute of Physics Belgrade, 11080 Belgrade, Serbia}
\author{Kari J.~Eskola}
\affiliation{University of Jyvaskyla, Department of Physics, P.O.\ Box 35,
  FI-40014 University of Jyvaskyla, Finland}
\affiliation{Helsinki Institute of Physics, P.O.\ Box 64, FI-00014 University of Helsinki, Finland}
\author{Pasi Huovinen}
\affiliation{Institute of Physics Belgrade, 11080 Belgrade, Serbia}
\affiliation{Institute of Theoretical Physics, University of Wroclaw, 50-204 Wroc\l aw, Poland}
\author{Harri Niemi}
\affiliation{University of Jyvaskyla, Department of Physics, P.O.\ Box 35,
  FI-40014 University of Jyvaskyla, Finland}
\affiliation{Helsinki Institute of Physics, P.O.\ Box 64, FI-00014 University of Helsinki, Finland}
\author{Risto Paatelainen}
\affiliation{Theoretical Physics Department, CERN, CH-1211 Geneve 23, Switzerland}
\author{P\'eter Petreczky}
\affiliation{Physics Department, Brookhaven National Laboratory, Upton, NY 11973, USA}

\begin{abstract}
We investigate the temperature dependence of the shear viscosity to
entropy density ratio $\eta/s$ using a piecewise linear
parametrization.  To determine the optimal values of the parameters
and the associated uncertainties, we perform a global Bayesian
model-to-data comparison on Au+Au collisions at $\sqrt{s_\NN}=200$ GeV
and Pb+Pb collisions at $2.76$ TeV and $5.02$ TeV, using a 2+1D
hydrodynamical model with the EKRT initial state. We provide three new
parametrizations of the equation of state (EoS) based on contemporary
lattice results and hadron resonance gas, and use them and the widely
used $s95p$ parametrization to explore the uncertainty in the analysis
due to the choice of the equation of state. We found that $\eta/s$ is
most constrained in the temperature range $T\approx 150$--$220$ MeV,
where, for all EoSs, $0.08 < \eta/s < 0.23$ when taking into account
the 90\% credible intervals. In this temperature range the EoS
parametrization has only a small $\approx 10\%$ effect on the favored
$\eta/s$ value, which is less than the $\approx 30\%$ uncertainty of the
analysis using a single EoS parametrization. Our parametrization of
$(\eta/s)(T)$ leads to a slightly larger minimum value of $\eta/s$
than the previously used parametrizations. When we constrain our
parametrization to mimic the previously used parametrizations, our
favored value is reduced, and the difference becomes statistically
insignificant.
\end{abstract}


\maketitle

\section{Introduction}

The main goal of the ultrarelativistic heavy-ion collisions at the
Large Hadron Collider (LHC) and the Relativistic Heavy-Ion Collider
(RHIC) is to understand the properties of the strongly interacting
matter produced in these collisions. In recent years the main interest
has been in extracting the dissipative properties of this QCD matter
from the experimental data (e.g.~\cite{Luzum:2008cw,Bozek:2009dw,
  Song:2011qa,Ryu:2015vwa,Karpenko:2015xea}), in particular its
specific shear viscosity: the ratio of shear viscosity to entropy
density $\eta/s$ (for a review, see Refs.~\cite{Heinz:2013th,Gale:2013da,
  Huovinen:2013wma,Shen:2020gef}). The field has matured to a level
where a global Bayesian analysis of the parameters can provide
statistically meaningful credibility ranges to the temperature
dependence of $\eta/s$~\cite{Bernhard:2016tnd,Bass:2017zyn,
  Bernhard:2019bmu}. These credibility ranges agree with earlier
results like those obtained using the EKRT model~\cite{Niemi:2015qia}.

However, with the exception of papers like Refs.~\cite{Pratt:2015zsa,
  Moreland:2015dvc,Alba:2017hhe}, the equation of state (EoS) is taken
as given in the models used to extract the $\eta/s$ ratio from the
data. Recent fluid dynamical studies generally use an EoS based on
contemporary lattice QCD results, but during the last decade many
studies in the literature used the EoS parametrization
$s95p$~\cite{Huovinen:2009yb}. This parametrization is based on by now
outdated lattice data~\cite{Bazavov:2009zn}, and recent studies have
reported an approximate 60\%~\cite{Alba:2017hhe} or 30\%
increases~\cite{Schenke:2019ruo} in the extracted value of $\eta/s$
when switching from $s95p$ to a contemporary lattice-based
EoS. Furthermore, even if the errors of the contemporary lattice QCD
calculations overlap, there is still a small tension between the trace
anomalies obtained using the HISQ~\cite{Bazavov:2014pvz,
  Bazavov:2017dsy} and stout~\cite{Borsanyi:2013bia, Borsanyi:2010cj}
discretization schemes. Consequently the EoSs differ, and if the
procedure to extract $\eta/s$ from the data is as sensitive to the
details of the EoS as Refs.~\cite{Alba:2017hhe,Schenke:2019ruo} claim,
this tension may lead to additional uncertainties in the $\eta/s$
values extracted from the heavy-ion collision data.

In the previously mentioned Bayesian analysis~\cite{Bernhard:2016tnd,
  Bass:2017zyn}, where the EoS is based on contemporary lattice data~\cite{Bazavov:2014pvz},
 the temperature dependence of $\eta/s$ was assumed to
be monotonously increasing above the QCD transition temperature
$T_c$. In a Bayesian analysis the slope parameter of such
parametrization is always constrained to be non-negative, and limiting
the final slope parameter to zero would require extremely strong
constraints from the experimental data. Therefore, by construction,
the analysis leads to an $\eta/s$ increasing with temperature above
$T_c$, even if there is no physical reason to exclude a scenario where
$\eta/s$ is constant in a broad temperature range above $T_c$. A more
flexible parametrization, which does not impose such constraints, is
thus needed to determine the temperature dependence of $\eta/s$.

In this work we address both the sensitivity of the extracted $\eta/s$
to the EoS used in the model calculation, and the temperature
dependence of $\eta/s$ in the vicinity of the QCD transition
temperature. We perform a Bayesian analysis of the results of EKRT +
hydrodynamics calculations~\cite{Niemi:2015qia,Niemi:2015voa}, and
the data obtained in $\sqrt{s_{\NN}}=200$ GeV Au+Au
collisions~\cite{Adler:2004zn,Adler:2003cb,Adams:2004bi}, and Pb+Pb
collisions at $2.76$ TeV~\cite{Aamodt:2010cz,Abelev:2013vea,
  Adam:2016izf} and $5.02$ TeV~\cite{Adam:2016izf,Adam:2015ptt}. To
study the temperature dependence of $\eta/s$ we use a piecewise linear
parametrization in three parts: linearly decreasing and increasing
regions at low and high temperatures are connected by a constant-value
plateau of variable range. With this parametrization, data favoring a
strong temperature dependence will lead to large slopes and a narrow
plateau; conversely, an approximately constant $\eta/s$ can be
obtained with small slope parameter values and a wide plateau. To
explore the sensitivity to the EoS, we use four different
parametrizations: the well-known $s95p$ parametrization, and three new
parametrizations based on contemporary lattice QCD results. A
comparison of the final probability distributions of the parameters
will tell whether the most probable parameter values depend on the EoS
used, and whether that difference is significant when the overall
uncertainty in the fitting procedure is taken into account.

\section{Equation of state}
 \label{sec:EoS}

In lattice QCD the calculation of the equation of state (EoS) usually
proceeds through the calculation of the trace anomaly,
$\Theta(T) = \epsilon(T) -3 p(T)$, where $\epsilon$ and $p$ are energy
density and pressure, respectively. Thermodynamical variables are
subsequently derived from it using so-called integral
method~\cite{Boyd:1996bx}. Therefore we base our EoS parametrizations
on the trace anomaly and obtain pressure from the integral
\begin{equation}
  \frac{p(T)}{T^4} - \frac{p(T_{\rm low})}{T^4_{\rm low}}
  =\int_{T_{\rm low}}^{T} \frac{\dif T'}{{T'}^5} \Theta(T').
 \label{eq:P-integral}
\end{equation}
Once the pressure is known, the energy and entropy densities can be
calculated, $\epsilon(T)=\Theta(T) + 3 p(T)$, and
$s(T)=[\epsilon(T)+p(T)]/T$, respectively. To make a construction of
chemical freeze-out at $T\approx 150$ MeV temperature possible, we use
the hadron resonance gas (HRG) trace anomaly at low temperatures
instead of the lattice QCD result. Equally important is that this
choice allows for energy and momentum conserving switch from fluid
degrees of freedom to particle degrees of freedom without any
non-physical discontinuities in temperature and/or flow
velocity\footnote{Energy and momentum conservation require that the
  fluid EoS is that of free particles, and that the degrees of freedom
  are the same in the fluid and particles~\cite{Laszlo}. If the
  dissipative corrections are small, switch from fluid consistent with
  the contemporary lattice QCD results~\cite{Huovinen:2018ziu} to
  particles in the UrQMD~\cite{Bass:1998ca} or SMASH~\cite{Weil:2016zrk}
  hadron cascades at $T=150$ MeV temperature leads to roughly 9--10\%
  or 6--7\% loss in both total energy and entropy, respectively.}.
Furthermore, it gives us a consistent value for the pressure at
$T_{\rm low}$ required for the evaluation of pressure (see
Eq.~(\ref{eq:P-integral})).

As a baseline, we use the $s95p$ parametrization~\cite{Huovinen:2009yb},
where HRG containing hadrons and resonances below $M<2$ GeV mass from
the 2004 PDG summary tables~\cite{Eidelman:2004wy} is connected to the
parametrized hotQCD data from Ref.~\cite{Bazavov:2009zn}. To explore
the effects of various developments during the last decade, we first
connect the HRG based on the PDG 2004 particle
list~\cite{Eidelman:2004wy} to parametrized contemporary lattice data
obtained using the HISQ discretization scheme~\cite{Bazavov:2014pvz,
  Bazavov:2017dsy}. The lattice spacing, $a$, is related to the
temperature and temporal lattice extent, $N_t$, as $a=1/(N_t T)$. Since
the lattice spacing ($N_t$) dependence is small for this action, we
use these results at fixed lattice spacing $N_t=8,10$ and 12. We name
our parametrizations according to the convention used to name $s95p$,
and label this parametrization $s87h_{04}$. '$s87$' signifies entropy
density reaching 87\% of its ideal gas value at $T=800$ MeV, the
letter '$h$' refers to the HISQ action, and the subscript '04' to the
vintage of the PDG particle list (2004). Note that even if our
parametrization differs from the lattice trace anomaly in the hadronic
phase, it agrees with the contemporary lattice calculations which show
that at $T=800$ MeV the entropy density reaches $87$--$88\%$ of the
ideal gas value (c.f.~Fig.~8 of Ref.~\cite{Bazavov:2017dsy}).

The number of well-established resonances has increased since 2004, so
we base our parametrization $s88h_{18}$ on HRG containing
all\footnote{With the exception of $f_0(500)$. See
  Refs.~\cite{Venugopalan:1992hy,Broniowski:2015oha}.} strange and
non-strange hadrons and resonances in the PDG 2018 summary
tables\footnote{Note that PDG Meson Summary Table and Baryon
  Summary Table contain (almost) all states listed by the PDG, and are
  different from the PDG Meson Summary Tables and Baryon Summary
  Tables we use~\cite{PDGtables}. The PDG Baryon Summary
  Tables contain the three and four star resonance states. The PDG
  does not assign stars to meson states, but the Meson Summary Tables
  contain the states not labeled ``Omitted from summary table'' in the
  individual listings.}\cite{Tanabashi:2018oca}, and on the same HISQ
lattice data~\cite{Bazavov:2014pvz,Bazavov:2017dsy} we used for
$s87h_{04}$. Furthermore, there is a slight tension in the trace
anomaly between the HISQ and stout discretization schemes. To explore
whether this difference has any effect on hydrodynamical modeling, we
construct the parametrization $s83s_{18}$ using PDG 2018 resonances,
and the continuum extrapolated lattice data obtained using the stout
discretization~\cite{Borsanyi:2013bia, Borsanyi:2010cj}. The second
letter '$s$' in the label refers now to the stout action, and the
subscript '18' to the vintage of the particle list. The details of
these parametrizations are shown in Appendix~\ref{appx:para}.

\begin{figure}[t]
 \centering
 \includegraphics[width=8cm]{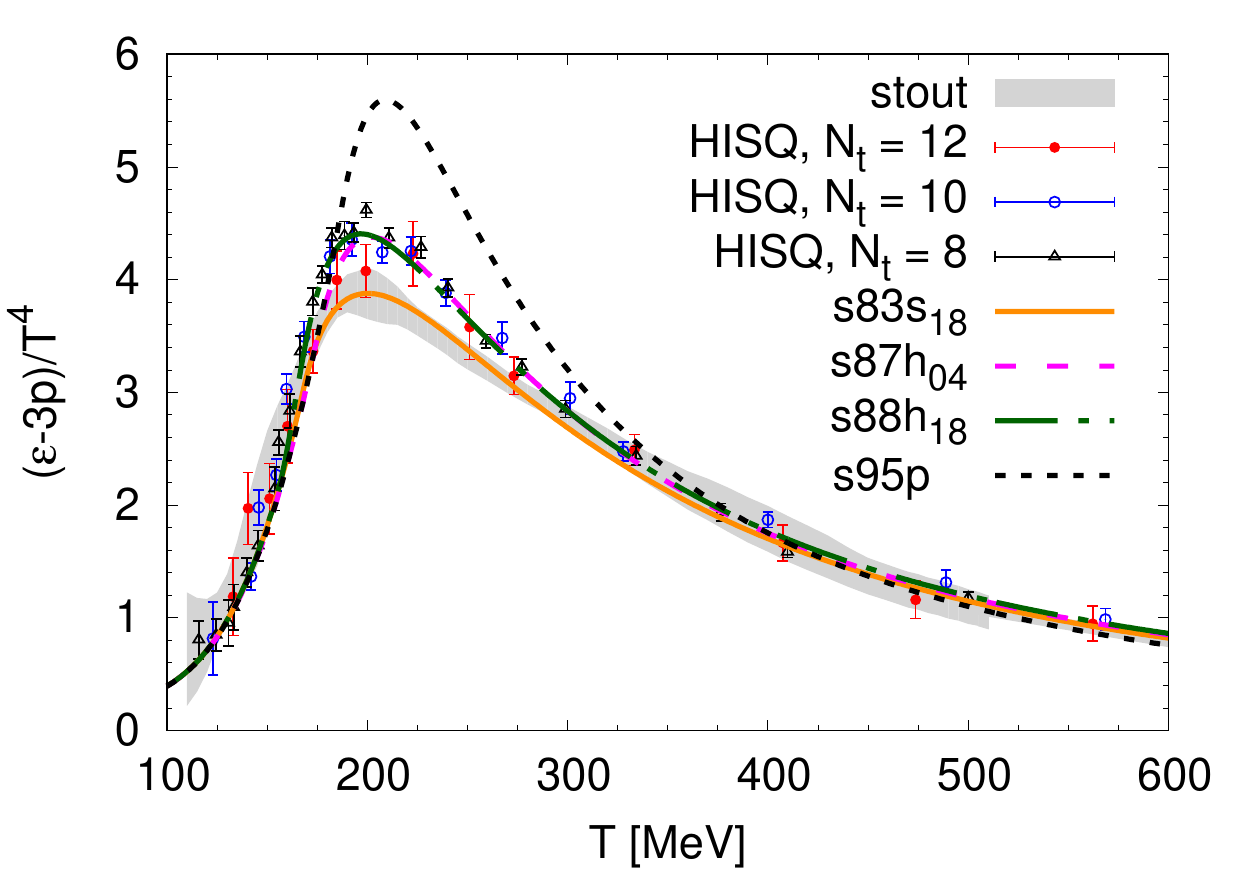}
 \includegraphics[width=8cm]{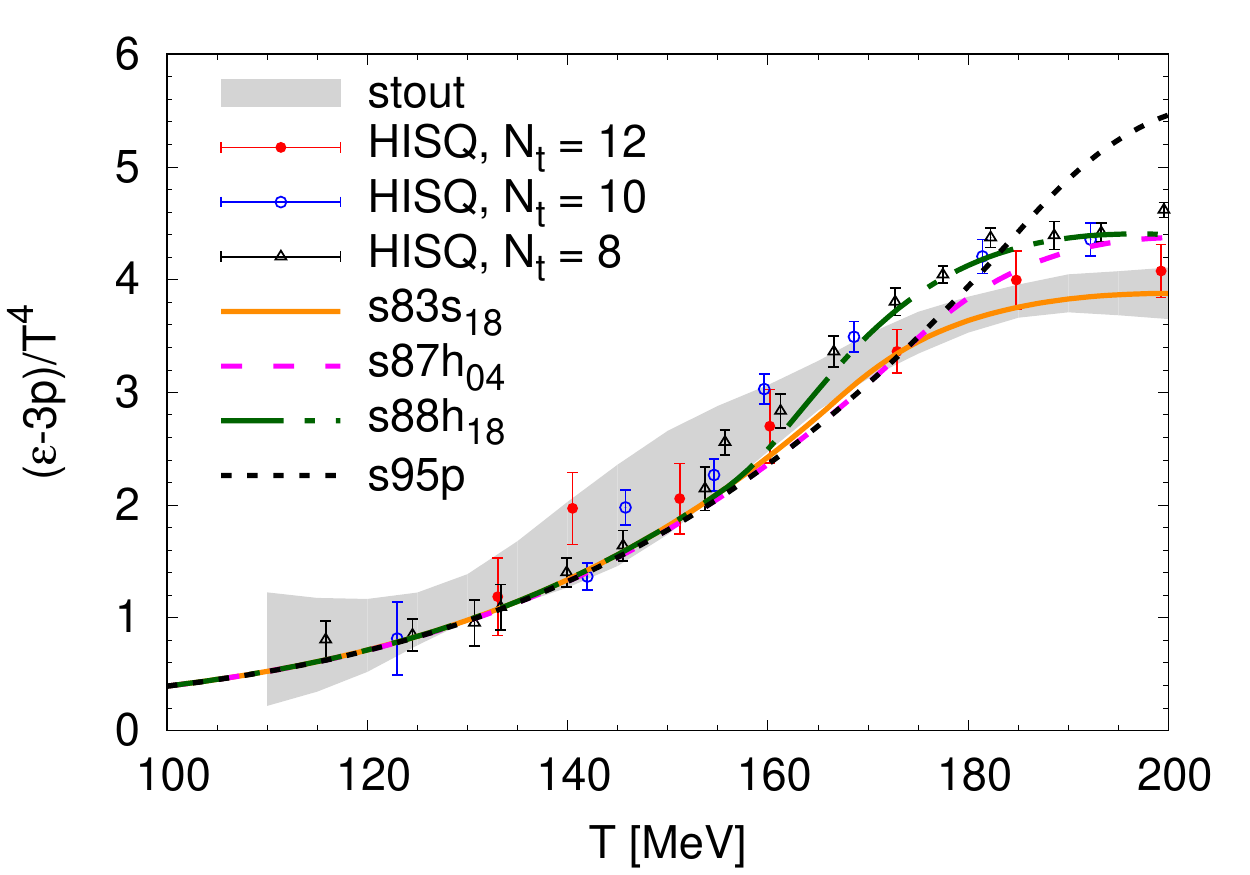}
 \hspace*{-5mm}\includegraphics[width=85mm]{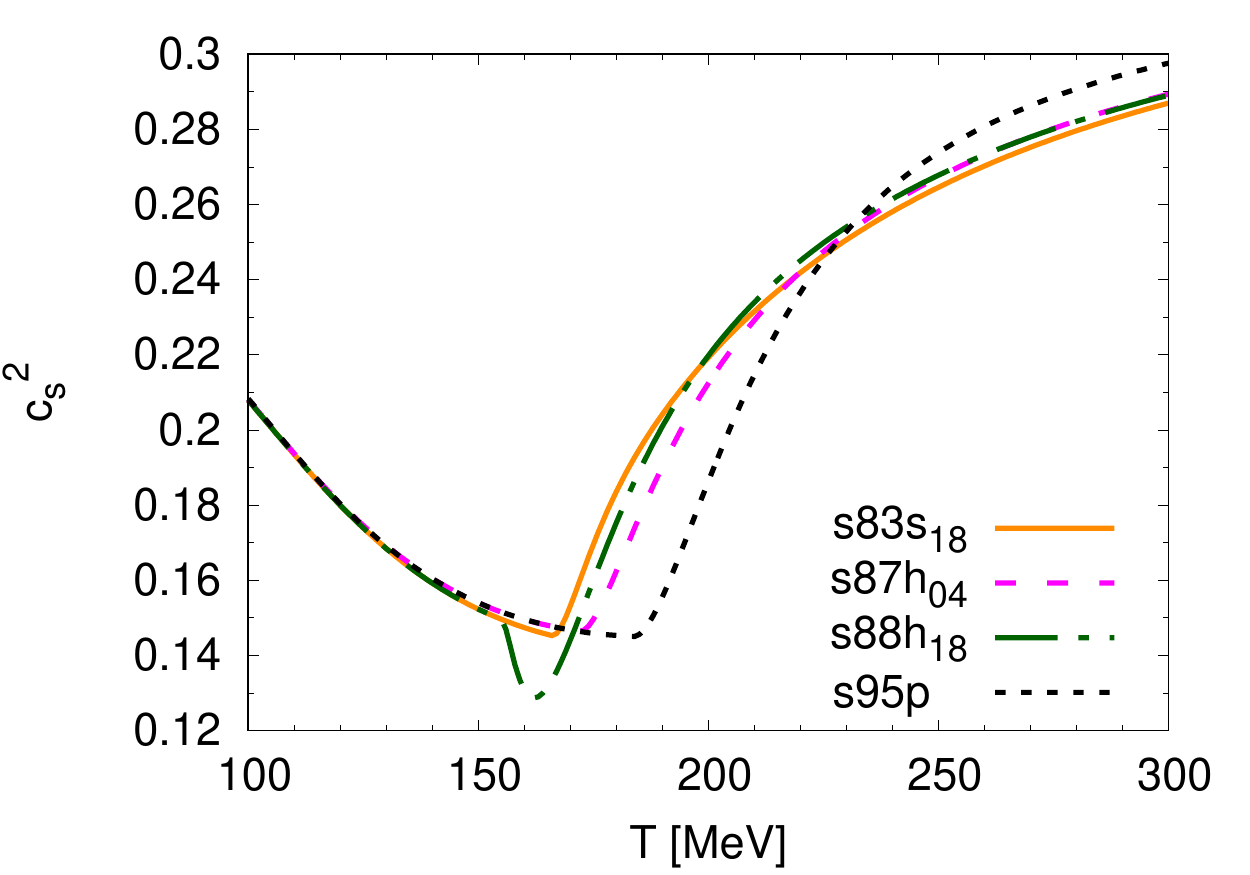}
 \caption{The trace anomaly (top and middle) and the speed of sound
   squared (bottom) as functions of temperature in the four
   parametrizations of the EoS compared to the lattice data obtained
   using the HISQ~\cite{Bazavov:2014pvz,Bazavov:2017dsy} and
   stout~\cite{Borsanyi:2013bia,Borsanyi:2010cj} discretization
   schemes.}
 \label{fig:eos}
\end{figure}

In the top and middle panels of Fig.~\ref{fig:eos}, we show the
parametrized trace anomalies, and the lattice data as used to make
them: continuum extrapolated for the stout action, and at fixed
lattice spacing for the HISQ
action, since its lattice spacing ($N_t$) dependence is small. As seen
in the topmost panel, the most noticable change in the lattice results
during the last decade is the reduction of the peak of the trace
anomaly (cf.~$s95p$ to others). Also, as mentioned, the lattice
results obtained using the HISQ and stout actions slightly differ
around the peak, and consequently $s83s_{18}$ differs from $s87h_{04}$
and $s88h_{18}$. The higher peak does not, however, mean a lower speed
of sound. As shown in the lowest panel of Fig.~\ref{fig:eos}, the
speed of sound in the $s95p$ parametrization is not significantly
lower than in the other parametrizations, but the temperature region
where it is low is broader than in the other parametrizations. Thus we
expect $s95p$ to be effectively softer than the other EoSs. On the
other hand, the speed of sound in $s88h_{18}$ depicts a characteristic
dip below the speed of sound in the other parametrizations. This is a
consequence of the parametrization of the trace anomaly in that
temperature region.

As known, the HRG trace anomaly is below the lattice
results~\cite{Borsanyi:2010cj,Borsanyi:2013bia,Bazavov:2014pvz} at low
temperatures. This difference has been interpreted to indicate the
existence of yet unobserved resonance states~\cite{Majumder:2010ik,
  Huovinen:2018ziu}. The need for further states has also been seen
in the study of the strangeness baryon correlations on the
lattice~\cite{Bazavov:2014xya}, and confirmed by the S-matrix based
virial expansion~\cite{Fernandez-Ramirez:2018vzu}. However, we do not
include predicted states from any model\footnote{As done in
  e.g.~Refs.~\cite{Huovinen:2018ziu, Alba:2020jir}.} in this work,
since we do not know how they would decay, but use the states from the
PDG summary tables only. Consequently the parametrized trace anomaly
is slightly below even the most generous error bars of the lattice
results around $T\approx 150$--160 MeV temperature, as shown in the
middle panel of Fig.~\ref{fig:eos}.

On the other hand, whether we use the PDG 2004 or 2018 particle list
causes only a tiny difference in the trace anomaly. The main
difference between the $s87h_{04}$ and $s88h_{18}$ parametrizations arises from
the connection of the HRG to the lattice parametrization. When
parametrizing $s88h_{18}$ we wanted the trace anomaly to reach its lattice
values soon above $T_c=155$ MeV, whereas we allowed $s87h_{04}$ to agree
with lattice at larger temperature where the lattice trace anomaly
drops below the HRG trace anomaly---for details, see
Appendix~\ref{appx:para}. Consequently the $s88h_{18}$ parametrization
rises above the HRG values leading to the characteristic dip in the
speed of sound (lowest panel in Fig.~\ref{fig:eos}). Note that the
$s83s_{18}$ parametrization does not depict a similar dip in the speed of
sound, since the lower peak and larger errors of the continuum
extrapolated stout action result allow the parametrized trace anomaly
to drop below the HRG values immediately.

\section{Hydrodynamical model}
\label{sec:model}

We employ a fluid dynamical model used previously in
Refs.~\cite{Niemi:2012ry, Niemi:2011ix, Niemi:2015qia, Niemi:2015voa,
  Eskola:2017bup}. The spacetime evolution is computed numerically in
(2+1) dimensions~\cite{Molnar:2009tx}, and the longitudinal expansion
is accounted for by assuming longitudinal boost invariance. We also
neglect here the bulk viscosity and the small net-baryon number. The
evolution of the shear-stress tensor $\pi^{\mu\nu}$ is described by
the second-order Israel-Stewart formalism~\cite{Israel:1979wp}, with
the coefficients of the non-linear second-order terms obtained by
using the 14-moment approximation in the ultrarelativistic
limit~\cite{Denicol:2012cn, Molnar:2013lta}. The shear relaxation time
is related to the shear viscosity by $\tau_{\pi} = 5\eta/(\epsilon + p)$,
where $\epsilon$ is energy density in the local rest frame, and $p$ is
thermodynamic pressure.

Transverse momentum spectra of hadrons are computed by using the
Cooper-Frye freeze-out formalism at a constant-temperature surface,
followed by all 2- and 3-body decays of unstable hadrons. The chemical
freeze-out is encoded into the EoS as described in Ref.~\cite{Huovinen:2007xh},
and the fluid evolves from chemical to kinetic freeze-out in partial
chemical equilibrium (PCE)\cite{Hirano:2002ds}. The kinetic and chemical freeze-out
temperatures $\Tdec$ and $\Tchem$ are left as free parameters to be
determined from the experimental data through the Bayesian analysis.
The dissipative corrections $\delta f$ to the momentum distribution at
the freeze-out are computed according to the usual 14-moment approximation
$\delta f_{\mathbf{k}} \propto f_{0\mathbf{k}} k^{\mu} k^{\nu} \pi_{\mu\nu}$,
where $f_{0\mathbf{k}}$ is the equilibrium distribution function, and
$k^{\mu}$ is the four-momentum of the hadron.

The remaining input to fluid dynamics are the EoS, initial conditions,
and the shear viscosity. The different options for EoS were discussed
in the previous section, and the initial conditions will be detailed
in the next section. The temperature dependence of the shear viscosity
$\eta/s$ is parametrized in three parts, controlled by $\Tmin$, the
lower bound of the temperature range where $\eta/s$ has its minimum
value, $\etasmin$, and the width of this temperature range, $\Wmin$:
\begin{equation} 
 (\eta/s)(T)= 
\begin{cases}
            \Shg (\Tmin-T) + \etasmin, & T < \Tmin \\
            \etasmin,  & \!\!\!\!\!\!\!\!\!\!\!\!\!\!\! \Tmin \leq T \leq T_{\rm Q} \\
            \Sqgp(T-T_{\rm Q}) + \etasmin, & T > T_{\rm Q},
             \end{cases} 
\end{equation}
where the additional parameters are the linear slopes below $\Tmin$
and above $T_{\rm Q} = \Tmin+\Wmin$, denoted by $\Shg$ and $\Sqgp$,
respectively.

We note that bulk viscosity and chemical non-equilibrium are
related~\cite{Paech:2006st,Dusling:2011fd}. Even if we ignore the bulk
viscosity, some of its effects are accounted for by the fugacities in
a chemically frozen fluid: At temperatures below $\Tchem$ the
isotropic pressure is reduced compared to the equilibrium pressure due
to the different chemical composition. Thus introducing the chemical
freeze-out changes not only the particle yields w.r.t.~evolution in
equilibrium, but similarly to the bulk viscosity, reduces the average
transverse momentum of hadrons too. However, this affects the
evolution only when temperature is below $\Tchem$, and in contrast to
the bulk viscosity, there is e.g.\ no entropy production associated
with the chemical freeze-out and subsequent chemical
non-equilibrium~\cite{Bebie:1991ij}.

Finally, we emphasize that we solve the spacetime evolution from the
hot QGP all the way to the kinetic freeze-out as a single continuous
fluid dynamical evolution. This is different from the hybrid models
used e.g.\ in Refs.~\cite{Bernhard:2016tnd,Bass:2017zyn,Bernhard:2019bmu}
where the evolution below some switching temperature is solved with a
microscopic hadron cascade. The advantage of the fluid dynamical
evolution without a cascade stage is that the transport properties are
continuous in the whole temperature range. Note that in the hybrid
models the switching from fluid dynamics to hadron cascade introduces
an unphysical discontinuity in e.g.\ $\eta/s$ that is $\mathcal{O}(1)$
in the cascade~\cite{Rose:2017bjz}, but $\mathcal{O}(0.1)$ in fluid
dynamical simulations at switching. Another advantage of our approach
is that we can freely parametrize the viscosity in the hadronic matter
too, and constrain it using the experimental data.

\section{Initial conditions}
\label{sec:ini}

The initial energy density profiles are determined using the EKRT
model \cite{Eskola:1999fc, Paatelainen:2012at, Paatelainen:2013eea}
based on the NLO perturbative QCD computation of the transverse
energy, and a gluon saturation conjecture. The latter controls the
transverse energy production through a local semi-hard scale
$p_{\rm sat}(T_A T_A, \sqrt{s_{\mathrm{NN}}}, A, K_{\rm sat})$, where
$T_A(x,y)$ is a nuclear thickness function at transverse location
$(x,y)$. The essential free parameters in the EKRT model are the
proportionality constant $K_{\rm sat}$ in the saturation condition,
and the constant $\beta$ controlling the exact definition of the
minijet transverse energy at NLO \cite{Paatelainen:2012at}. The setup
used here is identical to the one used in Refs.~\cite{Niemi:2015qia,
  Niemi:2015voa, Eskola:2017bup}, where $\beta=0.8$, and $K_{\rm sat}$
is left as a free parameter to be determined from the data. We note
that $K_{\rm sat}$ is independent of the collision energy
$\sqrt{s_{\rm NN}}$ and nuclear mass number $A$, so that once
$K_{\rm sat}$ is fixed the $\sqrt{s_{\rm NN}}$ and $A$ dependence of
the initial conditions is entirely determined from the QCD dynamics of
the EKRT model. With a given $p_{\rm sat}$ the local energy density at
the formation time $\tau_p = 1/p_{\rm sat}$ can be written as
\begin{equation}
 \epsilon(x, y, \tau_p) = \frac{\Ksat}{\pi}\left[p_{\rm sat}(x, y)\right]^4.
\label{eq:EKRT_ed}
\end{equation}
This we further evolve to the same proper time
$\tau_0 = 1/p_{\rm min}$, where $p_{\rm min} = 1$ GeV, at every point
in the transverse plane where $p_{\rm sat} > p_{\rm min}$ by using 0+1
dimensional Bjorken hydrodynamics with the assumption $\epsilon = 3p$.

In the EKRT model, fluctuations in the product of the nuclear
thickness functions, $T_A T_A$, give rise to the event-by-event
fluctuations in the energy density through $p_{\rm sat}$ in
Eq.~(\ref{eq:EKRT_ed}). Moreover, the centrality dependence of the
initial conditions arises from the centrality dependence of $T_A
T_A$. A full treatment of the dynamics in heavy-ion collisions would
take the event-by-event fluctuations into account by evolving each
event separately. However, to make the present study computationally
feasible, we omit the evolution of such fluctuations here; instead,
for each centrality class, we average a large number of these
fluctuating initial states, and compute the fluid dynamical evolution
only for the averaged initial distributions.

The computed energy densities are not linear in $K_{\rm sat}$ nor in
$T_A T_A$, and different averaging procedures can lead to
significantly different event-averaged initial conditions. In the
previous event-by-event EKRT studies~\cite{Niemi:2015qia,
  Niemi:2015voa,Eskola:2017bup} a fair agreement was obtained between
the data and the computed $\sqrt{s_{\rm NN}}$, $A$, and centrality
dependencies of the charged hadron multiplicity. To preserve as much
as possible of this agreement, we average the initial conditions by
averaging the initial entropy distributions: We compute first a large
set of initial energy density profiles using the procedure detailed in
Ref.~\cite{Niemi:2015qia}.  Each of the generated energy density
profiles is converted to an entropy density profile by using the EoS
which will be used later during the evolution. The entropy density
profiles are then averaged, and the average entropy density profile is
converted to an average energy density profile using the same EoS.

In the event-by-event framework the centrality classes were determined
from the final multiplicity distribution. However, this way of
classifying events is not available here, as it would require fluid
dynamical evolution of each of the fluctuating initial conditions.
Instead, we pre-determine the centrality classes according to the
number of wounded nucleons in the sampled Monte-Carlo nuclear
configurations, which were used to construct the event-by-event
initial conditions. The number of wounded nucleons are computed using
the nucleon-nucleon cross section $\sigma_{\rm NN} = 42$, $64$, and $70$
mb for $200$ GeV, $2.76$ TeV, and $5.02$ TeV collisions, respectively.
We note that the nucleon-nucleon cross section does not enter in the
computation of the initial conditions, but they are used here only in
the centrality classification. In the context of the full event-by-event
modeling we have tested that the final results are only weakly
sensitive to the precise way of the centrality classification.

\section{Statistical analysis}
 \label{sec:stats}
 
The eight free parameters of our model,
$\{\Ksat$, $\etasmin,\Tmin,\Wmin,\Shg,\Sqgp,\Tdec,\Tchem\}$, were
introduced in Secs.~\ref{sec:model} and~\ref{sec:ini}. We want to tune
them to achieve the best possible fit to an experimental data set of
90 data points. This set consists of the following observables at
(10--20)\%, (20--30)\%, (30--40)\%, (40--50)\% and (50--60)\%
centrality classes\footnote{Charged particle multiplicities at RHIC
  are averages over two adjacent PHENIX centrality classes; for
  example, at (10--20)\% centrality $N_{\mathrm{ch}}$ is an average over
  (10--15)\% and (15--20)\% classes, (20--30)\% is an average over
  (20--25)\% and (25--30)\% classes, and so on. This applies also for
  RHIC identified particle data at (10--20)\% centrality.}:
\begin{itemize}
 \item The charged particle multiplicity at midrapidity, $\dif
   N_{\ch}/\dif\eta$, and 4-particle cumulant $p_T$-averaged elliptic
   flow, $v_2\{4\}$, in Au+Au collisions at $\sqrt{s_\NN} = 200$ GeV
   (RHIC)~\cite{Adler:2004zn,Adams:2004bi} and Pb+Pb collisions at
   $\sqrt{s_{\NN}}=2.76$~\cite{Aamodt:2010cz,Adam:2016izf} and
   $\sqrt{s_{\NN}}=5.02 $ TeV~\cite{Adam:2016izf,Adam:2015ptt} (LHC).
\item The multiplicities at midrapidity, $\dif N_i/\dif y$, and
  average transverse momenta $\langle p_T\rangle_i$, of pions
  ($\pi^+$), kaons ($K^+$) and protons\footnote{We consider an average
    of measured protons and antiprotons as the target value for the
    proton multiplicity at RHIC.} ($p$) in Au+Au collisions at
  RHIC~\cite{Adler:2003cb} and in Pb+Pb collisions at the lower LHC
  energy~\cite{Abelev:2013vea}.
\end{itemize}

Let us consider each combination of the free parameters as a point $\vec{x}$
in the 8-dimensional input (parameter) space, the model output
$\vec{y}(\vec{x})$ as a corresponding point in the 90-dimensional
output space (space of observables), and the experimental data
$\vec{y}^{\text{\,exp}}$ as the target point in the space of
observables. With these definitions we can formulate the posterior
probability distribution $P(\vec{x}|\vec{y}^{\text{\,exp}})$ of the
best-fit parameter values by utilizing Bayes' theorem:
\begin{equation}
  P(\vec{x}|\vec{y}^{\text{\,exp}})\propto P(\vec{y}^{\text{\,exp}}|\vec{x})P(\vec{x}),
\end{equation}
where $P(\vec{x})$ is the prior probability distribution of input parameters and
$P(\vec{y}^{\text{\,exp}}|\vec{x})$ is the likelihood function
\begin{eqnarray}
\label{eq:likelihood}
\lefteqn{P(\vec{y}^{\text{\,exp}}|\vec{x})} \\
 & = &\frac{1}{\sqrt{|2\pi\Sigma|}}\exp\left(-\frac{1}{2}(\vec{y}(\vec{x})-\vec{y}^{\text{\,exp}})^T\Sigma^{-1}(\vec{y}(\vec{x})-\vec{y}^{\text{\,exp}}) \right).\nonumber
\end{eqnarray}
Here $\Sigma$ is the covariance matrix representing the uncertainties
related to the model-to-data comparison.

As a function with an eight-dimensional domain, the posterior
probability distribution $P(\vec{x}|\vec{y}^{\text{\,exp}})$ is too
complicated to evaluate and analyze fully. Instead, we produce samples
of it with a parallel tempered Markov chain Monte Carlo~\cite{Vousden:2016}
based on the \textsc{emcee} sampler~\cite{ForemanMackey:2012ig}.
An ensemble of random walkers is initialized in the input parameter
space based on the prior probability\footnote{In the present case, the
  shape of the prior is a uniform hypercube with an additional
  restriction $\Tdec < \Tchem$. The prior ranges are shown in
  Figs.~\ref{fig:posterior_ksat} and~\ref{fig:posterior_tmin}.} and
each proposed step in parameter space is accepted or rejected based on
the change in the value of the likelihood function.  At a large number
of steps, the distribution of the taken steps is expected to converge
to the posterior distribution.

Evaluating the output $\vec{y}(\vec{x})$ of the fluid dynamical model
at every point $\vec{x}$ where the random walker might enter is a
computationally impossible task. Therefore we approximate the output
using Gaussian process (GP) emulators \cite{Rasmussen:2006} (see
Appendix~\ref{appx:GP}). Each GP is able to provide estimates for only
one observable, so to keep the number of required emulators
manageable, we perform a principal component analysis (PCA) to reduce
the dimension of the output space from 90 observables into $k=6$ most
important principal components. Further details about the PCA are
described in Appendix~\ref{appx:PCA}. We utilize the
\texttt{scikit-learn} Python module \cite{Pedregosa:2012toh} and in
particular the submodules \texttt{sklearn.gaussian\_process} and
\texttt{sklearn.decomposition.PCA} in the model emulation.

Thus, in our likelihood function \eqref{eq:likelihood}, we replace
$\vec{y}(\vec{x})$ with the GP estimate in the principal component
space $\vec{z}^{\,\text{GP}}(\vec{x})$ (likewise $\vec{y}^{\text{\,exp}}$
is transformed to $\vec{z}^{\text{\,exp}}$), and include the emulator
estimation error into the covariance matrix:
\begin{equation}
\Sigma_z=\Sigma_{z}^{\text{exp}}+\Sigma_z^{\text{GP}},
\end{equation}
where $\Sigma_{z}^{\text{exp}}$ is the (originally diagonal) experimental
error matrix transformed to principal component space, and
\begin{equation}
\Sigma_z^{\text{GP}}=\text{diag}(\sigma_{z,1}^{\text{GP}}(\vec{x})^2,\sigma_{z,2}^{\text{GP}}(\vec{x})^2,...,\sigma_{z,k}^{\text{\,GP}}(\vec{x})^2)
\end{equation}
is the GP emulator covariance matrix obtained from the emulator
(see Appendix~\ref{appx:GP}).

\begin{figure}[h]
 \centering
 \includegraphics[width=7cm]{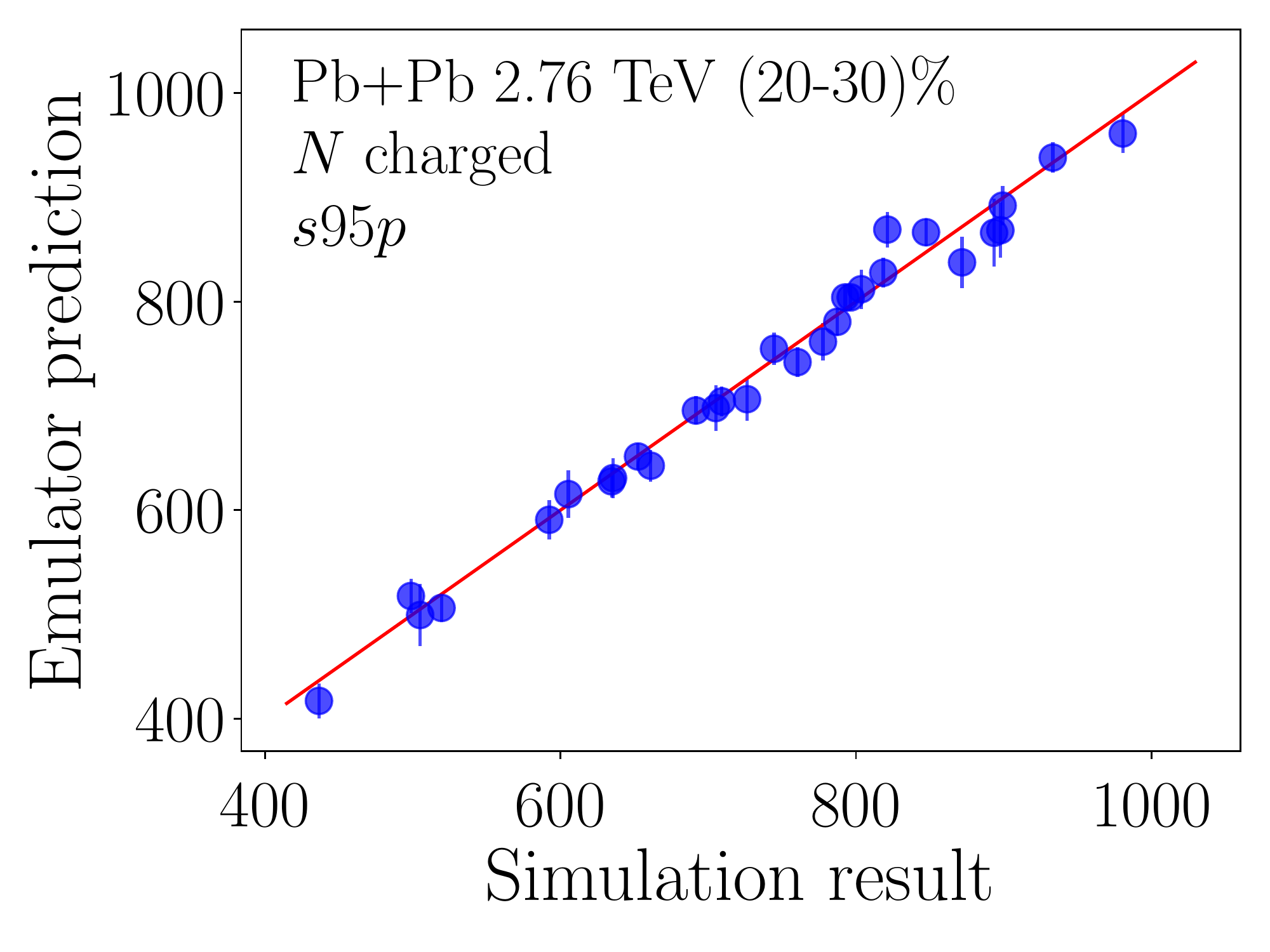}
 \includegraphics[width=7cm]{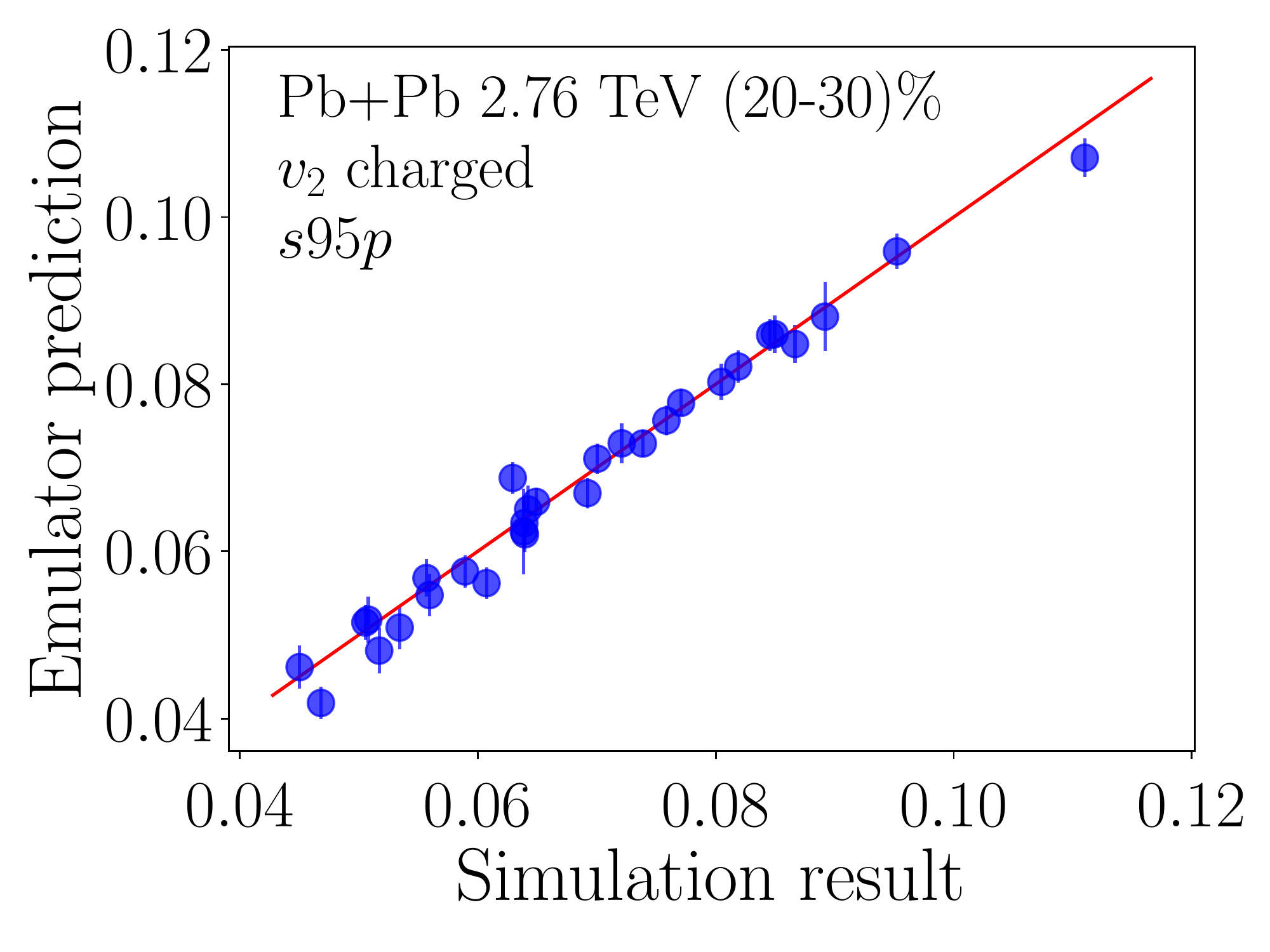}
 \caption{Illustration of the quality of the Gaussian process emulation for 30 test points
 for simulations with the $s95p$ EoS.
 Upper panel: Charged particle multiplicity in (20-30)\% most central Pb+Pb collisions at $\sqrt{s_{\NN}}=2.76$ TeV.
 Lower panel: Elliptic flow $v_2\{\text{RP}\}$ in (20-30)\% most central Pb+Pb collisions at $\sqrt{s_{\NN}}=2.76$ TeV.}
 \label{fig:verification}
\end{figure}

To work, the GP emulators must be conditioned with a set of training
points, $\{\vec{z}(\vec{x}_i)\}$, created by running the fluid
dynamical model with several different parameter combinations
$\{\vec{x}_i\}$. For the present investigation, we have produced 170
training points for each EoS, distributed evenly in the input
parameter space\footnote{The restriction $\Tdec < \Tchem$ does not
  apply to the training points.}  using minmax Latin hypercube
sampling~\cite{pyDOE:LHS}. The emulation quality was then checked by
using the trained emulator to predict the results at 30 additional
test points, which were not part of the training data.  An example of
the results of this confirmation process is shown in
Fig.~\ref{fig:verification} for 2.76 TeV Pb+Pb collisions using the
$s95p$ parametrization.

\section{Results}
   \label{sec:results}

\begin{figure*}[h]
 \centering
 \includegraphics[width=6.5cm]{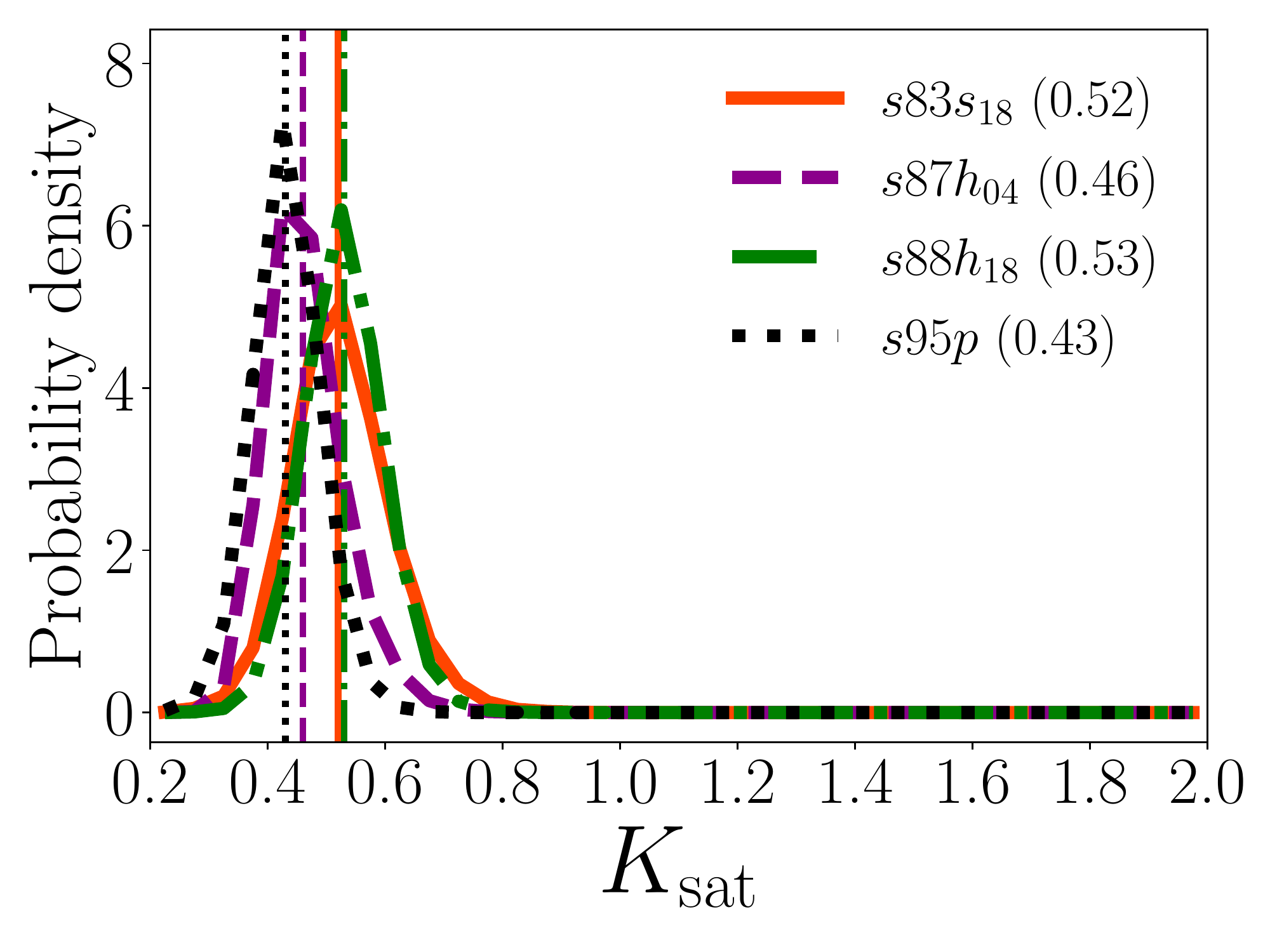}
 \includegraphics[width=6.5cm]{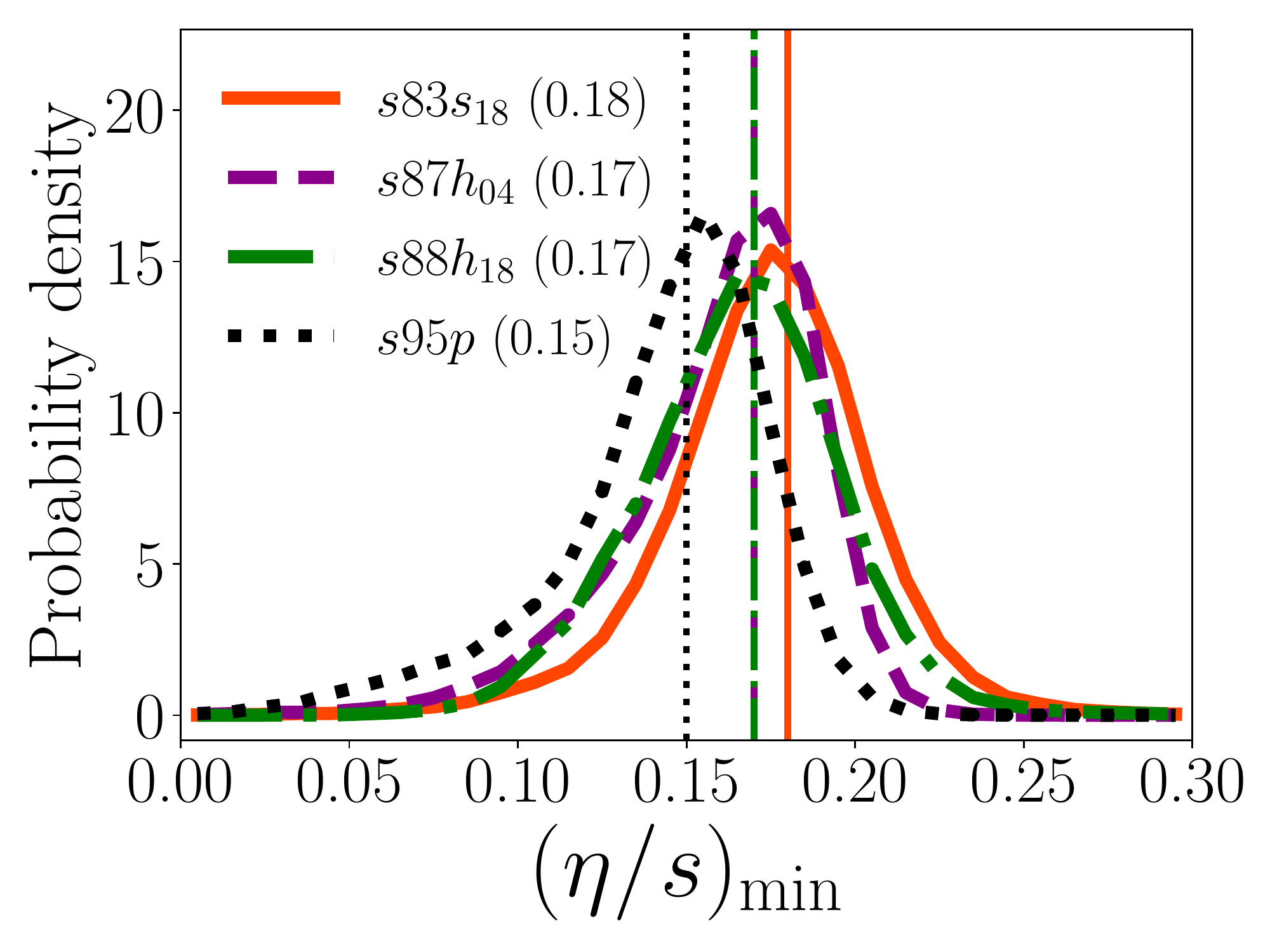}
 \includegraphics[width=6.5cm]{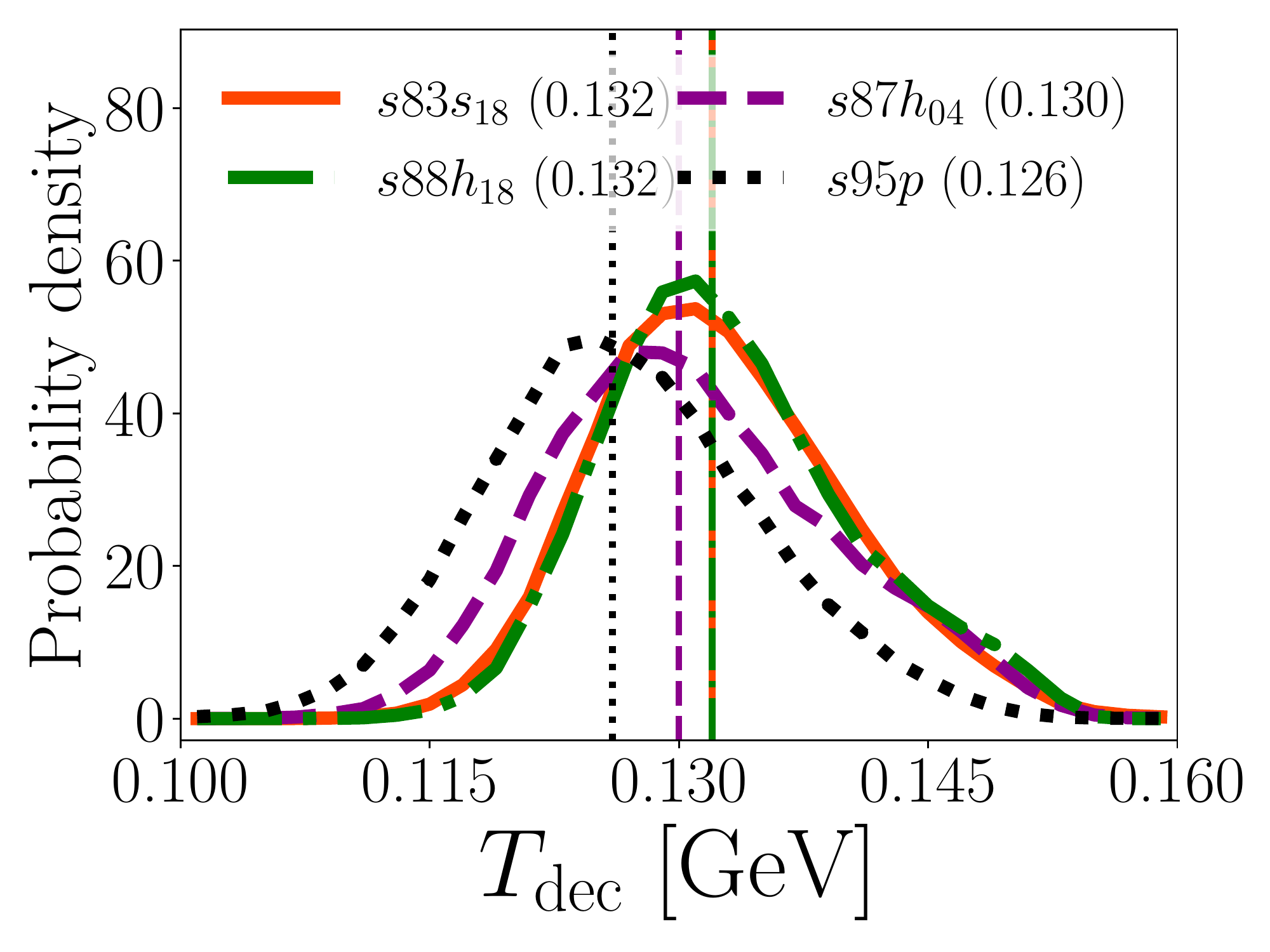}
 \includegraphics[width=6.5cm]{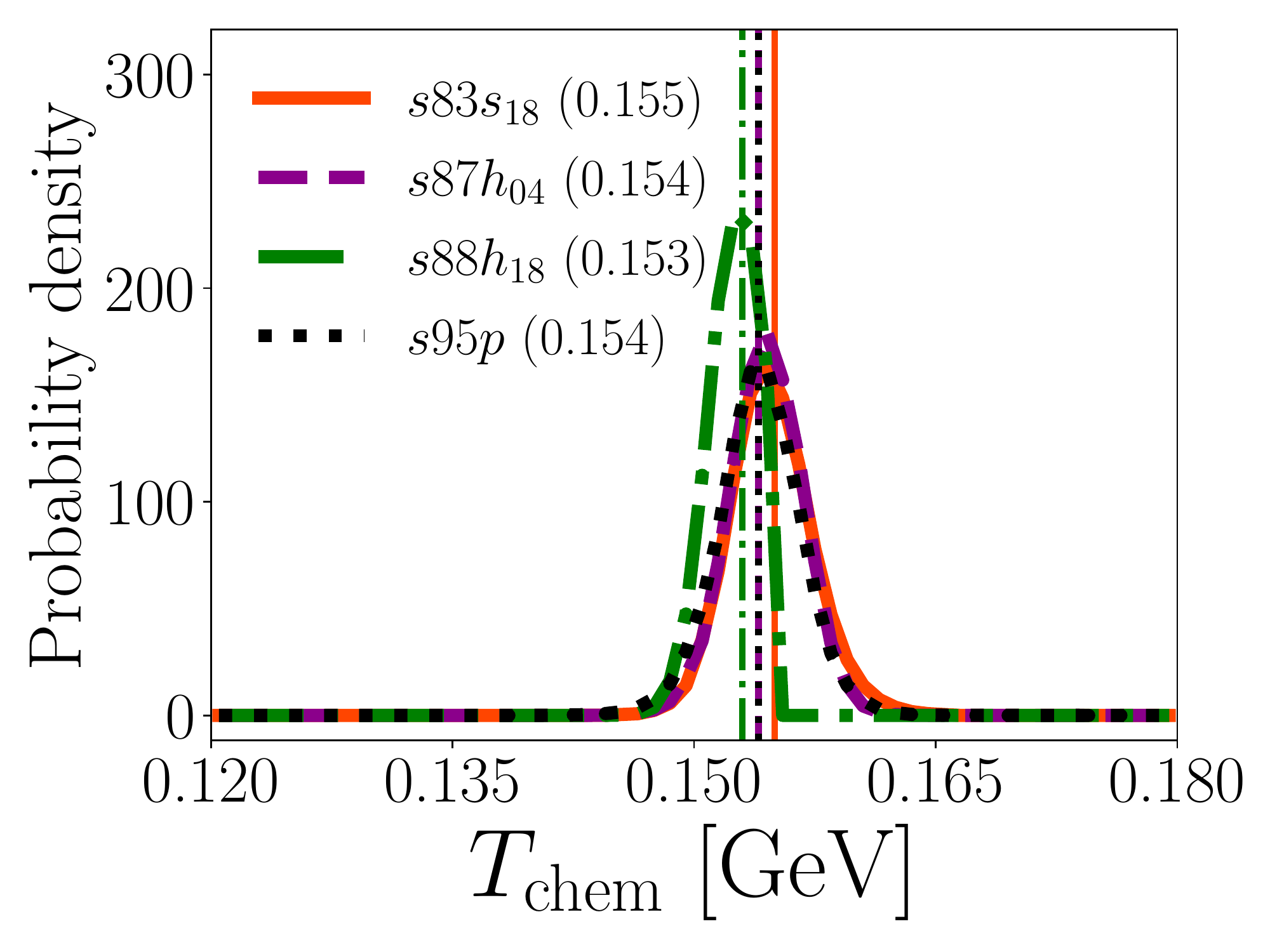}
 \caption{Comparison of $\Ksat$, $\etasmin$, $\Tdec$, and $\Tchem$
   marginal posterior probability distributions for the four
   investigated EoSs.  Vertical lines (bracketed numbers in legend)
   indicate median values for the distributions. With the exception of
   $\Tchem$, the range of the x-axis in the plots is the original
   prior range.}
 \label{fig:posterior_ksat}
\end{figure*}

\begin{figure*}[h]
 \centering
 \includegraphics[width=6.5cm]{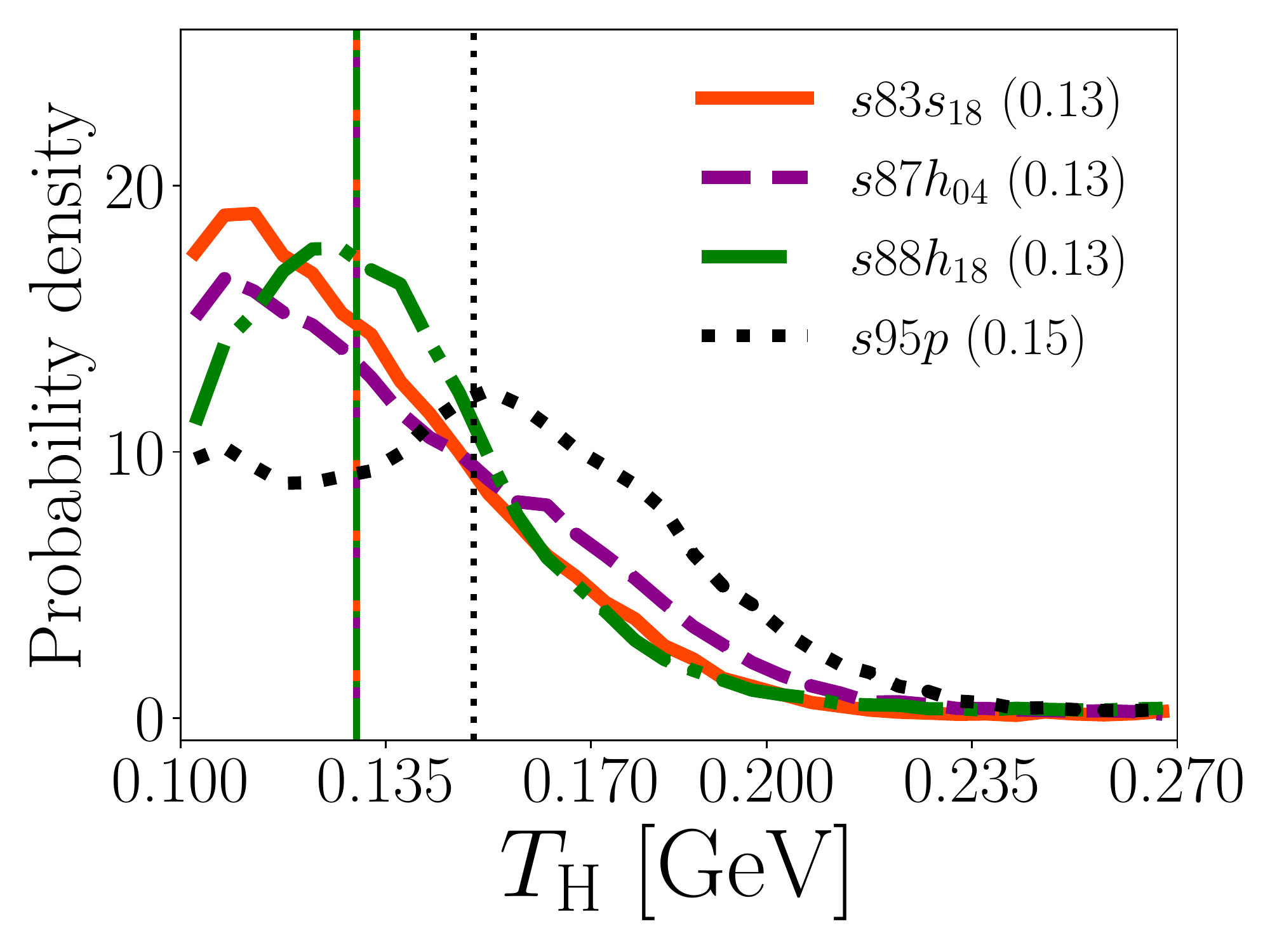}
 \includegraphics[width=6.5cm]{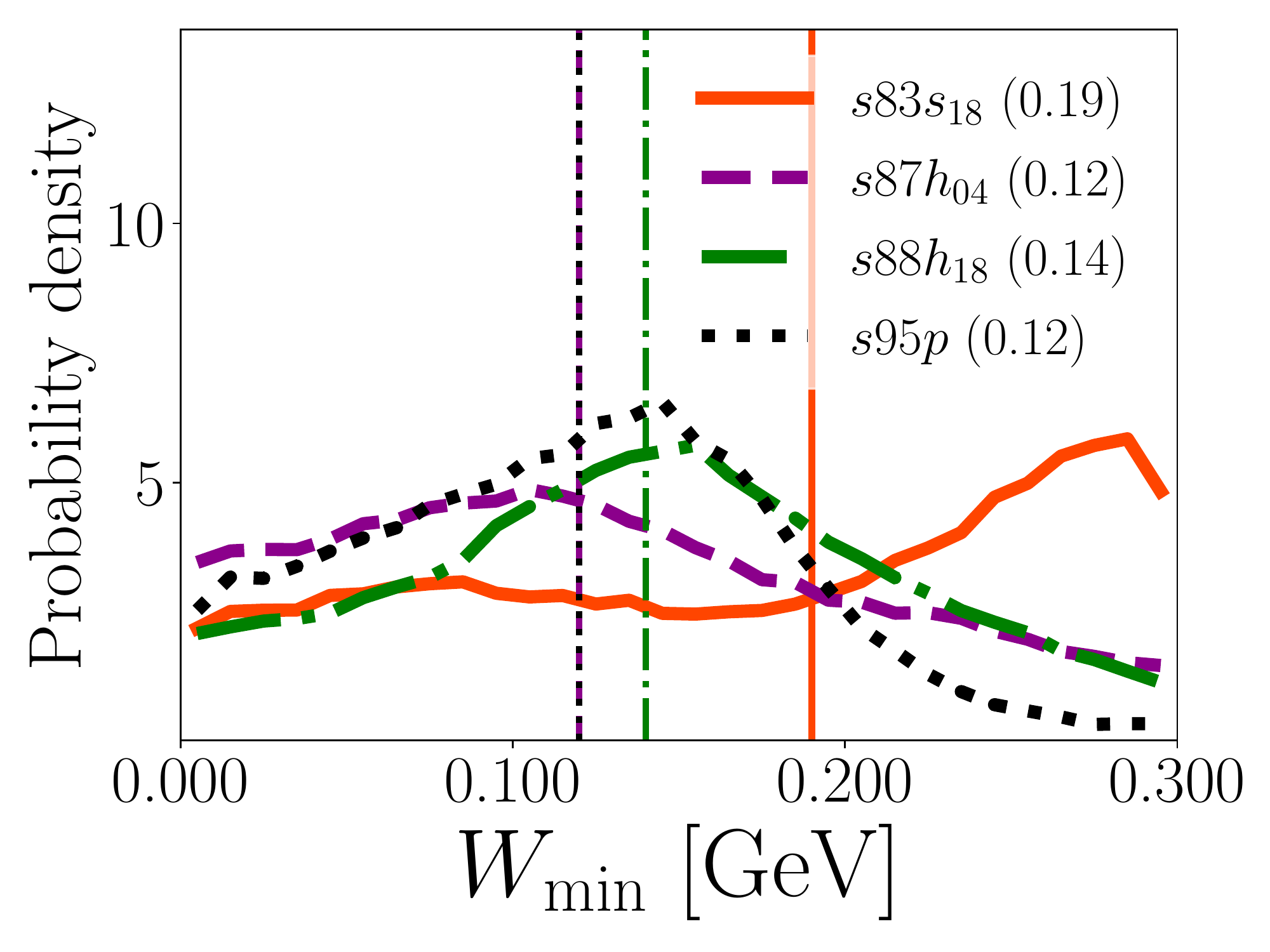}
 \includegraphics[width=6.5cm]{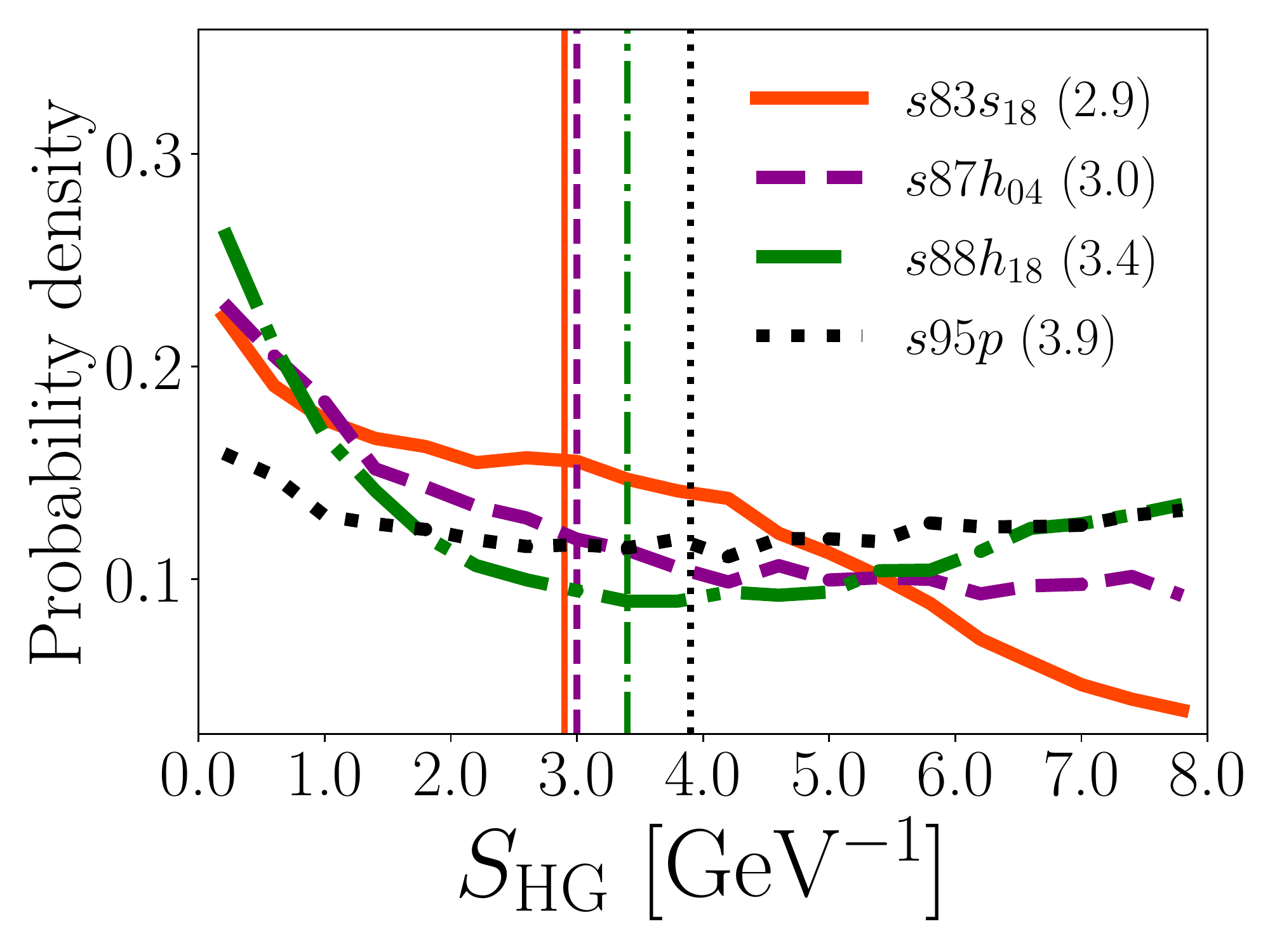}
 \includegraphics[width=6.5cm]{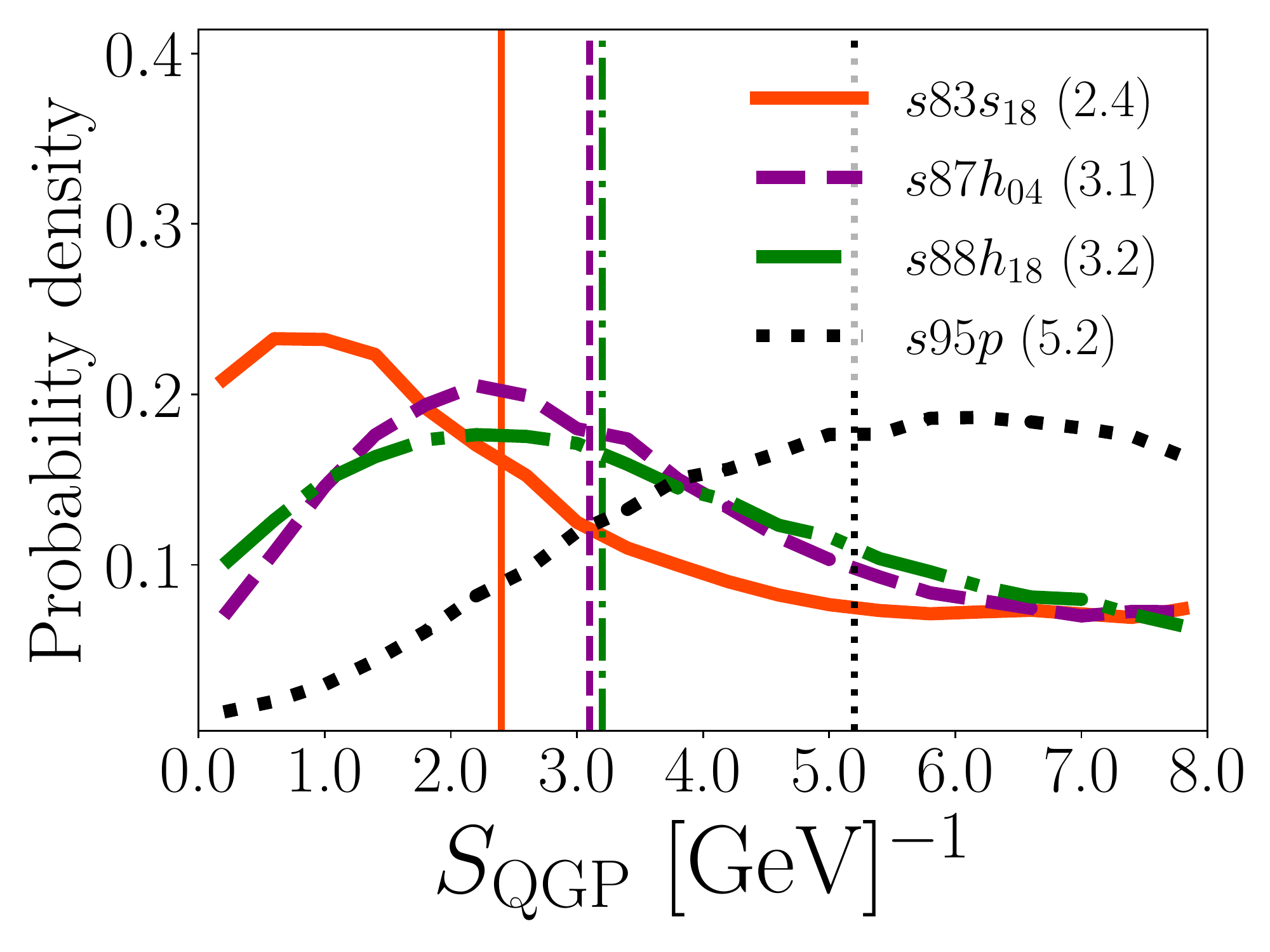}
 \caption{As Fig.~\ref{fig:posterior_ksat}, but showing the marginal
   posterior probability distributions of $\Tmin$, $\Wmin$, $\Shg$, and $\Sqgp$.}
 \label{fig:posterior_tmin}
\end{figure*}

The marginal posterior probability distributions for each parameter
are obtained from the full 8-dimensional probability distribution (see
Section \ref{sec:stats}) by integrating over the other seven
parameters. The resulting distributions when using the four
investigated EoSs are shown in Figs.~\ref{fig:posterior_ksat}
and~\ref{fig:posterior_tmin}. In these figures the range of the
x-axis illustrates the prior range of values, with the exception of
$\Tchem$ which range depends on the EoS\footnote{For $s83s_{18}$,
  $s87h_{04}$, and $s88h_{18}$, the prior range is $120\ \mathrm{MeV} < \Tchem < T_0$, where
  $T_0$ is the temperature where the parametrization deviates from the
  HRG (see Appendix~\ref{appx:para}). For $s95p$ the range is
  $120 < \Tchem/\mathrm{MeV} < 180$.}. The median values of these
distributions provide a good approximation for the most probable
values, and are listed both in the legends of the figures, and in
Table~\ref{tab:param}. The 90\% credible intervals---i.e.~the range
which covers 90\% of the distribution around the median---are shown as
errors in Table~\ref{tab:param}. Two dimensional projections of the
probability distribution depicting correlations between parameter
pairs are shown in Appendix~\ref{appx:posterior}.

\subsection{Nuisance parameters}

The analysis involves three parameters which are not directly related
to the transport properties of produced matter: $\Ksat$, $\Tdec$, and
$\Tchem$. The probability distributions for these three "nuisance"
parameters, shown in Fig.~\ref{fig:posterior_ksat}, are nicely peaked,
and the parameters have well defined constraints. For the chemical
freeze-out temperature, the median is $\Tchem = 153$--$155$ MeV, which
is compatible with the values obtained using the statistical
hadronization model~\cite{Andronic:2017pug}. Note that the difference
in the median is not due to the resonance content of the EoS, but due
to a complicated interplay of the softness of the EoS, shear, and
build-up of the flow. Nevertheless, the particle ratios are the
dominant factor in constraining $\Tchem$.

For $\Ksat$ and $\Tdec$, we see a common trend where $s95p$ gives a
distribution which peaks at the lowest value of the four EoSs,
followed by $s87h_{04}$, and the highest peak values are shared by
$s83s_{18}$ and $s88h_{18}$ with almost identical distributions. The
obtained values for the EKRT normalization parameter, $\Ksat\approx
0.5$, are compatible with the values found
previously~\cite{Niemi:2015qia}, and the small differences between
different EoS parametrizations result from slightly different entropy
production during the evolution. Differences seen in the kinetic
freeze-out temperature $\Tdec = 126$--$132$ MeV are also small, and
seem to follow the conventional rule of thumb: a softer EoS requires a
lower freeze-out temperature to create hard enough proton
$p_\mathrm{T}$ distributions. On the other hand, differences in the
median values of all these three parameters are smaller than the
credibility intervals, and thus not statistically meaningful.

\begin{table} \renewcommand{\arraystretch}{1.3}
   \caption{Estimated parameter values (medians) and uncertainties (90\% 
credible intervals) from the posterior distributions.}
\begin{tabular}{l@{\hspace{1mm}}l|r@{}l@{\hspace{3mm}}r@{}l@{\hspace{3mm}}r@{}l@{\hspace{3mm}}r@{}l}
     \hline
     \hline
     \multicolumn{2}{l|}{Parameter} & \multicolumn{2}{c}{$s83s_{18}$} & \multicolumn{2}{c}{$s87h_{04}$} & \multicolumn{2}{c}{$s88h_{18}$} & 
\multicolumn{2}{c}{$s95p$}  \\
     \hline
   $\Ksat$ & & $0.52$ & $^{+0.15 }_{-0.12 }$ & $0.46$ & $^{+0.12 }_{-0.09 }$ & $0.53$ & $^{+0.11 }_{-0.10 }$ & $0.43$ & $^{+0.10 }_{-0.09 }$ \\
   $\etasmin$ & & $0.18$ & $^{+0.04 }_{-0.06 }$ & $0.17$ & $^{+0.03 }_{-0.07 }$ & $0.17$ & $^{+0.04 }_{-0.06 }$ & $0.15$ & $^{+0.03 }_{-0.07 }$ \\
   $T_H $ & [GeV] & $0.13$ & $^{+0.05 }_{-0.03 }$ & $0.13$ & $^{+0.06 }_{-0.03 }$ & $0.13$ & $^{+0.06 }_{-0.03 }$ & $0.15$ & $^{+0.06 }_{-0.04 }$ \\
   $\Wmin$ & [GeV] & $0.19$ & $^{+0.10 }_{-0.17 }$ & $0.12$ & $^{+0.15 }_{-0.11 }$ & $0.14$ & $^{+0.13 }_{-0.12 }$ & $0.12$ & $^{+0.10 }_{-0.10 }$ \\
   $\Shg$  & [GeV$^{-1}$] & $2.9$ & $^{+4.0 }_{-2.7 }$ & $3.0$ & $^{+4.5 }_{-2.8 }$ & $3.4$ & $^{+4.2 }_{-3.2 }$ & $3.9$ & $^{+3.7 }_{-3.6 }$ \\
   $\Sqgp$ & [GeV$^{-1}$] & $2.4$ & $^{+4.9 }_{-2.1 }$ & $3.1$ & $^{+4.2 }_{-2.5 }$ & $3.2$ & $^{+4.1 }_{-2.7 }$ & $5.2$ & $^{+2.5 }_{-3.5 }$ \\
   $\Tdec$ & [MeV] & $132$ & $^{+14 }_{-11 }$ & $130$ & $^{+16 }_{-12 }$ & $132$ & $^{+15 }_{-10 }$ & $126$ & $^{+15 }_{-12 }$ \\
   $\Tchem$ & [MeV] & $155$ & $^{+4 }_{-4 }$ & $154$ & $^{+4 }_{-3 }$ & $153$ & $^{+2 }_{-3 }$ & $154$ & $^{+4 }_{-4 }$ \\
     \hline
     \hline
   \end{tabular}
   \label{tab:param}
\end{table}

\subsection{$(\eta/s)(T)$}

\begin{figure}[h]
 \centering
 \includegraphics[width=8.5cm]{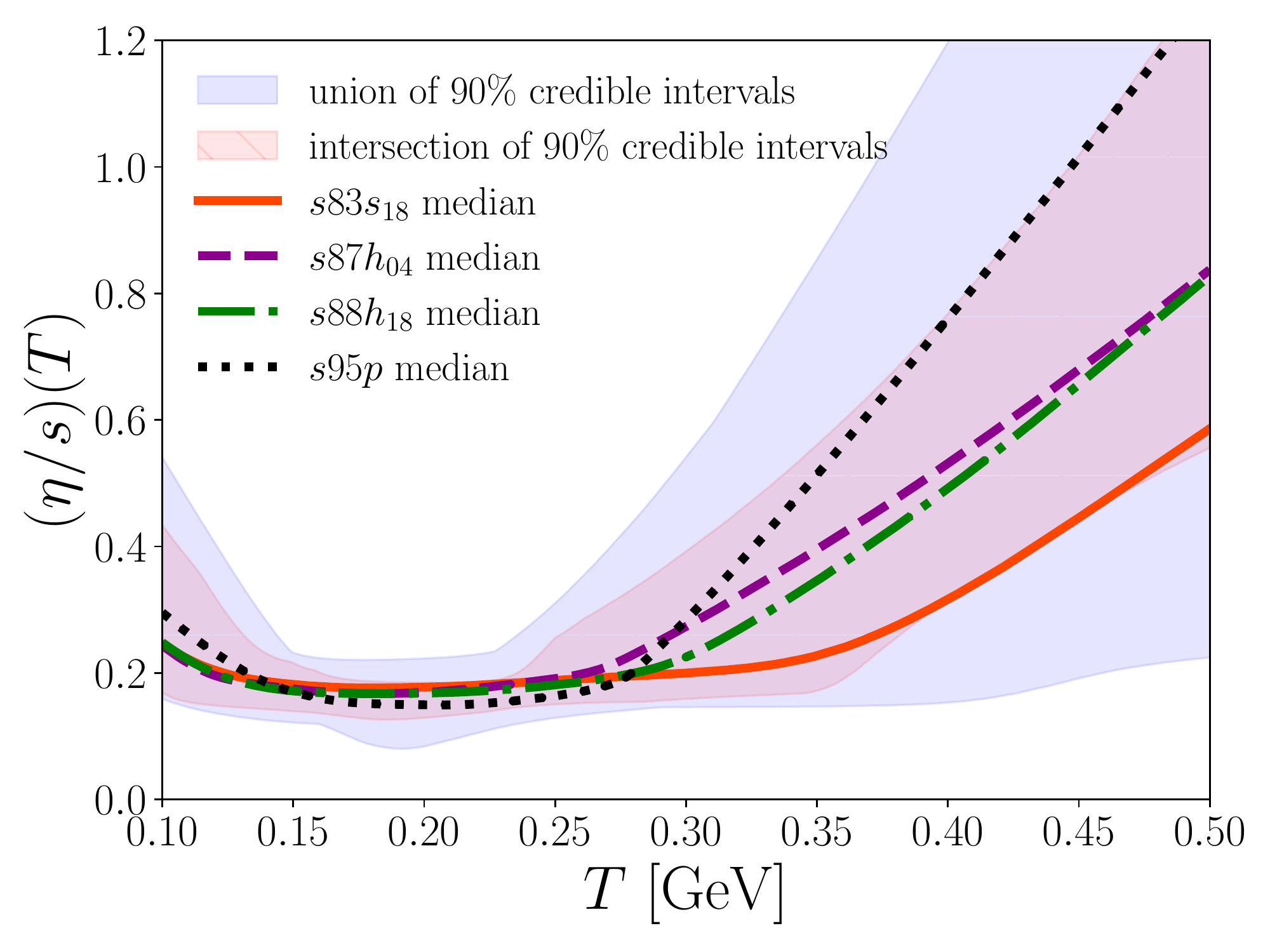}
 \includegraphics[width=8.5cm]{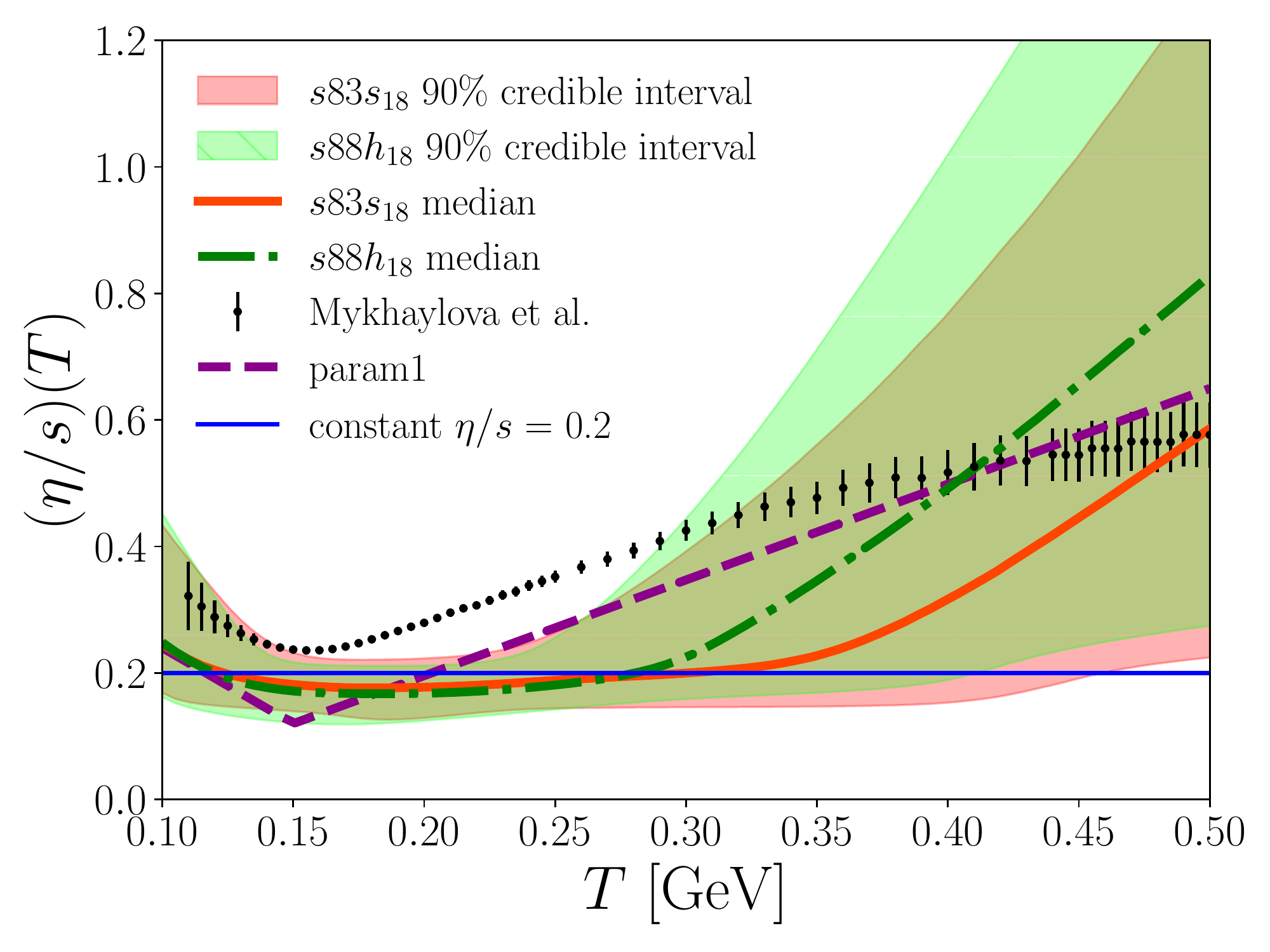}
 \caption{Temperature dependence of $\eta/s$.
   Upper figure: Median of $\eta/s$ w.r.t.~$T$ for each EoS with
   the union and intersection of the 90\% credible intervals of the
   distributions.
   Lower figure: Median of $\eta/s$ w.r.t.~$T$ for the $s83s_{18}$ and
   $s88h_{18}$ parametrizations with corresponding credibility intervals
   compared with two results from Ref.~\cite{Niemi:2015qia}
   ($\eta/s = 0.2$ and param1) and a recent quasiparticle model prediction
   by Mykhaylova \emph{et al.}~\cite{Mykhaylova:2019wci}.}
 \label{fig:etas_vs_t}
\end{figure}

\begin{figure}[t]
 \centering
 \includegraphics[width=4.2cm]{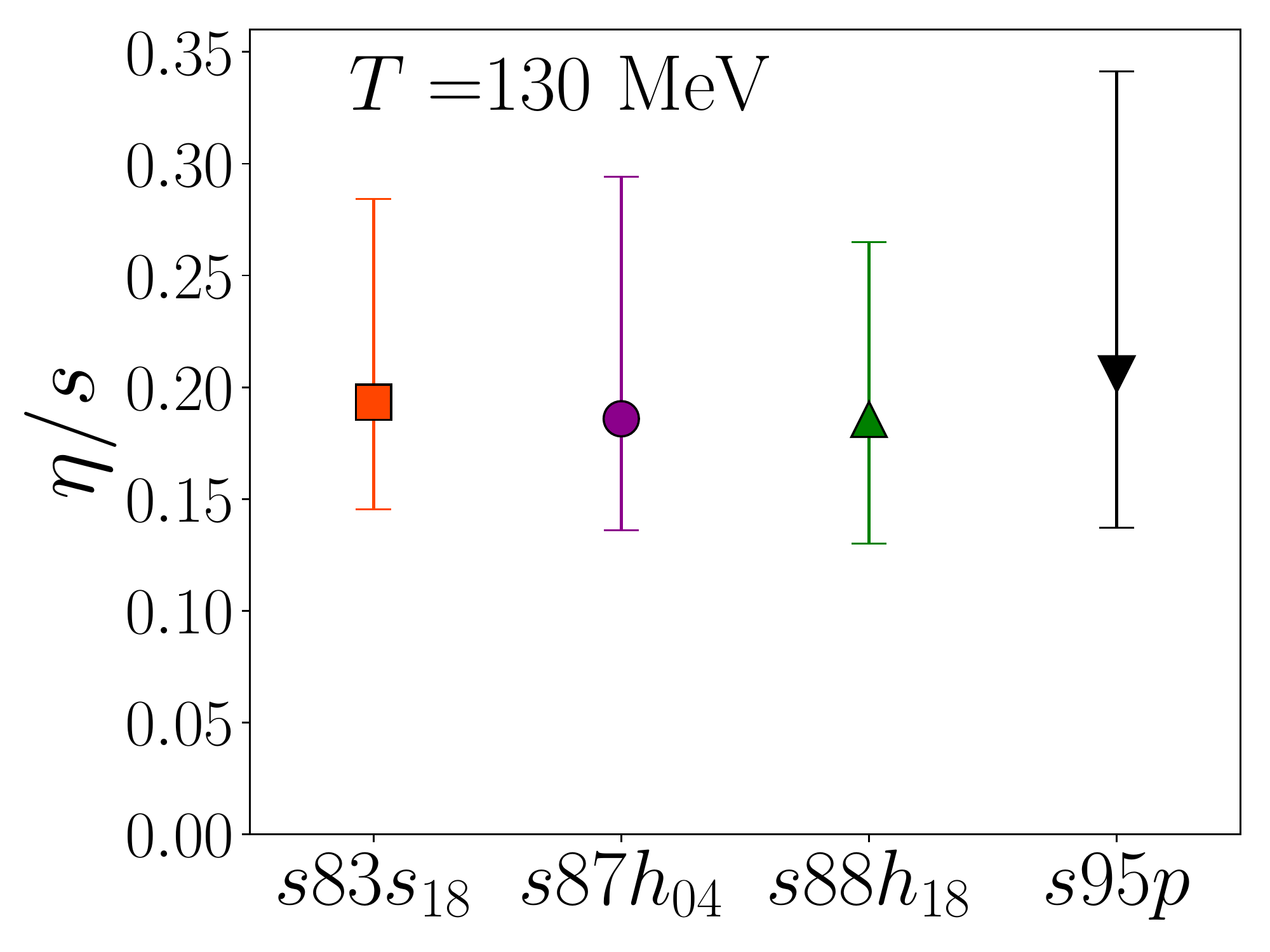}
 \includegraphics[width=4.2cm]{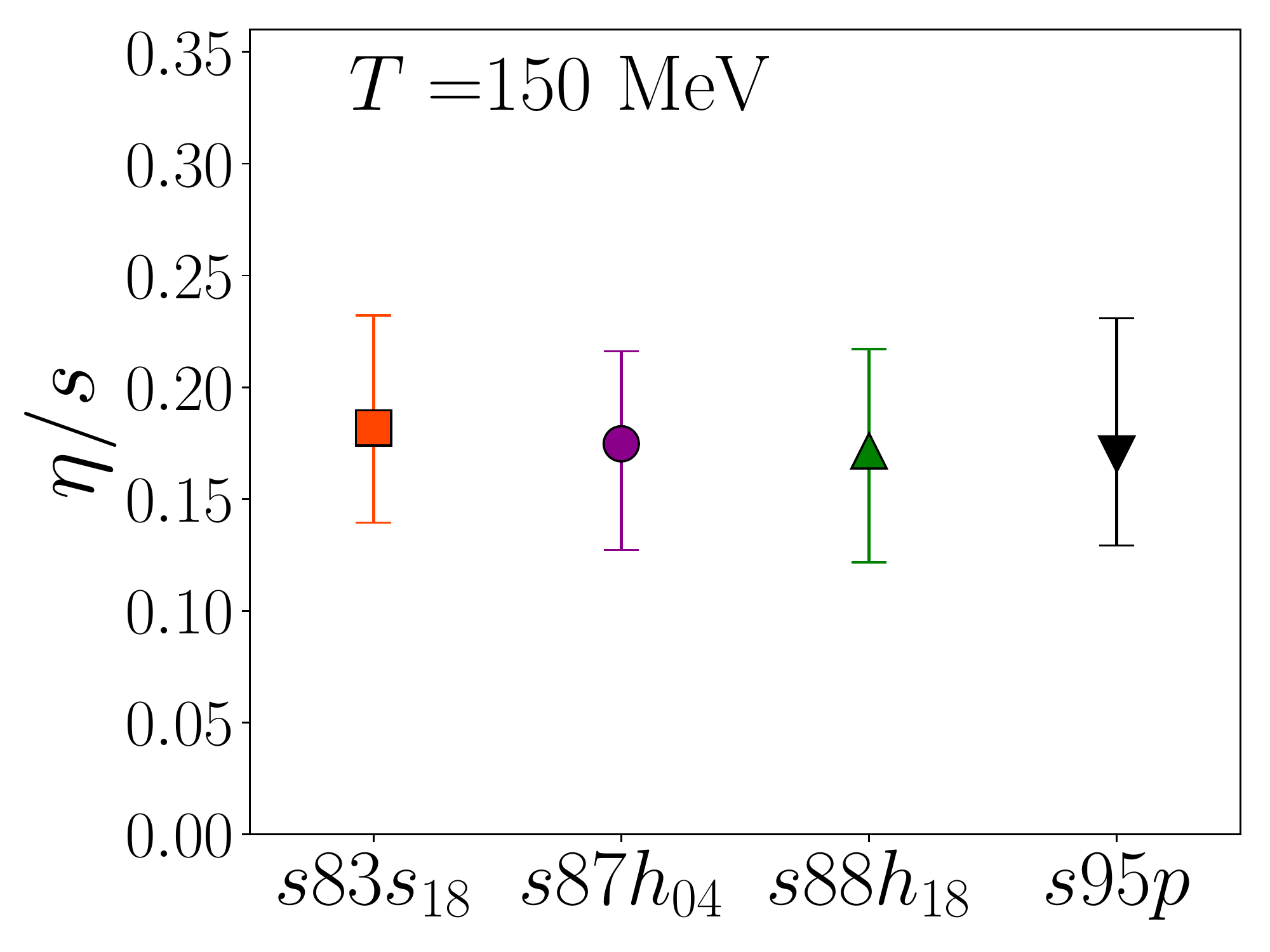}
 \includegraphics[width=4.2cm]{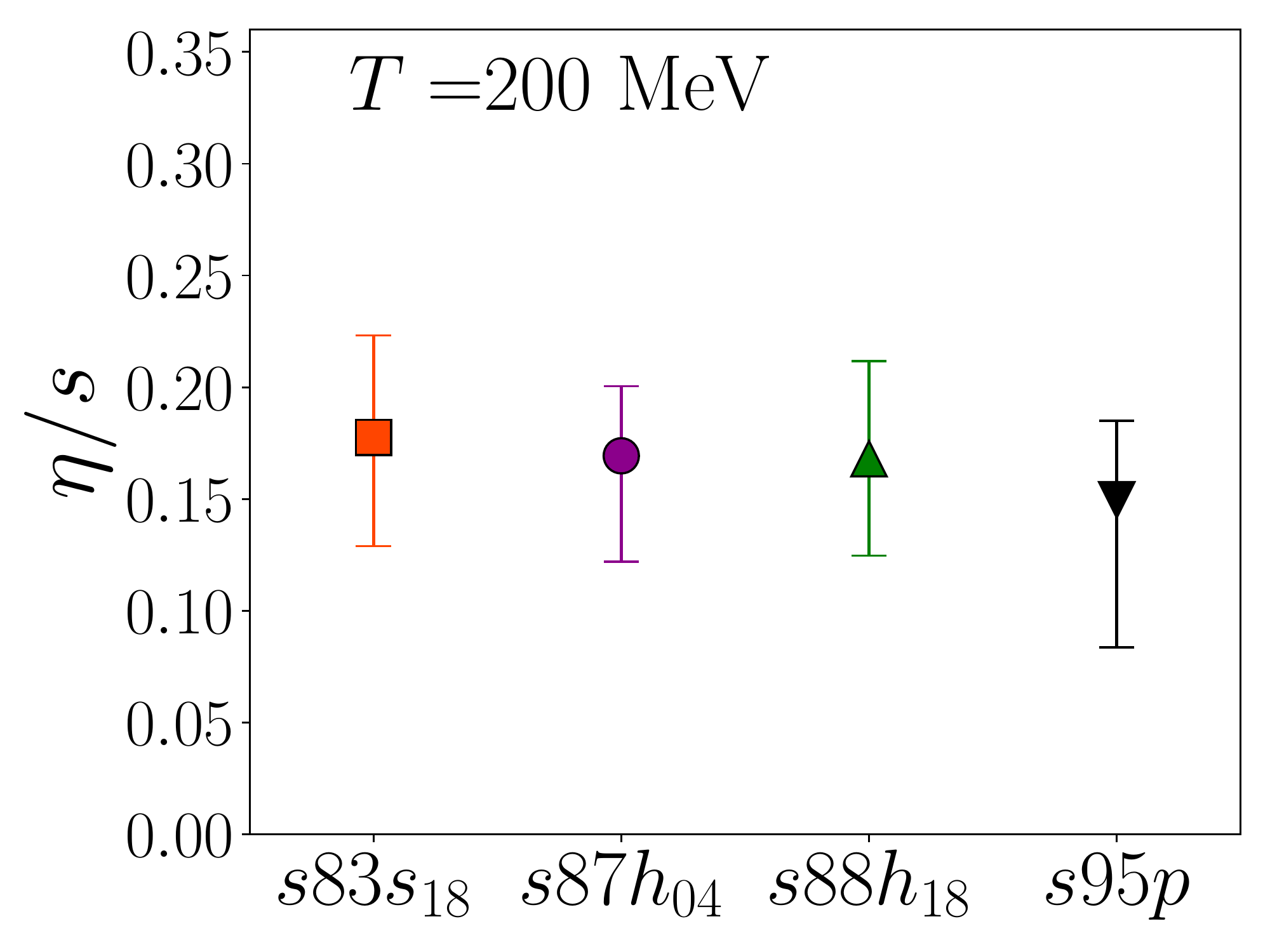}
 \includegraphics[width=4.2cm]{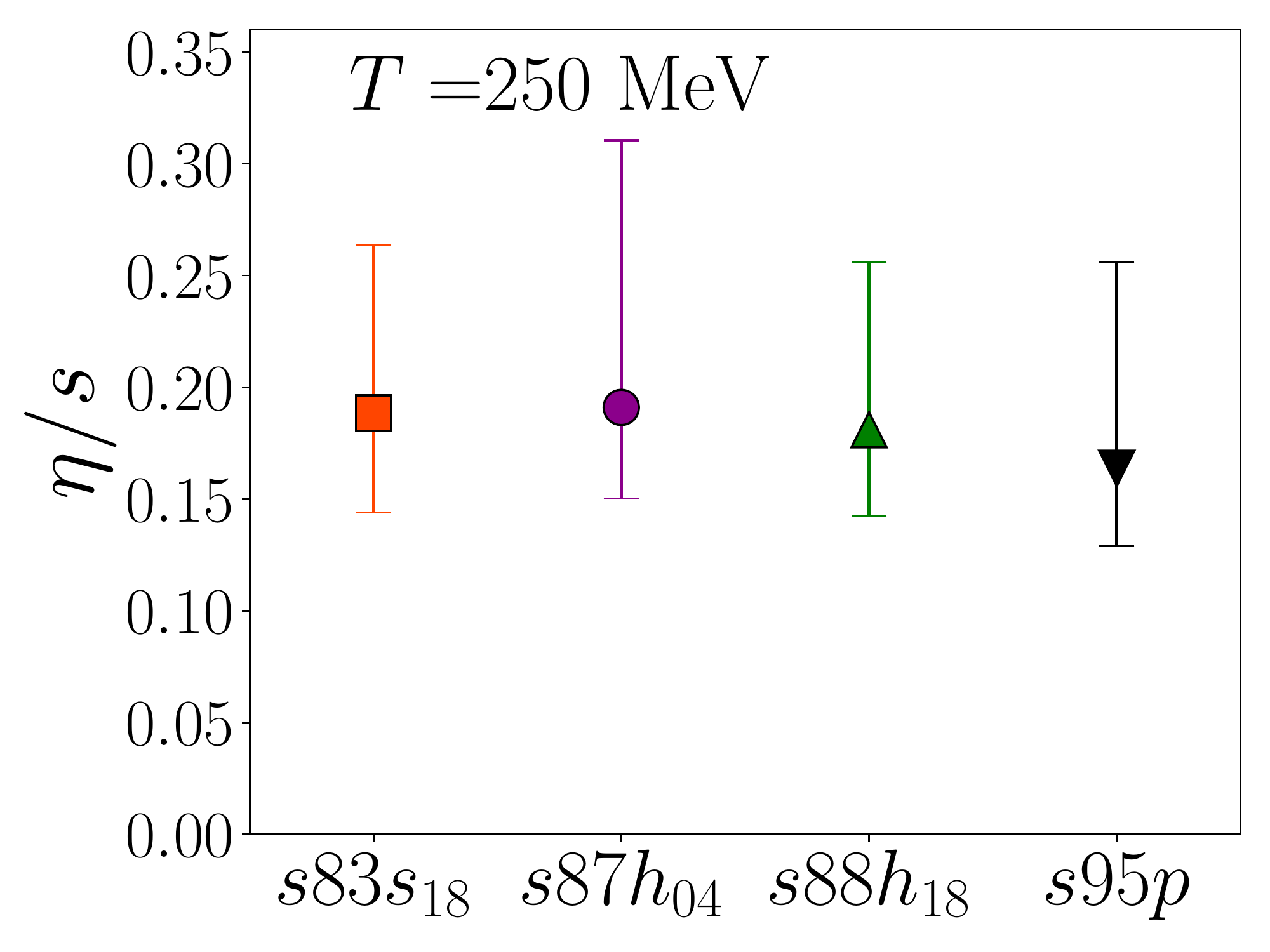}
 \includegraphics[width=4.2cm]{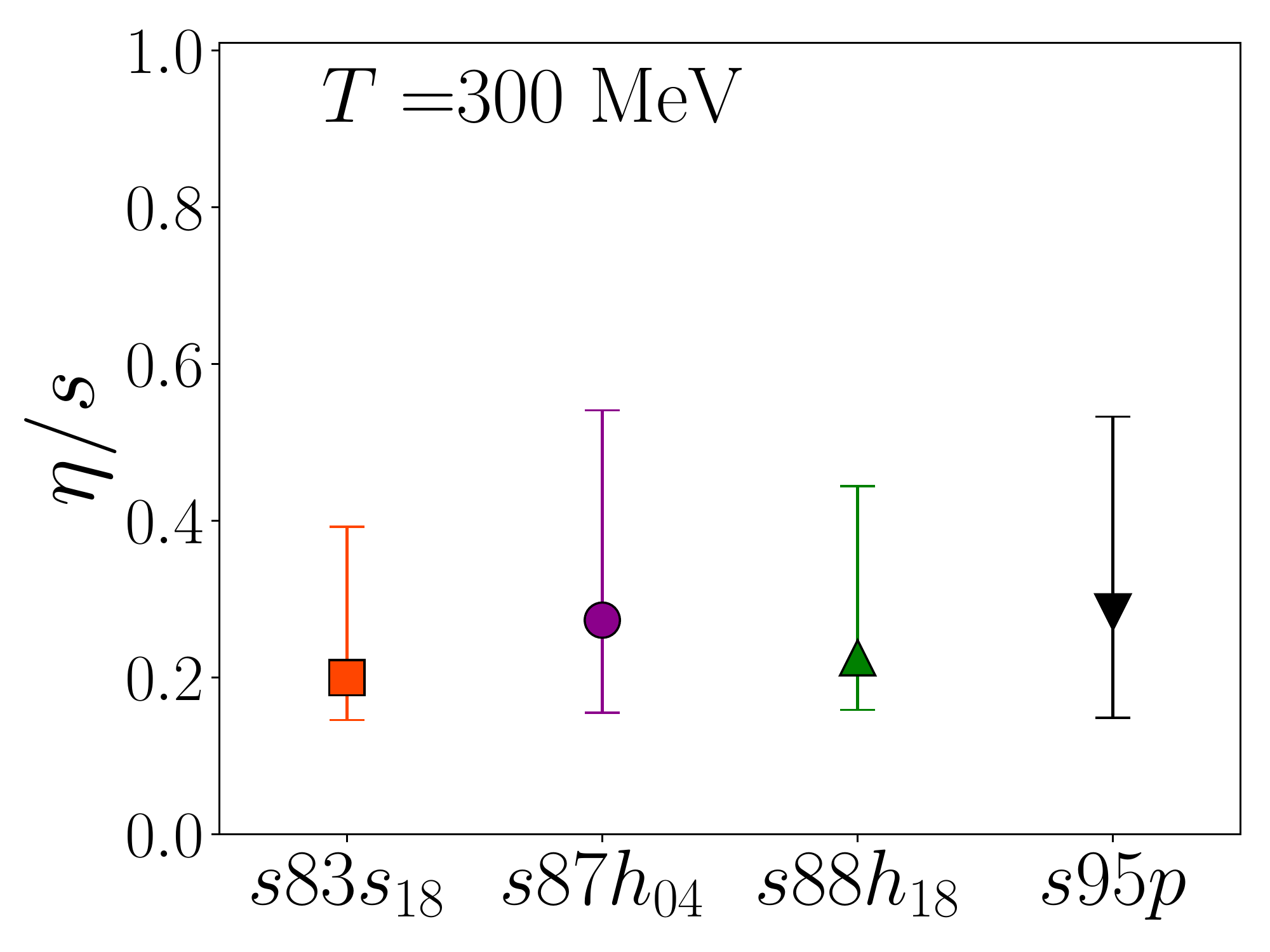}
 \includegraphics[width=4.2cm]{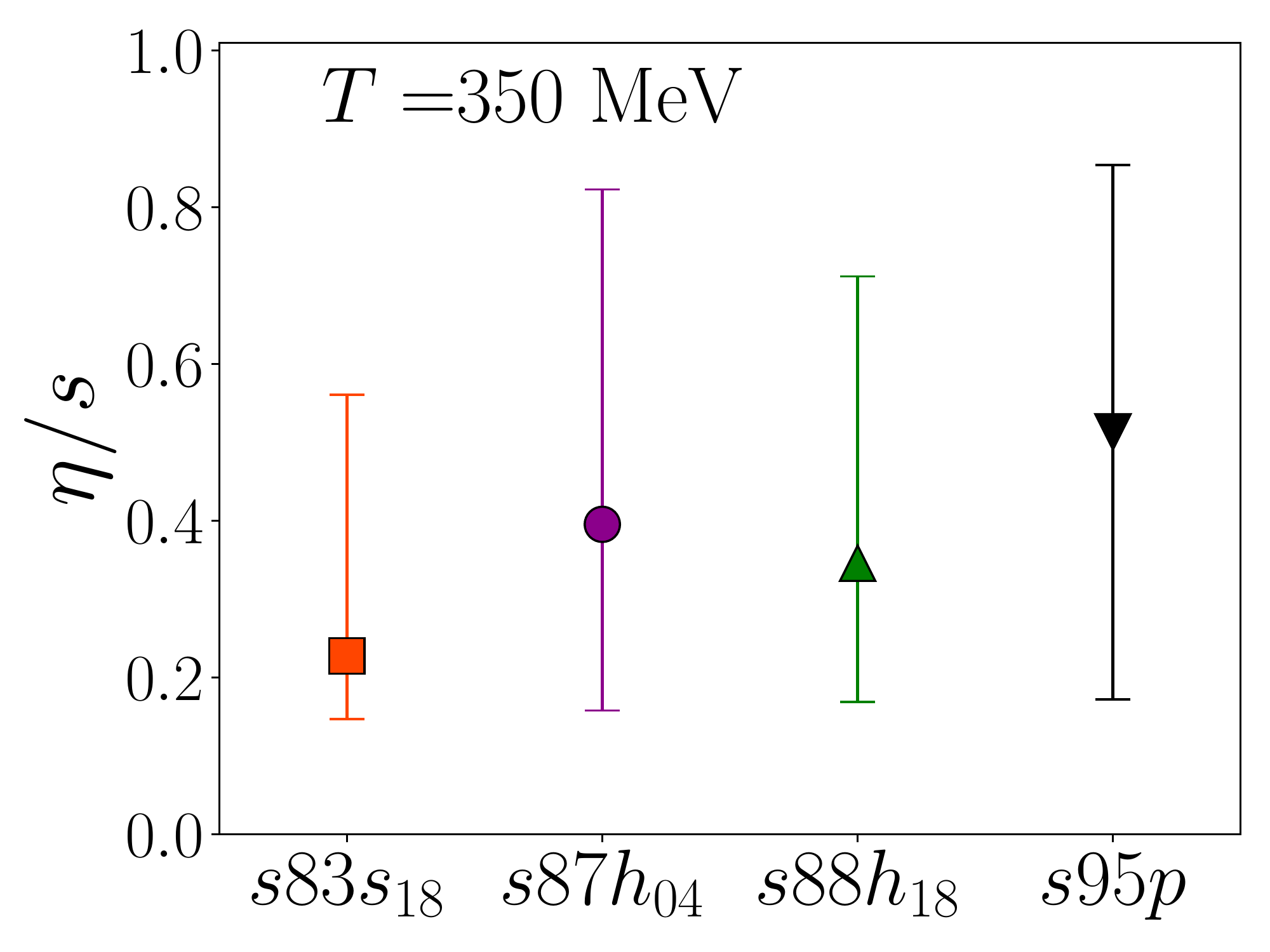}
 \includegraphics[width=4.2cm]{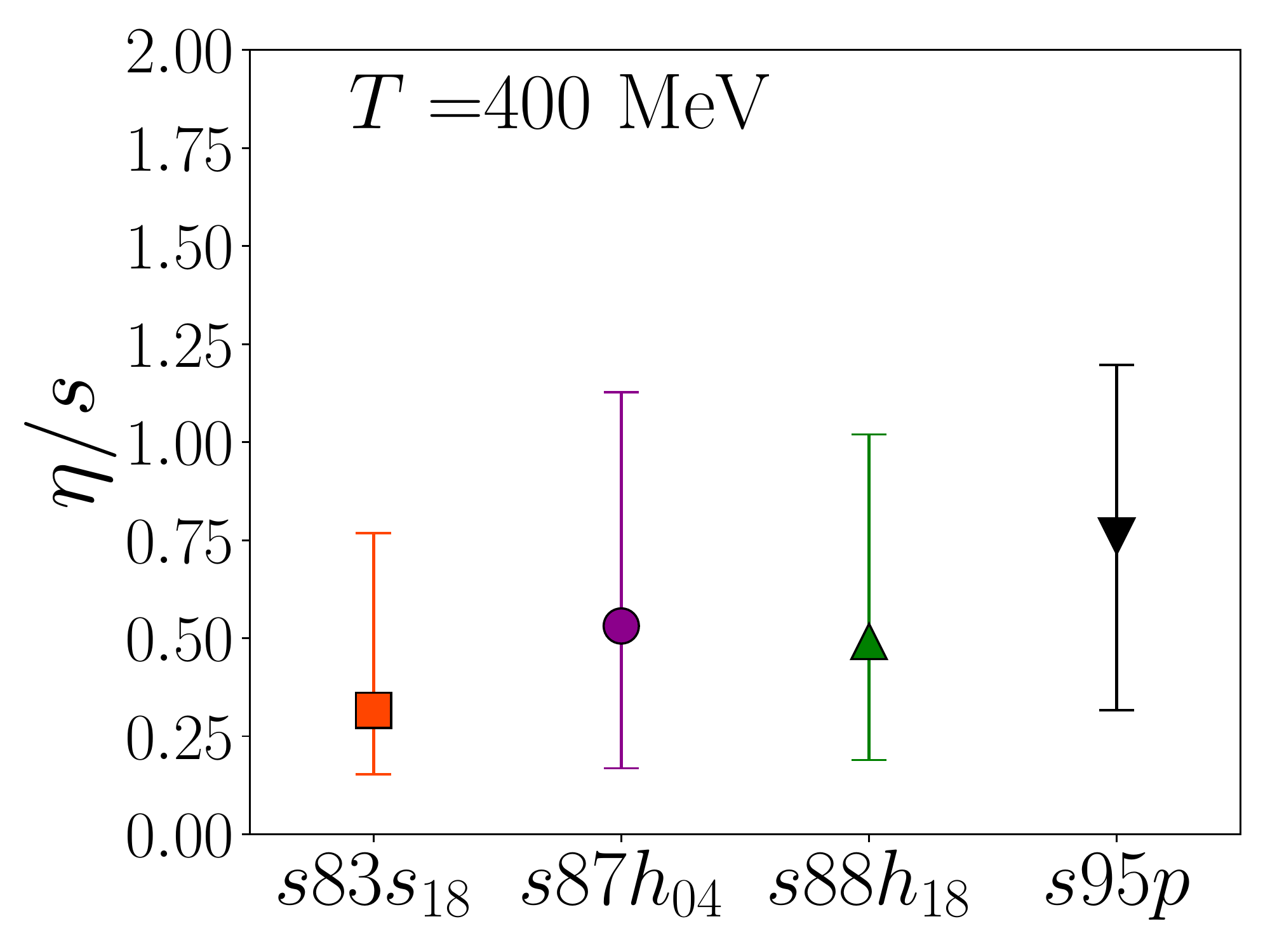}
 \includegraphics[width=4.2cm]{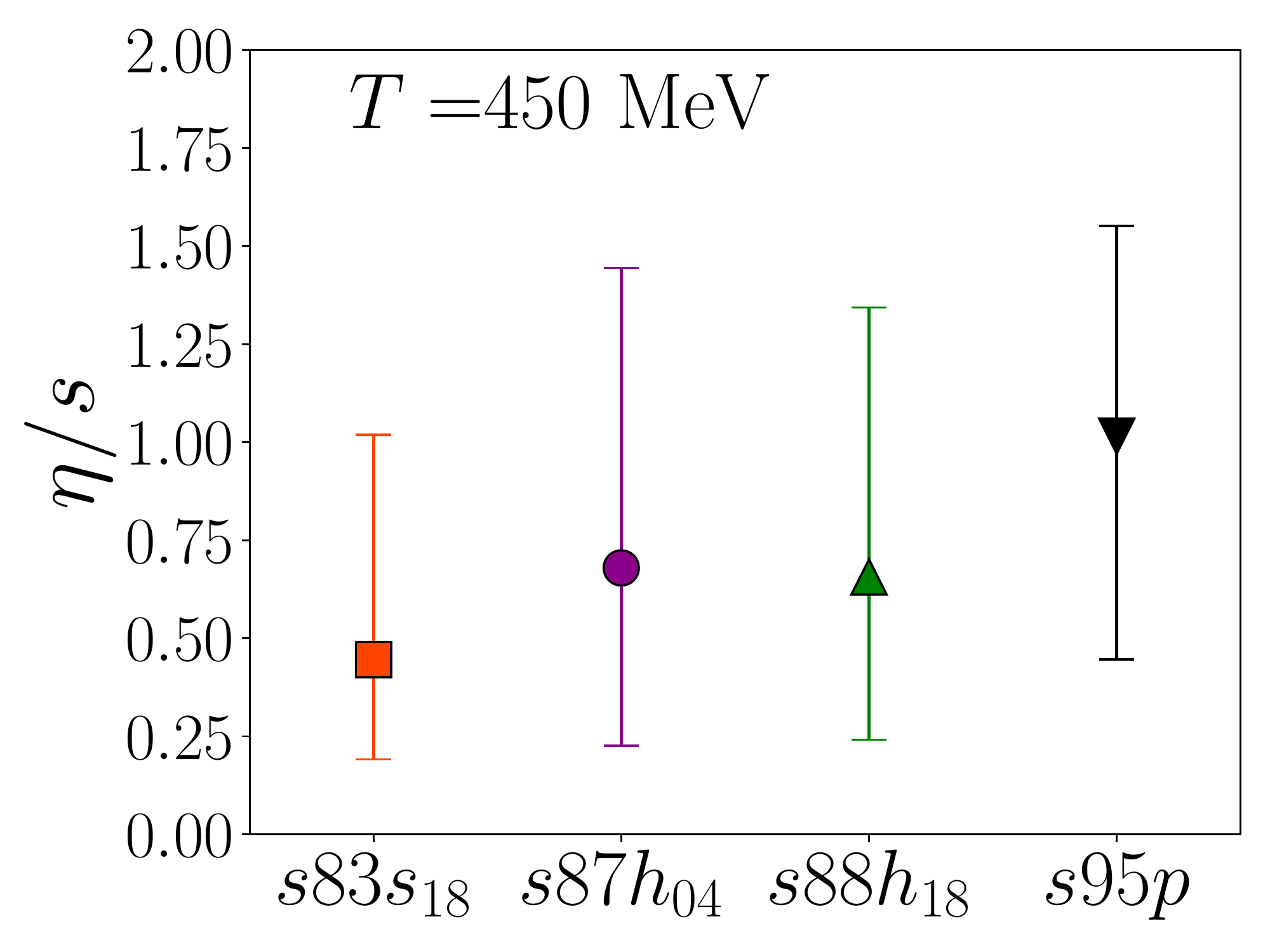}
 \caption{Median values (filled markers)
 and 90\% credible intervals (error bars) for $\eta/s$ at temperatures $T=130,150,200,250,300,350,400$, and $450$ MeV.}
 \label{fig:etasmin_comparison}
\end{figure}

At first sight $\etasmin$ depicts the behavior described in
Refs.~\cite{Alba:2017hhe,Schenke:2019ruo}: the favored value is lower
for $s95p$ than for the newer parametrizations (see
Fig.~\ref{fig:posterior_ksat} and Table~\ref{tab:param}). However, the
effect is noticeably smaller than seen in those studies---only $\approx
10$--$20\%$---and well within the 90\% credible intervals ($\pm\approx
30\% $) of the analysis. The comparison of $\eta/s$ for different EoSs
is further complicated by the large number of parameters controlling
the temperature dependence of $\eta/s$. The probability distributions
of parameters $\Tmin,\Wmin,\Shg,$ and $\Sqgp$, shown in
Fig.~\ref{fig:posterior_tmin}, are very broad extending to the whole
prior range in most cases, and thus do not possess any clearly favored
values. However, the wide posterior distributions of the $(\eta/s)(T)$
parameters are partly caused by the inherent ambiguity in the chosen
parametrization: for a given temperature $T$, multiple parameter
combinations can generate similar values of $\eta/s$. For example, at
low temperatures $(\eta/s)(T)$ is mostly determined by $\Shg$ and
$\Tmin$, but it is better constrained than either of these parameters.
The reason is that $\Shg$ and $\Tmin$ are not independent, but slightly
anti-correlated---the correlations between the pairs of parameters are
shown in Appendix~\ref{appx:posterior}. Thus it is more illustrative
to construct the probability distribution for $\eta/s$ values
w.r.t.~temperature, and plot the median and credibility intervals of
this distribution as shown in Figs.~\ref{fig:etas_vs_t}
and~\ref{fig:etasmin_comparison}.

In the upper panel of Fig.~\ref{fig:etas_vs_t} we show the median
of $(\eta/s)(T)$ for each EoS parametrization, and the union
and intersection of the 90\% credible intervals of all four
distributions. The union of the credibility intervals provides insight on the
total uncertainty in the analysis including the uncertainty from the
EoS parametrization, whereas the difference between the union and
intersection illustrates how much of the uncertainty comes from the EoS
parametrizations. To emphasize the result using state-of-the-art EoSs,
the lower panel of Fig.~\ref{fig:etas_vs_t} depicts the median and
credibility intervals for the parametrizations $s83s_{18}$ and $s88h_{18}$ only. In the
same panel two older results from Ref.~\cite{Niemi:2015qia}, and a
recent theoretical prediction from Ref.~\cite{Mykhaylova:2019wci} are
shown as well. To make it possible to distinguish the credibility
intervals for each EoS separately, $\eta/s$ for each EoS at various
temperatures with associated uncertainties is shown in
Fig.~\ref{fig:etasmin_comparison}.

We obtain well constrained $\eta/s$ in a temperature range $150 \lton
T/\mathrm{MeV} \lton 220$, where the median values of $\eta/s$ are
practically constant for all contemporary EoSs, and $s95p$ leads to
modest temperature dependence well within the credibility intervals.
Within this temperature range $\eta/s$ is constrained between 0.08 and
0.23 by the 90\% credible intervals. In particular, for the
state-of-the-art EoSs ($s83s_{18}$ and $s88h_{18}$), we obtain even
tighter limits $0.12 < \eta/s < 0.23$ within this range, and the well
constrained region extends to slightly higher temperature. For further
details see Fig.~\ref{fig:etas_vs_t} and Table~\ref{tab:etasT}.
Interestingly $\eta/s$ at 130 MeV (or at 150 MeV in case of $s95p$)
temperature differs from the favored value (median) of the $\etasmin$
parameter (compare Tables~\ref{tab:param} and~\ref{tab:etasT}), even
if the favored value of the $\Tmin$ parameter is 130 MeV (or 150 MeV)
(see Fig.~\ref{fig:posterior_tmin} and Table~\ref{tab:param}).  This
seemingly counterintuitive behavior is due to the fat tails of $\Tmin$
distributions extending to larger temperatures, and thus broadening
the region where $\Shg$ affects the $\eta/s$ values.  Consequently we
see the lowest $\eta/s$ values at $T\approx 200$ MeV temperature
(Fig.~\ref{fig:etas_vs_t} and Table~\ref{tab:etasT}), where the effect
of the lower $\etasmin$ value of the $s95p$ parametrization is also
visible.

\begin{table} \renewcommand{\arraystretch}{1.3}
  \caption{Median values of $\eta/s$ at various temperatures with associated
    uncertainties (90\% credible intervals) from the posterior distributions. Values rounded to two significant figures.}
  \begin{tabular}{c|r@{}l@{\hspace{3mm}}r@{}l@{\hspace{3mm}}r@{}l@{\hspace{3mm}}r@{}l}
    \hline
    \hline
    $T$ [MeV] & \multicolumn{2}{c}{$s83s_{18}$} & \multicolumn{2}{c}{$s87h_{04}$} & \multicolumn{2}{c}{$s88h_{18}$} & \multicolumn{2}{c}{$s95p$}  \\
     \hline
   $130$ & $0.19$ & $^{+0.09 }_{-0.04 }$ & $0.19$ & $^{+0.10 }_{-0.05 }$ & $0.19$ & $^{+0.07 }_{-0.06 }$ & $0.21$ & $^{+0.13 }_{-0.07 }$ \\
   $150$ & $0.18$ & $^{+0.05 }_{-0.04 }$ & $0.17$ & $^{+0.05 }_{-0.04 }$ & $0.17$ & $^{+0.05 }_{-0.05 }$ & $0.17$ & $^{+0.06 }_{-0.04 }$ \\
   $200$ & $0.18$ & $^{+0.04 }_{-0.05 }$ & $0.17$ & $^{+0.03 }_{-0.05 }$ & $0.17$ & $^{+0.04 }_{-0.05 }$ & $0.15$ & $^{+0.04 }_{-0.07 }$ \\
   $250$ & $0.19$ & $^{+0.07 }_{-0.05 }$ & $0.19$ & $^{+0.12 }_{-0.04 }$ & $0.18$ & $^{+0.08 }_{-0.04 }$ & $0.16$ & $^{+0.10 }_{-0.03 }$ \\
   $300$ & $0.20$ & $^{+0.19 }_{-0.05 }$ & $0.27$ & $^{+0.27 }_{-0.11 }$ & $0.23$ & $^{+0.21 }_{-0.07 }$ & $0.28$ & $^{+0.25 }_{-0.13 }$ \\
   $350$ & $0.23$ & $^{+0.33 }_{-0.08 }$ & $0.40$ & $^{+0.42 }_{-0.24 }$ & $0.35$ & $^{+0.36 }_{-0.18 }$ & $0.51$ & $^{+0.34 }_{-0.34 }$ \\
    \hline
     \hline
   \end{tabular}
   \label{tab:etasT}
\end{table}

It is not surprising that we get the best constraints on $\eta/s$ in
the temperature range $150 \lton T/\mathrm{MeV} \lton 220$. As was
shown in Ref.~\cite{Niemi:2012ry}, the temperature range where $v_2$
is most sensitive to the shear viscosity is only slightly broader than
this, and higher order anisotropies are sensitive to shear at even
narrower temperature ranges\footnote{Note that the studies in
  Ref.~\cite{Niemi:2012ry} were carried out using the $s95p$ EoS. We
  haven't checked how sensitive those results are to the EoS
  parametrization.}. 

Even if the uncertainties remain large, we can see qualitative
differences in the high temperature behavior of $\eta/s$, where $s95p$
seems to favor earlier and more rapid rise of $\eta/s$ with increasing
temperature (Figs.~\ref{fig:etas_vs_t}
and~\ref{fig:etasmin_comparison}), a difference which is visible in
the $\Sqgp$ parameter as well (Fig.~\ref{fig:posterior_tmin}).

Considering earlier results in the literature this is intriguing. Alba
\emph{et al.}~\cite{Alba:2017hhe} used an EoS based on contemporary
stout action data called PDG16+/WB2+1, and observed that the
reproduction of the LHC data ($\sqrt{s_{\NN}}=5.02 $ TeV) required
larger constant $\eta/s$ for this EoS than for $s95p$. On the other
hand, they were able to use the same value of constant $\eta/s$ for
both EoSs to reproduce the RHIC data. They interpreted this to mean
that at large temperatures $s95p$ would necessitate lower values of
$\eta/s$, but we see an opposite behavior. In a similar fashion we see
a difference between the high temperature behavior obtained using the
HISQ ($s88h_{18}$ and $s87h_{04}$) and stout action based EoSs
($s83s_{18}$), but the differences are way smaller than the
credibility intervals, and thus cannot be considered meaningful.

At temperatures below 150 MeV we again see expanding credibility
intervals, and a tendency of $\eta/s$ to increase with decreasing
temperature, but hardly any sensitivity to the EoS. Anisotropies
measured at RHIC energy are sensitive to the shear viscosity in the
hadronic phase~\cite{Niemi:2012ry,Niemi:2011ix}, and since Schenke
\emph{et al.}~in Ref.~\cite{Schenke:2019ruo} saw sensitivity to the
EoS using RHIC data only, we would have expected some sensitivity to
the EoS at low temperatures. The difference may arise from the bulk
viscosity which depended on the EoS as well in Ref.~\cite{Schenke:2019ruo},
or from a different EoS in the hadronic phase. As mentioned, our EoSs
are based on known resonance states, whereas the EoSs in
Refs.~\cite{Alba:2017hhe,Schenke:2019ruo} follow the lattice results
closely. Better fit to lattice results can be obtained by including
predicted but unobserved resonance states in the HRG. We plan to study
how the inclusion of these states might affect the results, once we
have concocted a plausible scheme for their decays, so that we can
evaluate their contribution to the EoS after chemical freeze-out in a
consistent manner.

Furthermore, unlike in Ref.~\cite{Schenke:2019ruo} where a hadron
cascade was used to describe the evolution in the hadronic phase, in
our approach the change in the EoS can also be partly compensated by a
change in the freeze-out temperature instead of shear viscosity. As
shown in Appendix~\ref{appx:posterior}, there is indeed an
anti-correlation between $\Tdec$ and $\etasmin$. Therefore forcing the
system to freeze out at the same temperature, independent of the EoS,
would increase the difference in $\etasmin$. However, the
anti-correlation is rather weak $\approx -0.4(-0.2)$ for $s88h_{18}$ ($s95p$),
and thus requiring EoS independent $\Tdec$ would not change $\etasmin$ a lot.

Our result of a very slowly rising $\eta/s$ with decreasing
temperature in the hadronic phase (i.e., below $T\approx 150$ MeV) may
look inconsistent with microscopic calculations predicting relatively
large $\eta/s\sim 1$ in the hadronic phase~\cite{Rose:2017bjz,Prakash:1993bt,
  Csernai:2006zz}. However, our result is for a chemically frozen HRG, while
the microscopic calculations usually give $\eta/s$ in chemical equilibrium.
In the transport model study of Ref.~\cite{Demir:2008tr}, it was shown
that nonunit pion and kaon fugacities, i.e.~chemical non-equilibrium,
can significantly reduce $\eta/s$ in hadron gas. Since, as a first
order approximation, $\eta$ depends only weakly on the chemical
non-equilibrium~\cite{Wiranata:2014jda}, the main effect is due to
$s$: At a given temperature the entropy density $s_{\mathrm{PCE}}$ in
a chemically frozen HRG can be significantly larger than the entropy
density in chemical equilibrium $s_{\mathrm{CE}}$, and as a
consequence $(\eta/s)_{\mathrm{PCE}}$ can be way smaller than
$(\eta/s)_{\mathrm{CE}}$. We may thus obtain an approximation for the
$\eta/s$ in a chemically equilibrated system as
$(\eta/s)_{\mathrm{CE}} =
(\eta/s)_{\mathrm{PCE}}(s_{\mathrm{PCE}}/s_{\mathrm{CE}})$~\cite{Niemi:2015qia}.
In our case, where $\Tchem = 154$ MeV, the ratio of entropies in a
chemically frozen to a chemically equilibrated system is $\approx 3.5$ at
$T=100$ MeV ($\approx 1.8$ at $T=130$ MeV) which is sufficient to bring
our results to the level described in Ref.~\cite{Csernai:2006zz}.

In Fig.~\ref{fig:etas_vs_t} we also made a comparison to the earlier
results of Ref.~\cite{Niemi:2015qia} and the recent quasiparticle
model prediction from Ref.~\cite{Mykhaylova:2019wci}. As expected, the
earlier results from Ref.~\cite{Niemi:2015qia, Niemi:2015voa} are not
far from the present analysis, and param1 
is practically within the 90\% credible interval in the whole temperature range. 
On the other hand, constant $\eta/s = 0.2$ is below the $s95p$ limits at high
temperatures, but as discussed, the overall sensitivity to $\eta/s$ at
high temperatures is low. Interestingly the prediction of the
quasiparticle model of Ref.~\cite{Mykhaylova:2019wci} comes very close
to our values for $\eta/s$ around $T_c$, although the region
where $\eta/s$ is low is narrower than what we found here. This is
intriguing, since the quasiparticle model was tuned to reproduce the
stout action EoS, i.e., our EoS $s83s_{18}$, which in our analysis leads to
the broadest region where $\eta/s$ is almost constant.

The small value of $\eta/s$ and its weak temperature dependence in the
temperature range $150 \lton T/\mathrm{MeV} \lton 220$ may indicate
that the QGP is strongly coupled not only in the immediate vicinity of
$T_c$, but in a broader temperature region. This was first proposed in
Ref.~\cite{Shuryak:2003ty}, and agrees with the lattice QCD
calculations that indicate the presence of hadronlike resonances in
QGP in a similar or slightly broader temperature interval~\cite{Wetzorke:2001dk,Karsch:2002wv,
  Asakawa:2002xj,Mukherjee:2015mxc}. The strongly coupled nature of
QGP can also be seen in the large value of the coupling constant
defined in terms of the free energy of static quark anti-quark
pairs~\cite{Bazavov:2018wmo}. In any case, our result for $(\eta/s)(T)$
is compatible with the lattice QCD calculations, which indicate that
weakly coupled QGP picture may be applicable only for $T>350$
MeV~\cite{Bazavov:2013uja,Bellwied:2015lba,Ding:2015fca,Bazavov:2018wmo}.

\subsection{The effect of the parametric form}

\begin{figure}[h]
 \centering
 \includegraphics[width=8.5cm]{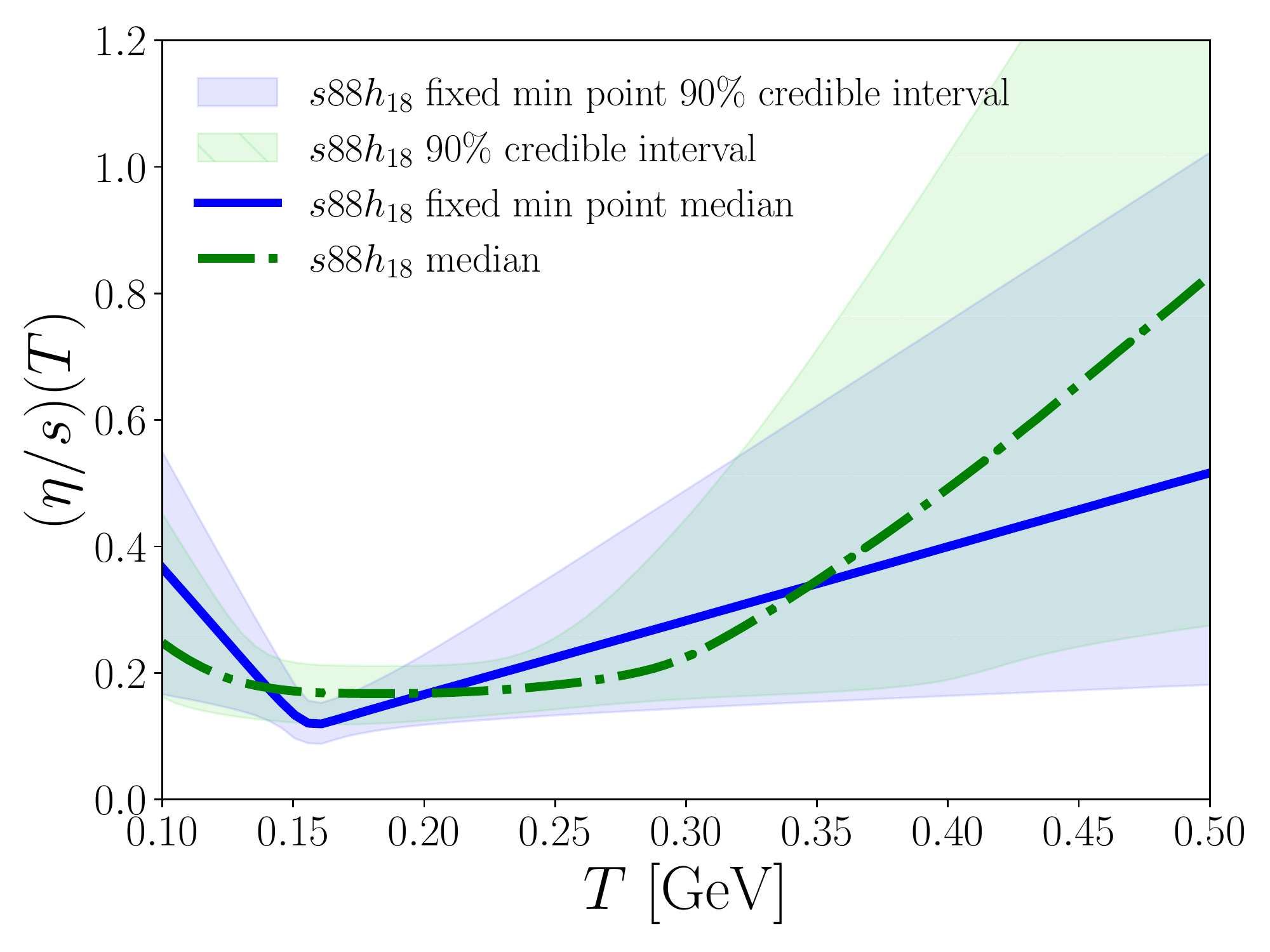}
 \includegraphics[width=8.5cm]{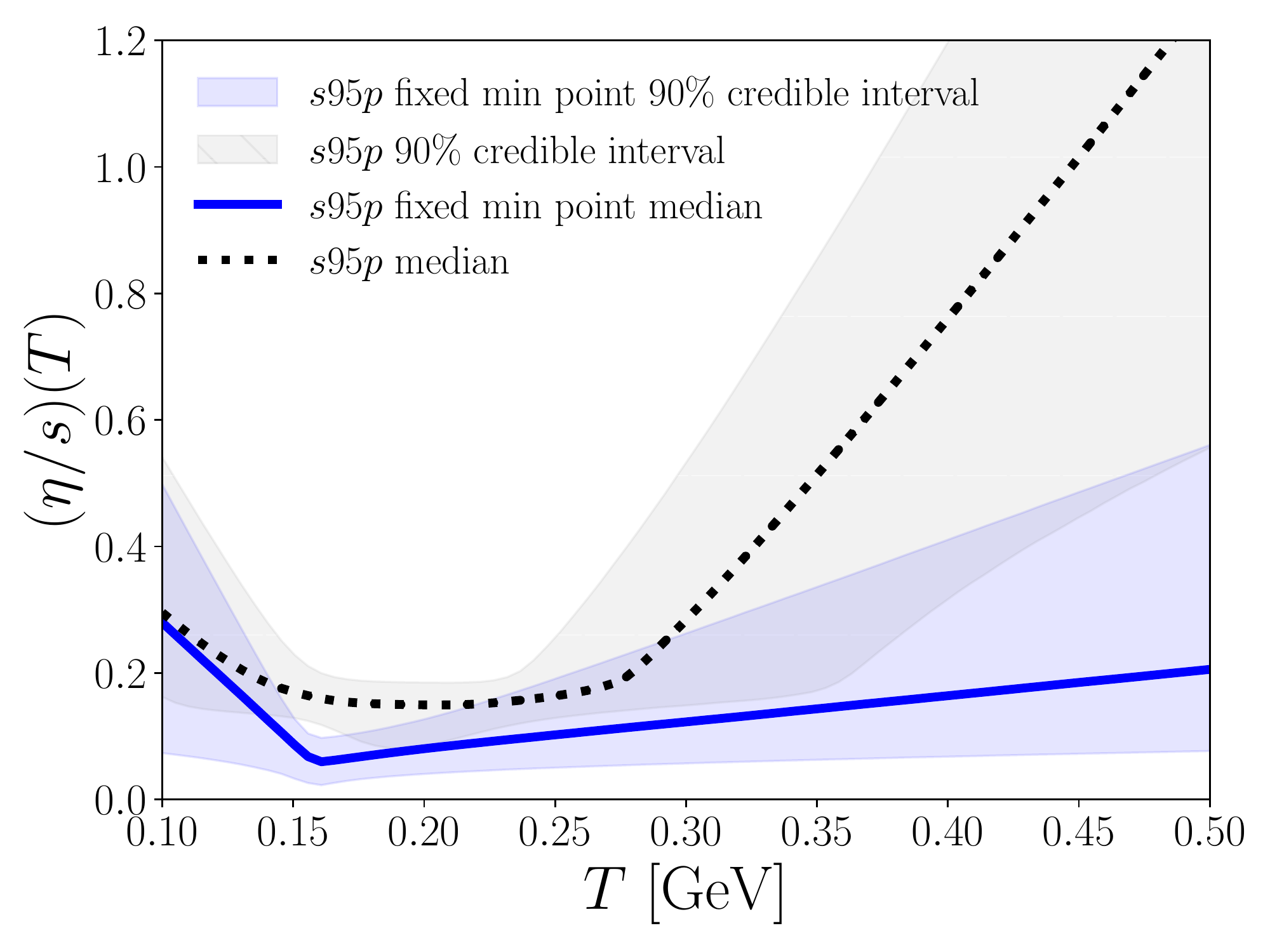}
 \caption{Temperature dependence of $\eta/s$ for the $s88h_{18}$ (top) and
   $s95p$ (bottom) EoSs using the full parametrization and a
   parametrization constrained to have a minimum at a fixed point in
   temperature (``fixed min point'').}
 \label{fig:etas_duke}
\end{figure}

When we use the state-of-the-art EoSs ($s88h_{18}$ and $s83s_{18}$), our result for
the minimum value of $\eta/s$ is higher than the result obtained in an
earlier Bayesian analysis of Ref.~\cite{Bernhard:2016tnd}:
$0.12 < \eta/s < 0.23$ vs.~$\eta/s = 0.07^{+0.05}_{-0.04}$.
While the equations of state in both analyses are based on the latest lattice results,
an important difference is that Ref.~\cite{Bernhard:2016tnd}
assumed the minimum of $\eta/s$ to occur at fixed $T=154$ MeV
temperature, and $\eta/s$ to rise linearly above that temperature.
Moreover, below $T=154$ MeV they used a hadron cascade to model the
evolution, and the transport properties of the hadronic phase were 
thus fixed. 

To explore how much the results depend on the form of the
$(\eta/s)(T)$ parametrization, we mimic the parametrization used
in Ref.~\cite{Bernhard:2016tnd} by constraining the plateau in our
parametrization to be very small ($0<\Wmin/\mathrm{MeV}<2$), and the
minimum to appear close to $T_c$ ($150 < \Tmin/\mathrm{MeV} < 160$).
The resulting temperature dependence of $\eta/s$ for the $s88h_{18}$ and
$s95p$ parametrizations is shown in Fig.~\ref{fig:etas_duke}, and
compared to our full result (the behavior of the $s87h_{04}$ and $s83s_{18}$
parametrizations is similar to $s88h_{18}$).

The change in parametrization reduces the minimum value of $\eta/s$ to
$0.12^{+0.03}_{-0.03}$ for $s88h_{18}$, which is closest to the EoS
used in Ref.~\cite{Bernhard:2016tnd}. The credibility interval now
overlaps with the result from the earlier analysis~\cite{Bernhard:2016tnd},
and the results are thus consistent. The remaining difference may
result from the bulk viscosity, event-by-event-fluctuations,
differences in the EoS parametrization scheme, or the transport
description of the hadronic phase. As mentioned earlier, switching to
hadron cascade creates a discontinuity in $(\eta/s)(T)$. Enforcing
a similar discontinuity in the $(\eta/s)(T)$ parametrization might bring
closer the minimum values of $\eta/s$ obtained using hybrid models
and continuous fluid dynamics.
For $s95p$ the minimum value drops to $0.06^{+0.04}_{-0.04}$, but
since the $s95p$ parametrization is based on the older lattice data,
comparing this value against Ref.~\cite{Bernhard:2016tnd} is not
straightforward. However, due to the significant overlap of the
crediblity intervals, we consider both results consistent with
Ref.~\cite{Bernhard:2016tnd}, demonstrating the weak sensitivity of
the extracted $\eta/s$ to the EoS used in the calculations.

Another interesting change is seen in the high-temperature
behavior. In the full analysis the $s95p$ parametrization leads to the
largest $\eta/s$ at large temperatures, but the restricted
parametrization causes $s95p$ to favor the lowest $\eta/s$ at large
temperatures. As seen previously, $s95p$ favors the lowest $\eta/s$ at
$200 < T/\mathrm{MeV} < 250$ temperatures (see
Fig.~\ref{fig:etasmin_comparison} and Table \ref{tab:etasT}), which in
the restricted parametrization dictates the behavior at much higher
temperatures as well.

Nevertheless, even if the results depend on the form of the
parametrization, the credibility intervals overlap and the results
are consistent. The only deviation from this rule is for the $s95p$
parametrization around $T=160$ MeV temperature where the difference is
statistically significant (see Fig.~\ref{fig:etas_duke}).  We have
also checked that when we use the favored parameter values, the
typical differences in the fit to the data due to different parametric
forms are only $\approx (1-3)\%$.

Similarly, we can mimic temperature independent $\eta/s$ by
constraining the priors of the $\Shg$ and $\Sqgp$ parameters close to
zero.  We have checked that such a choice does not increase the
sensitivity of $\eta/s$ to the EoS parametrization, and that the
median values of the constant $\eta/s = \etasmin$ were only $\approx
10\%$ larger than the median values for $\eta/s$ at $T = 200$ MeV for
the full parametrization. Again, a sign of $v_2$ being most sensitive
to shear viscosity in the $150 \lton T/\mathrm{MeV} \lton 220$
temperature range~\cite{Niemi:2012ry}.

Thus, in the Bayesian analysis the parametric form of $\eta/s$ does
affect the results, and is therefore a kind of prior whose effects are
difficult to quantify. On the other hand, the credibility intervals overlap in
all the cases, which emphasizes their importance: The
``true'' value could be anywhere within the credibility interval, and
there is still a 10\% chance it is outside of it.

\subsection{Comparison with the data}
  \label{sec:data}

\begin{figure}[h]
 \centering
 \includegraphics[width=6.5cm]{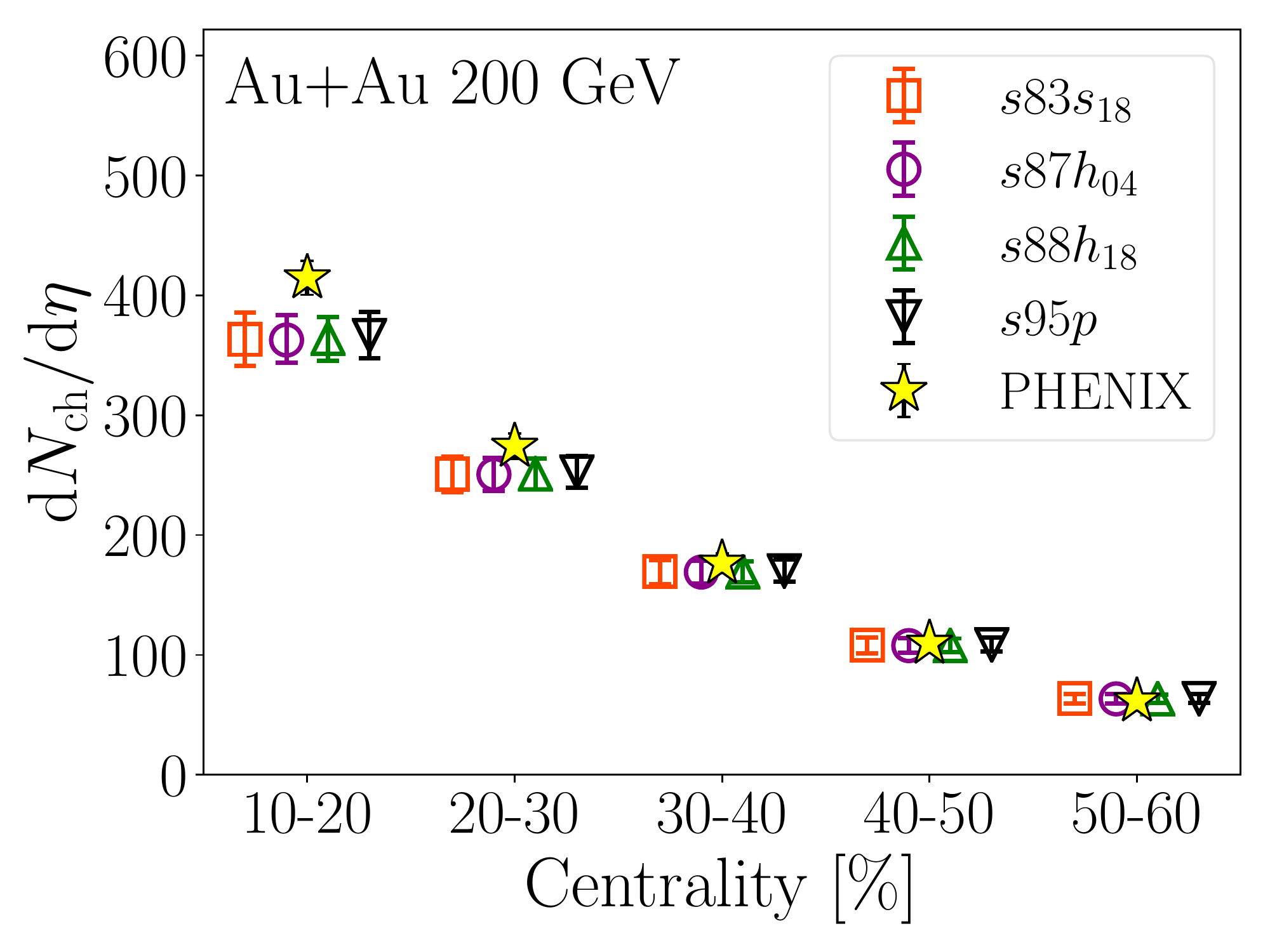}\\
 \vspace{1mm}
 \includegraphics[width=6.5cm]{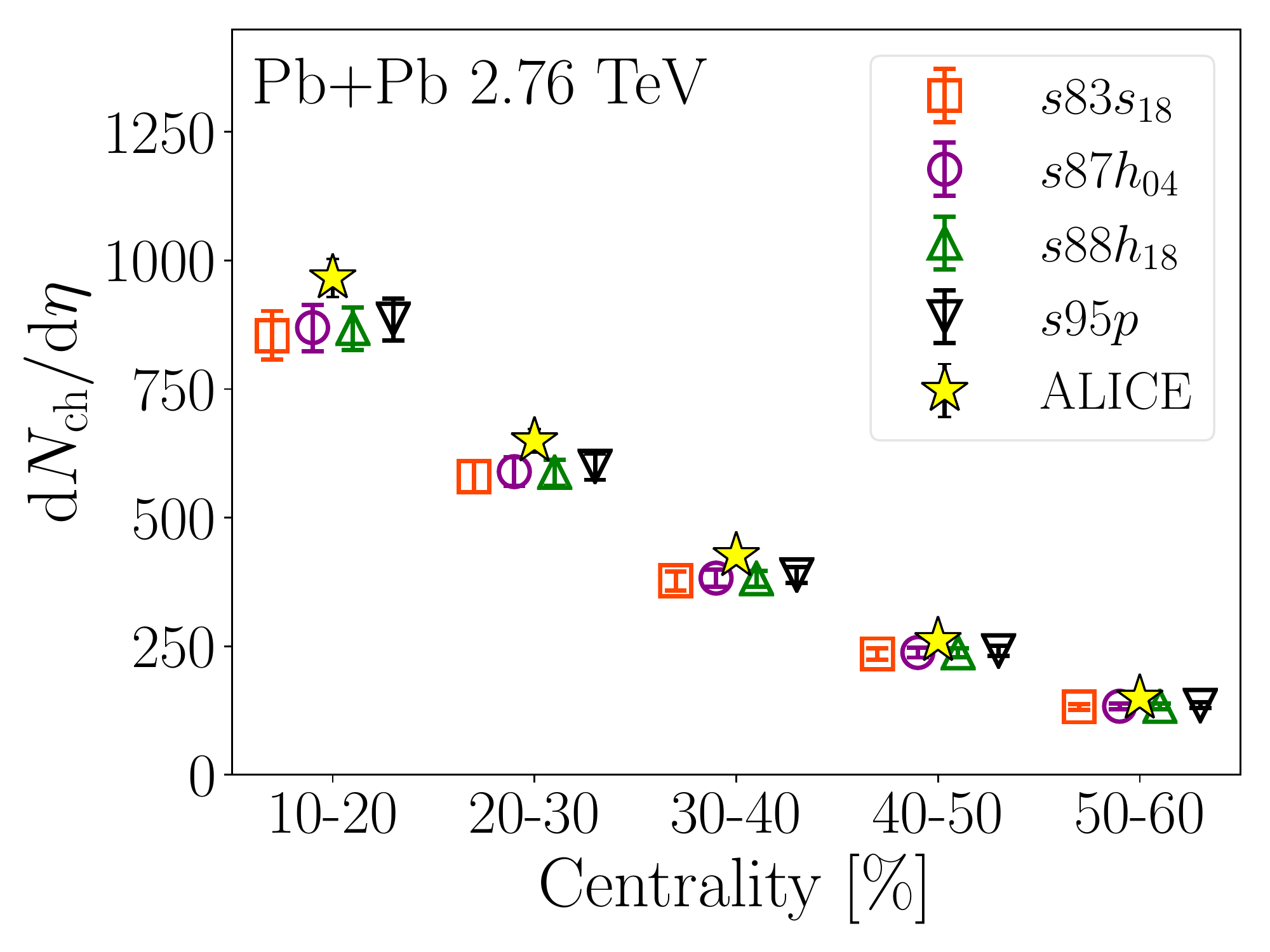}\\
 \vspace{1mm}
 \includegraphics[width=6.5cm]{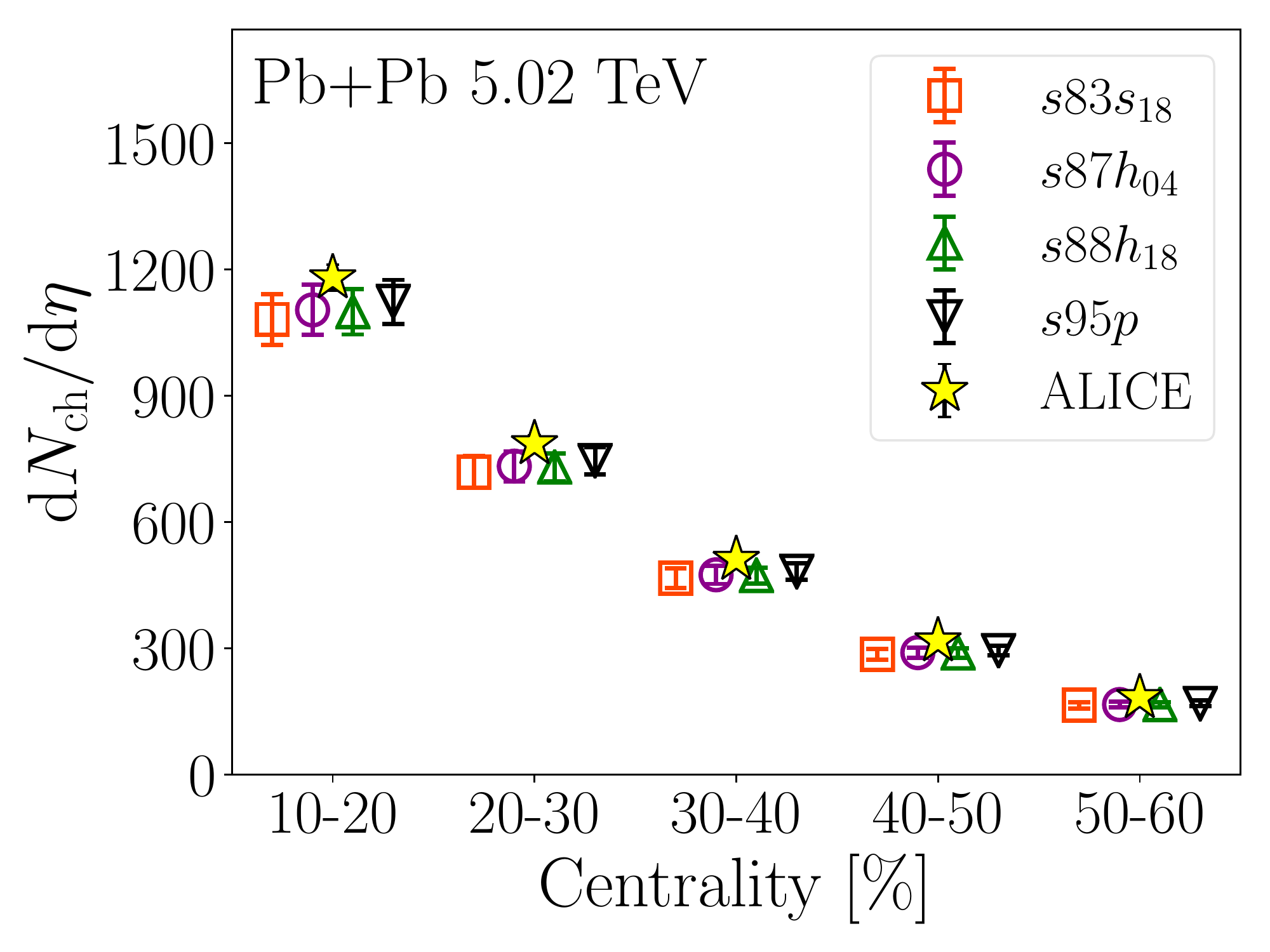}
 \caption{Charged particle multiplicity at various centralities
 using 1000 samples from the posterior distribution of each EoS.
 Marker centers indicate median values, and error bars 90\% credible intervals.
 Top panel: Au+Au at $\sqrt{s_{\NN}}=200$ GeV compared to PHENIX data \cite{Adler:2004zn}.
 Middle panel: Pb+Pb at $\sqrt{s_{\NN}}=2.76$ TeV compared to ALICE data \cite{Aamodt:2010cz}.
 Bottom panel: Pb+Pb at $\sqrt{s_{\NN}}=5.02$ TeV compared to ALICE data \cite{Adam:2015ptt}.}
 \label{fig:nch_posterior}
\end{figure}

\begin{figure}[h]
 \centering
 \includegraphics[width=4.25cm]{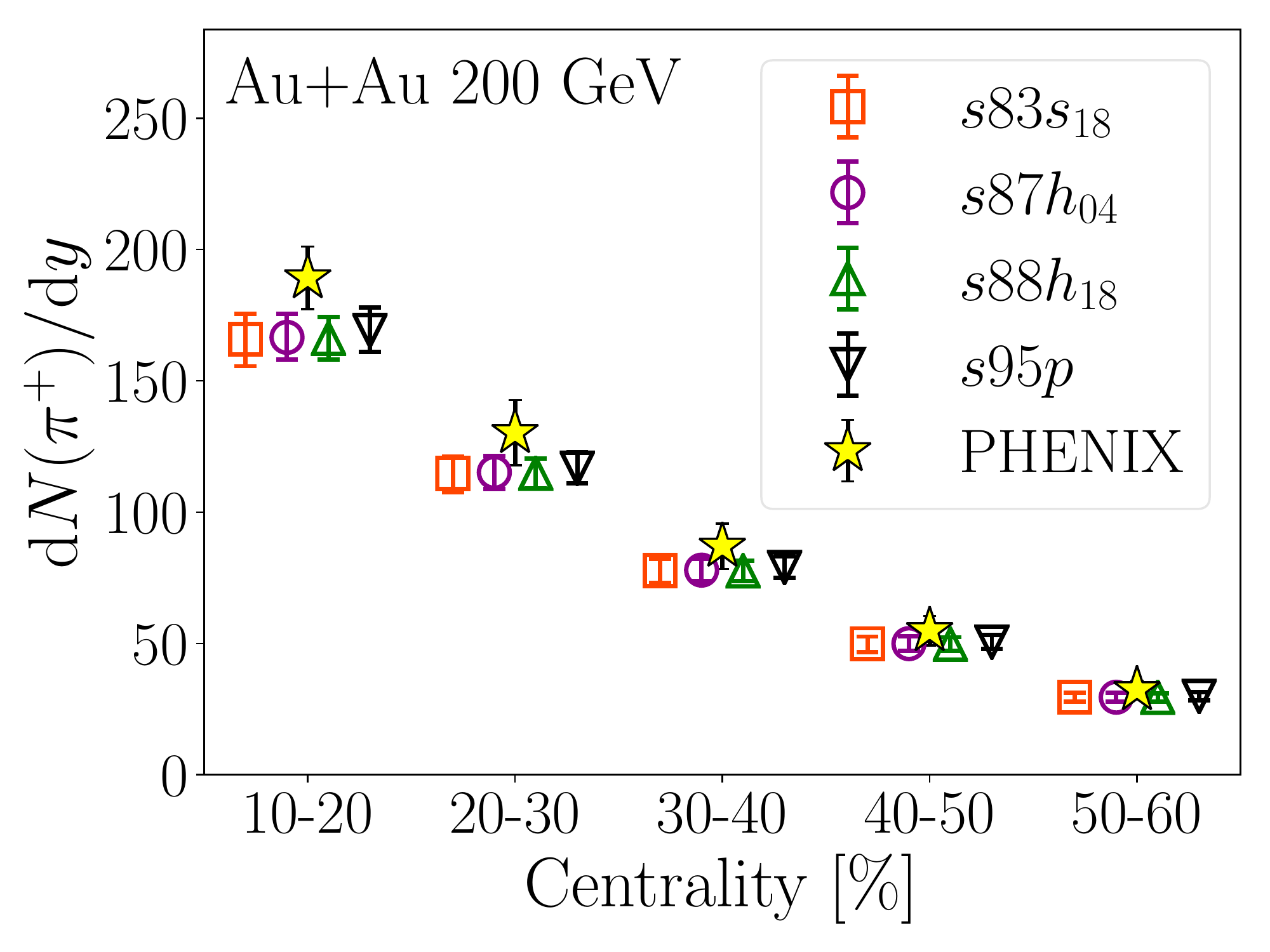}
 \includegraphics[width=4.25cm]{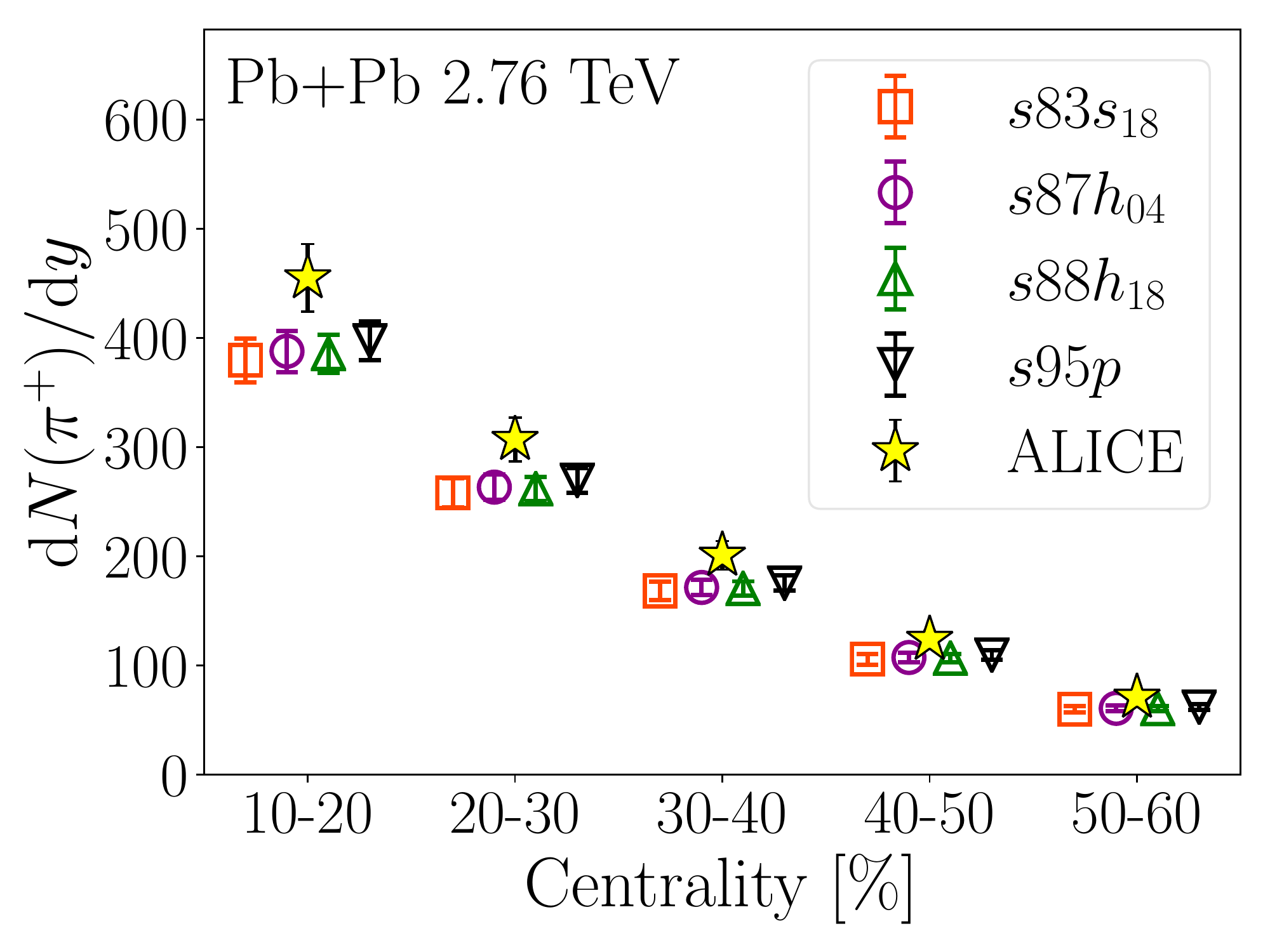}\\
 \vspace{2mm}
 \includegraphics[width=4.25cm]{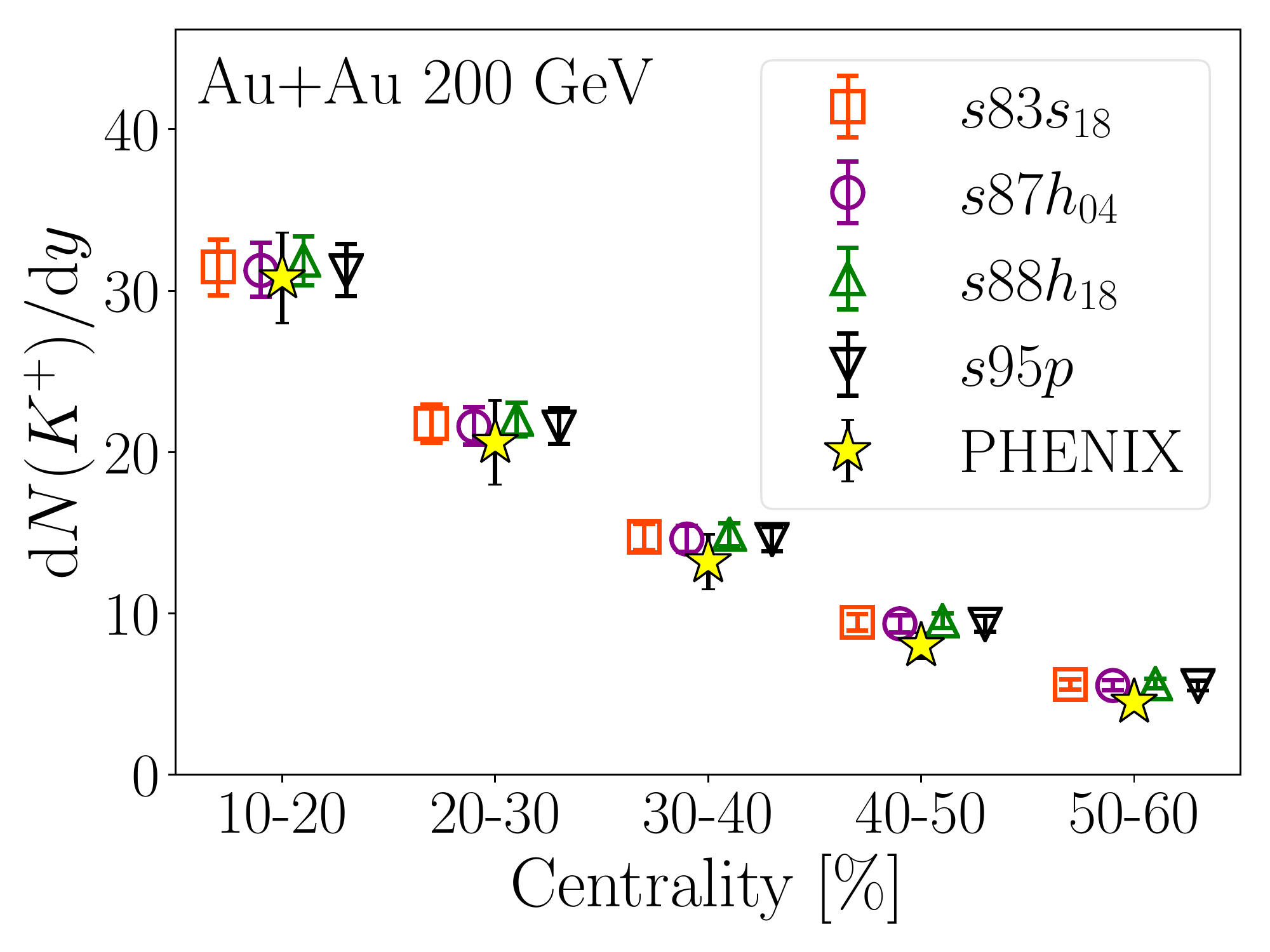}
 \includegraphics[width=4.25cm]{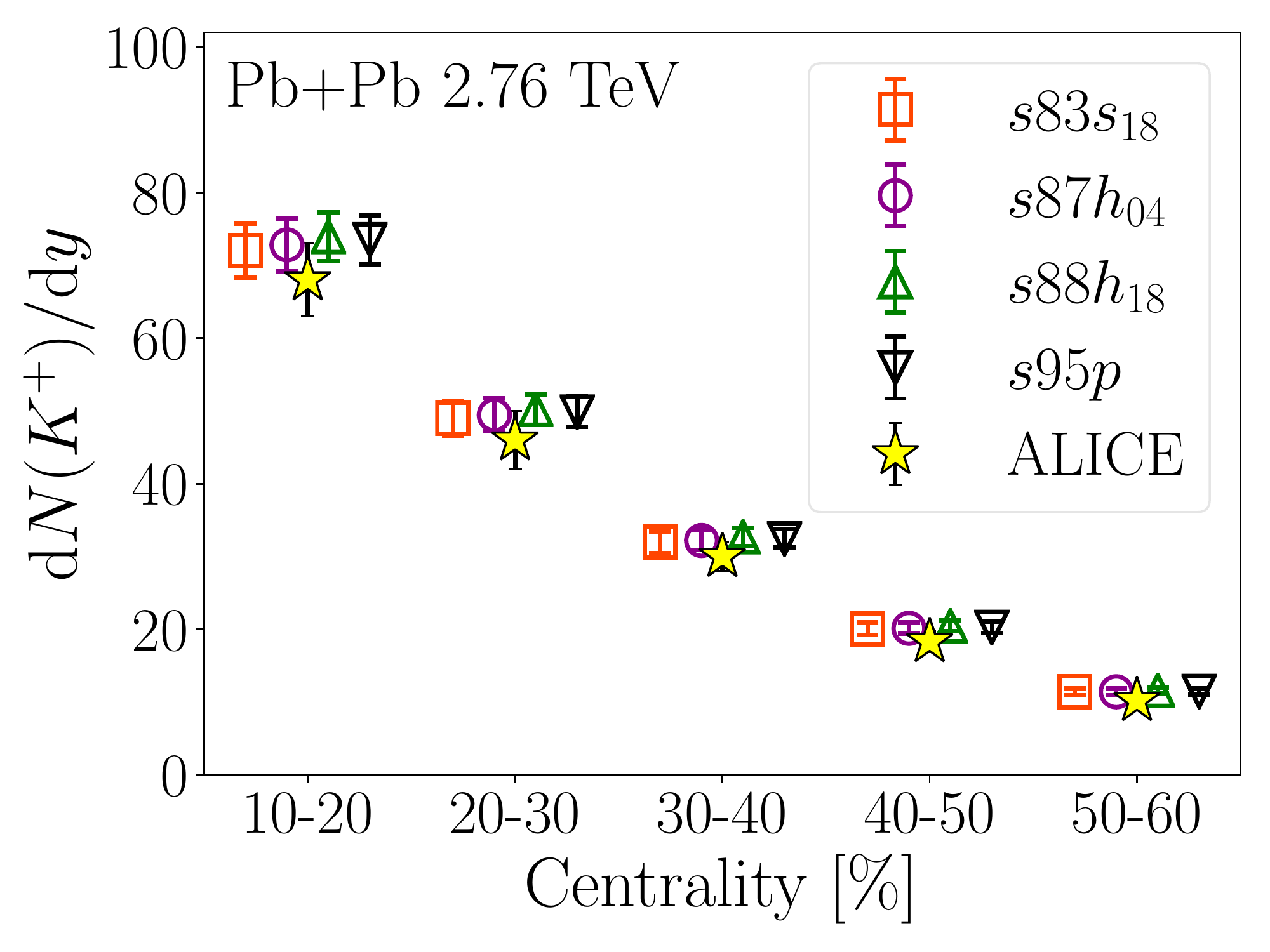}\\
 \vspace{2mm}
 \includegraphics[width=4.25cm]{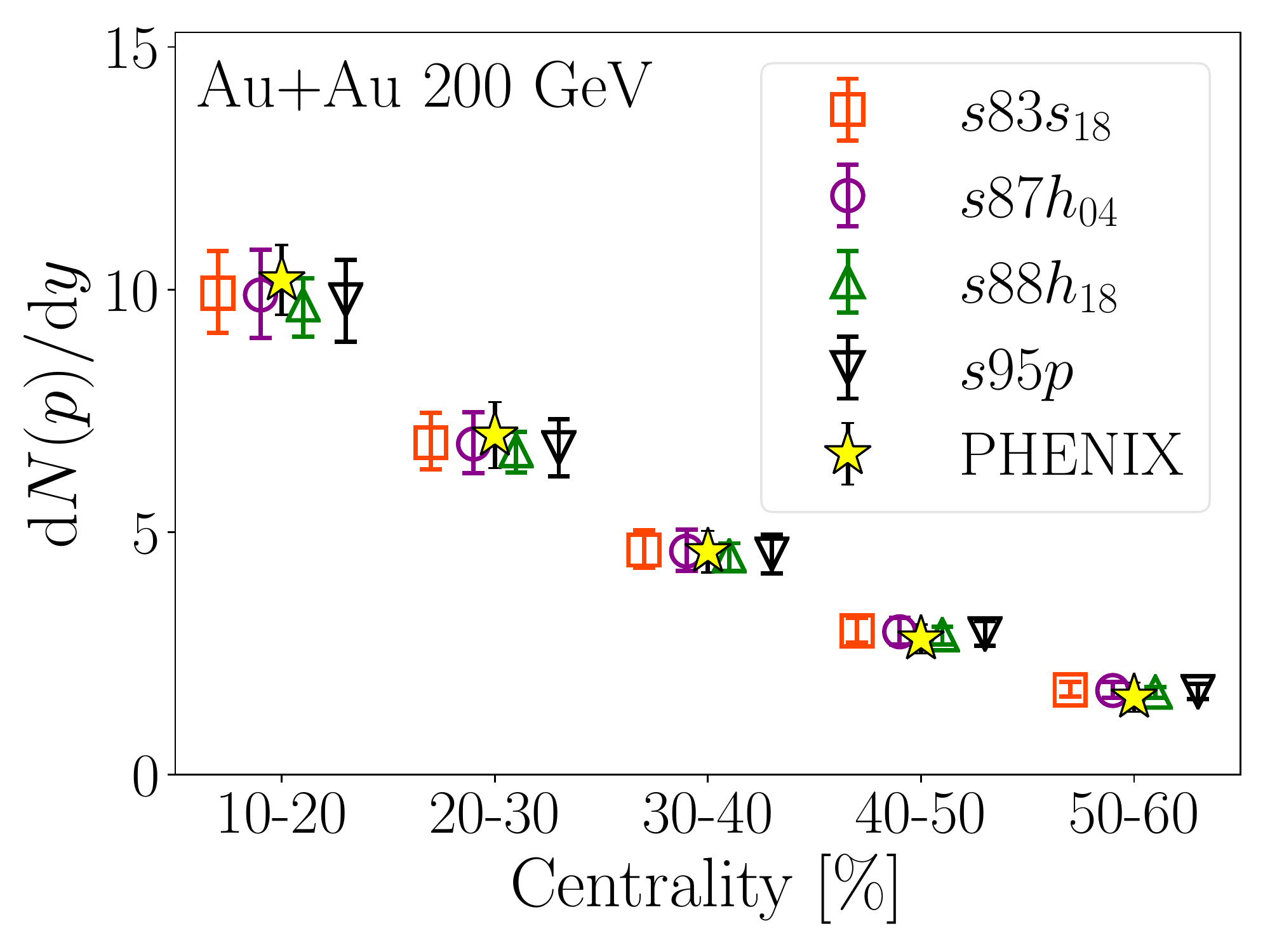}
 \includegraphics[width=4.25cm]{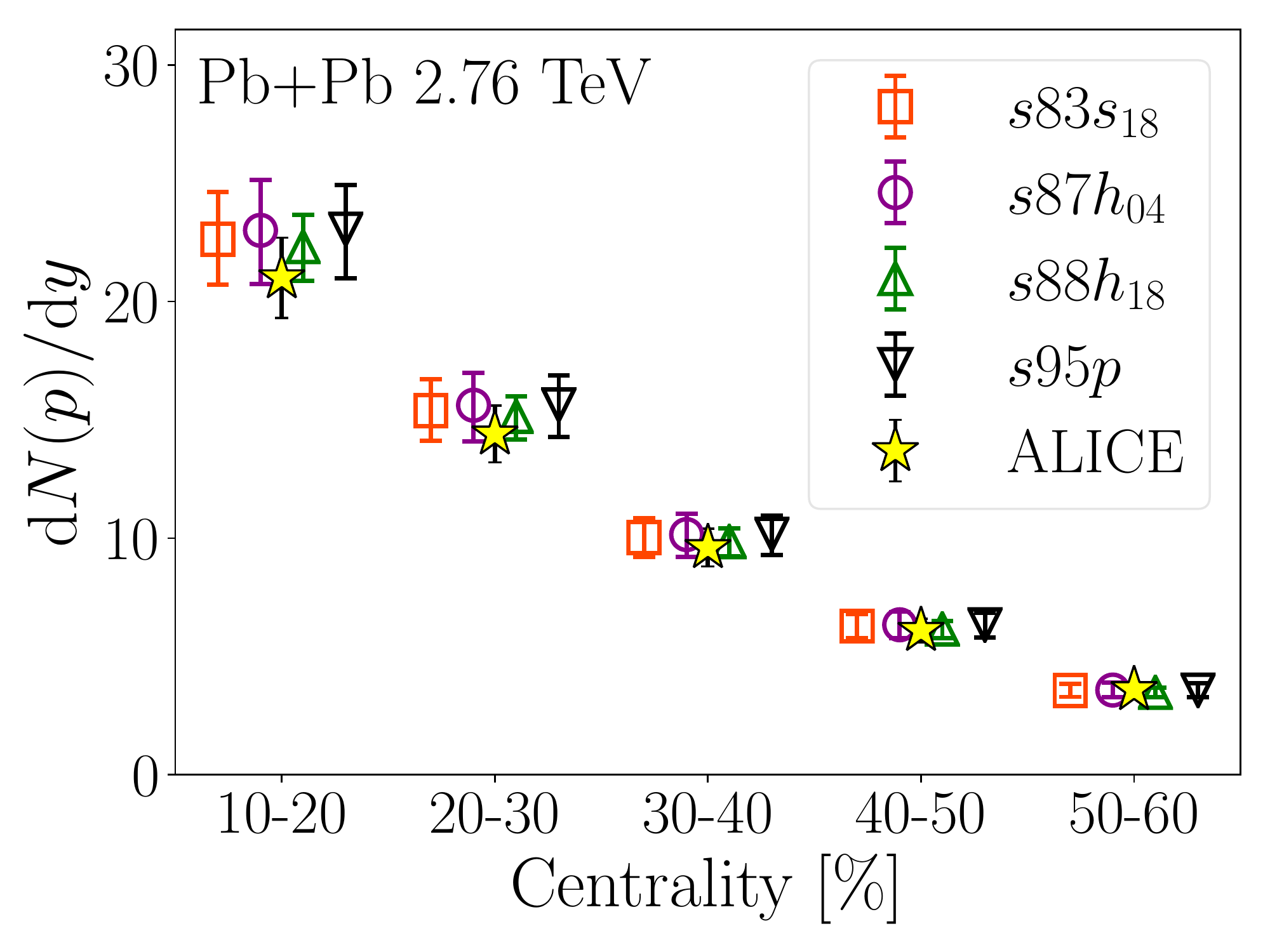}
 \caption{Pion (upper panels), kaon (middle panels), and proton (lower
   panels) multiplicities at various centralities using 1000 samples
   from the posterior distribution of each EoS.  Marker
   centers indicate median values, and error bars 90\% credible
   intervals.  Left panels: Au+Au at $\sqrt{s_{\NN}}=200$ GeV compared
   to PHENIX data \cite{Adler:2003cb}.  Right panels: Pb+Pb at
   $\sqrt{s_{\NN}}=2.76$ TeV compared to ALICE data
   \cite{Abelev:2013vea}}.
 \label{fig:nid_posterior}
\end{figure}

\begin{figure}[h]
 \centering
 \includegraphics[width=4.25cm]{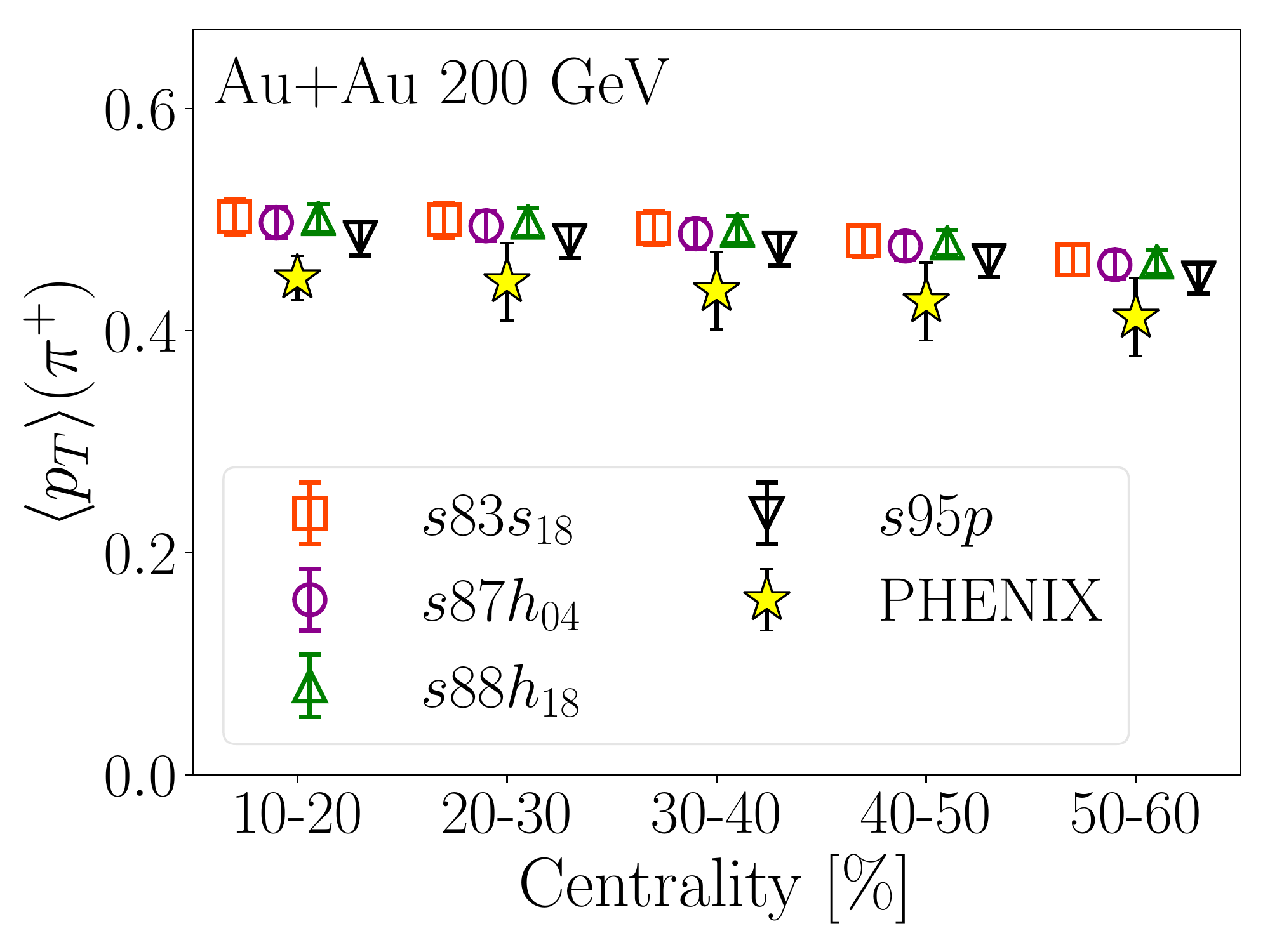}
 \includegraphics[width=4.25cm]{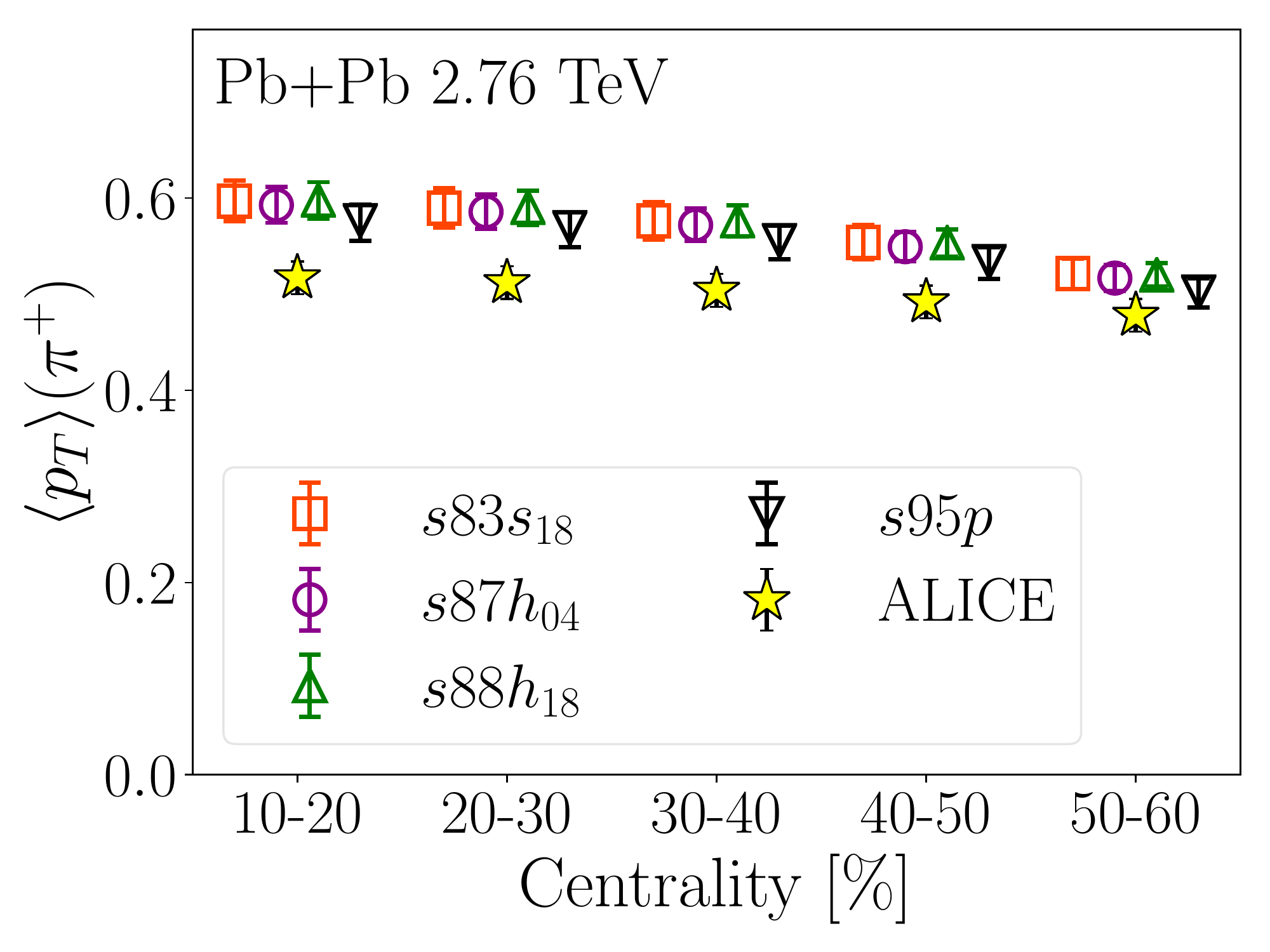}\\
 \vspace{2mm}
 \includegraphics[width=4.25cm]{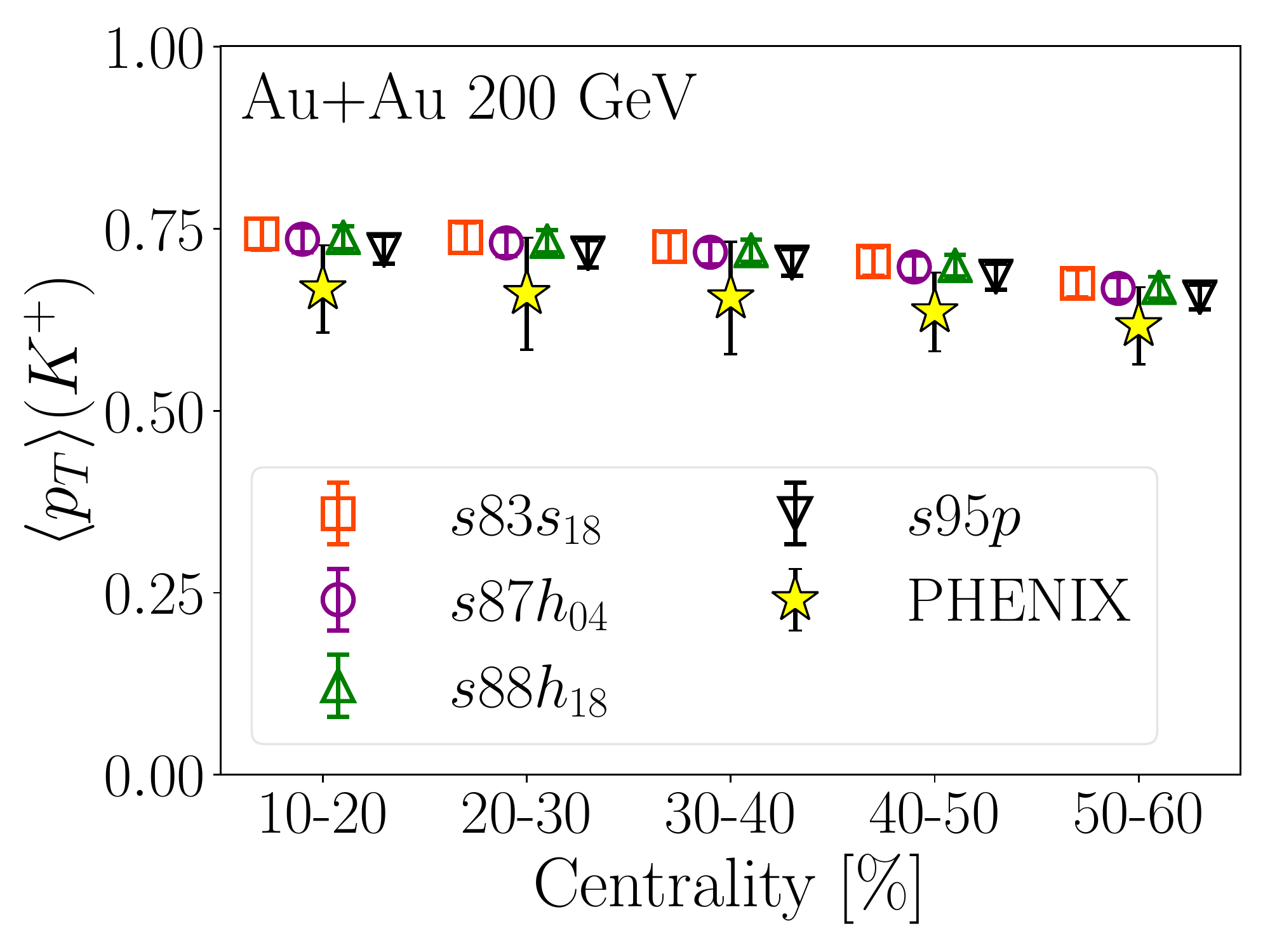}
 \includegraphics[width=4.25cm]{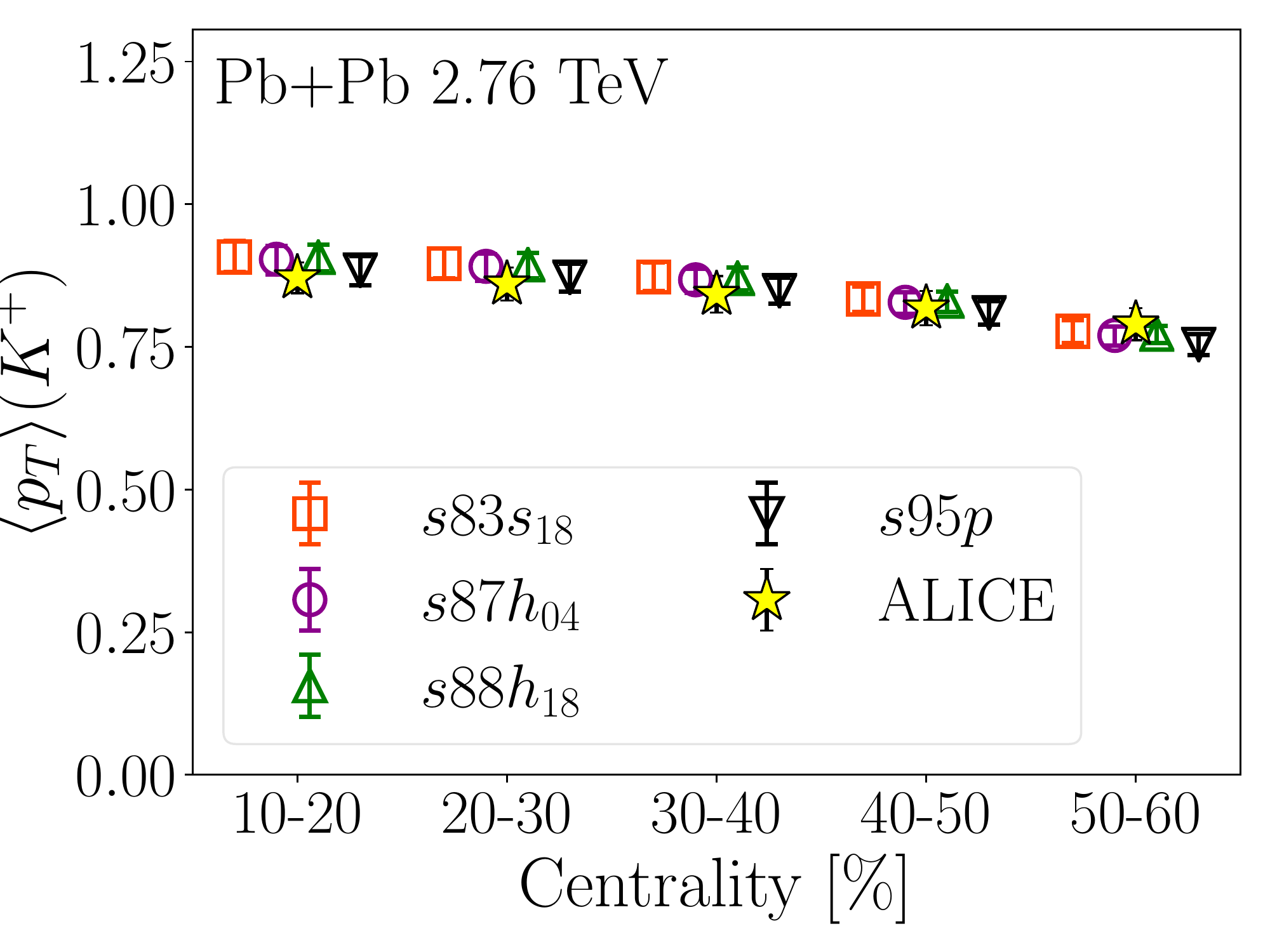}\\
 \vspace{2mm}
 \includegraphics[width=4.25cm]{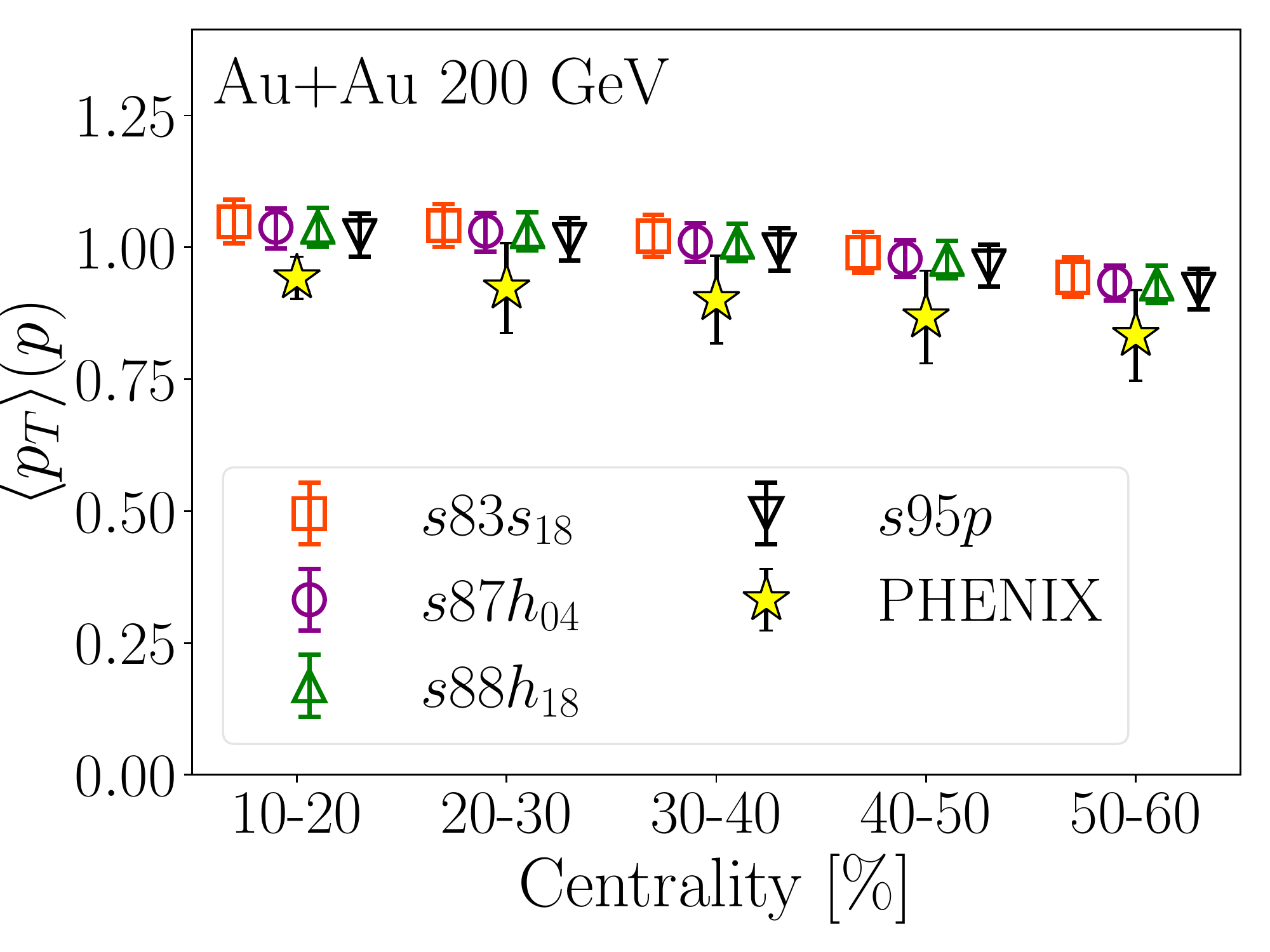}
 \includegraphics[width=4.25cm]{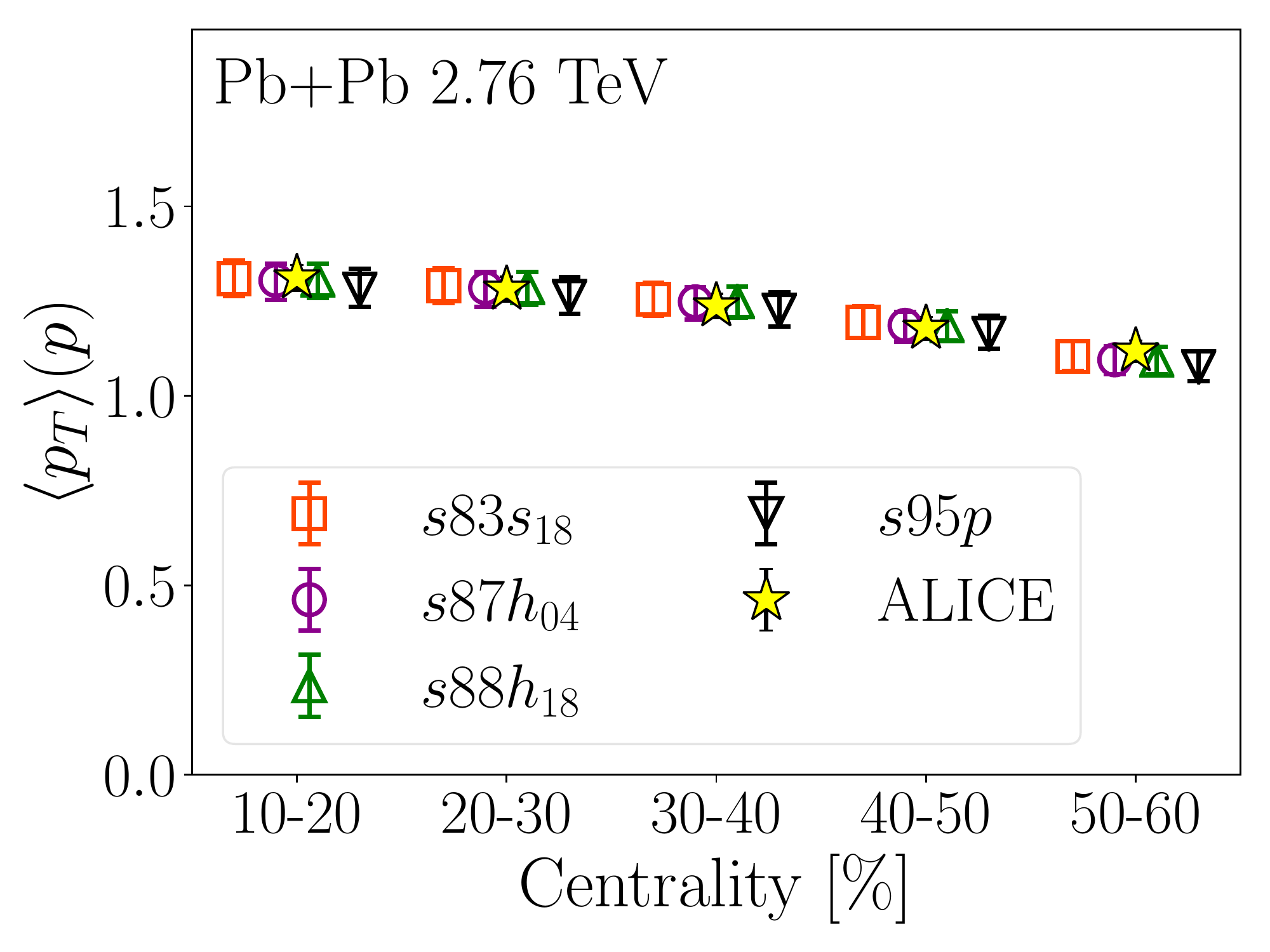}
 \caption{As Fig.~\ref{fig:nid_posterior} but for the mean transverse momentum.}
 \label{fig:meanptid_posterior}
\end{figure}

\begin{figure*}[ht]
 \centering
 \includegraphics[width=5.9cm]{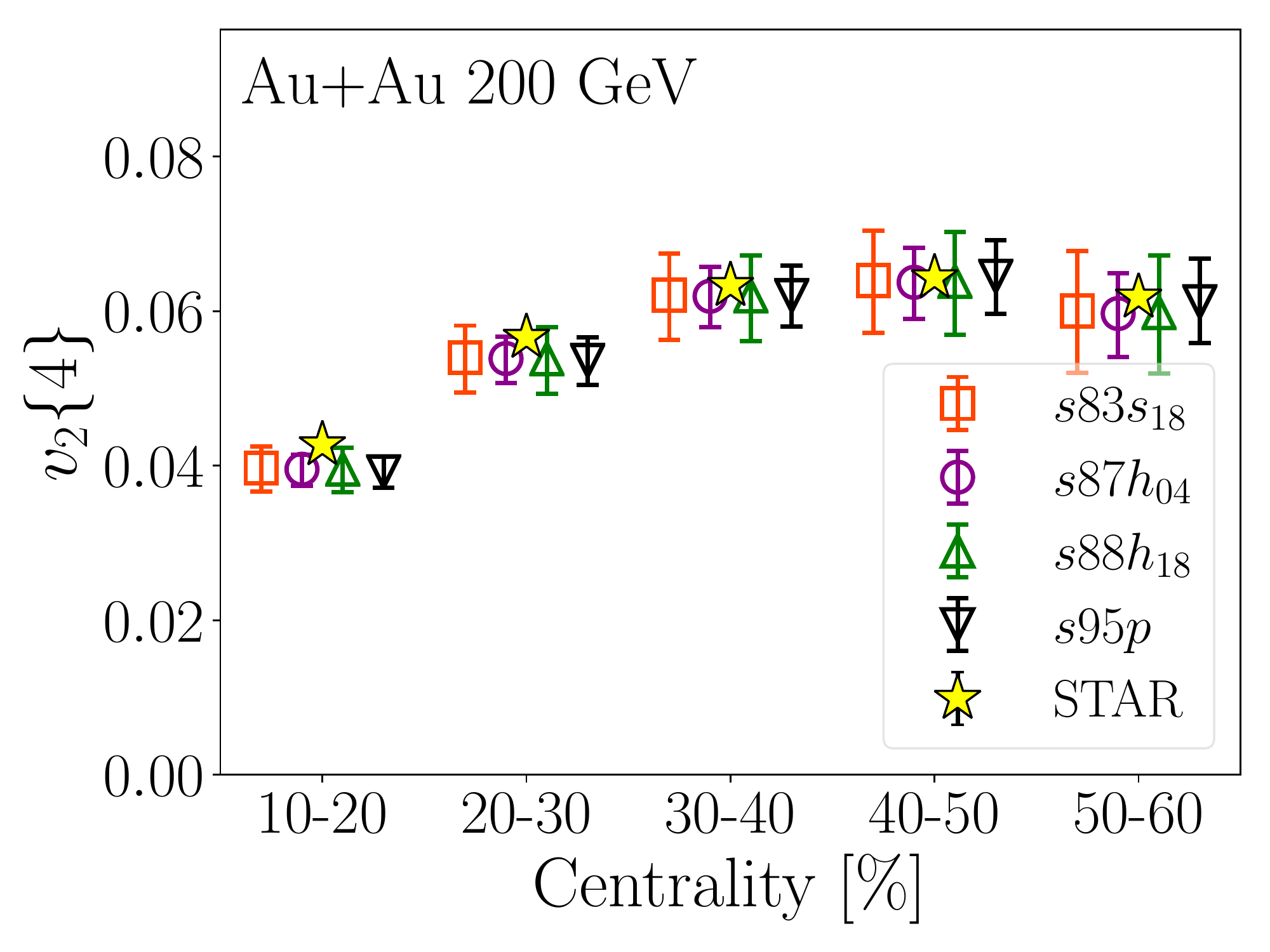}
 \includegraphics[width=5.9cm]{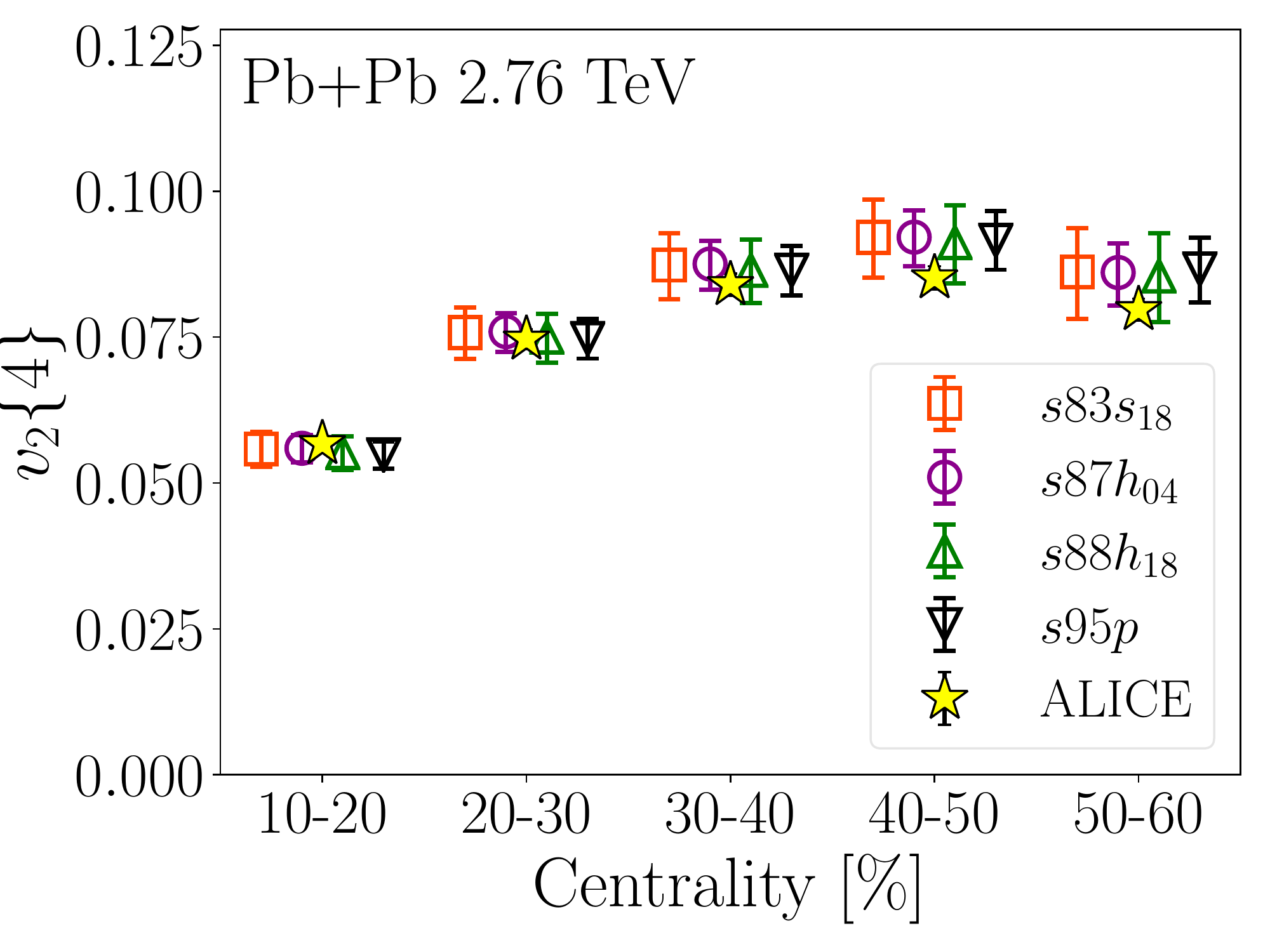}
 \includegraphics[width=5.9cm]{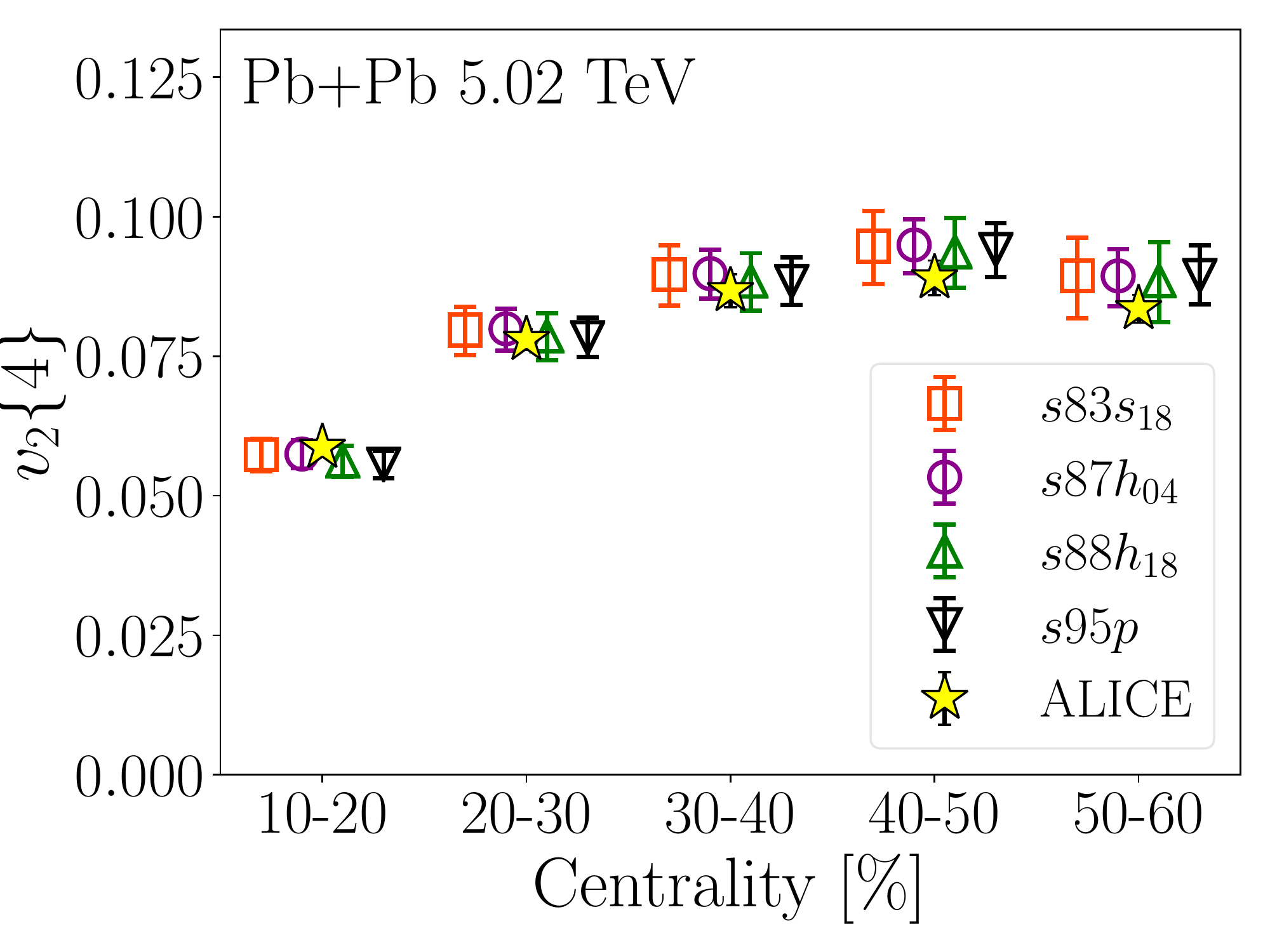}
 \caption{Charged particle elliptic flow $v_2\{4\}$ at various centralities
 using 1000 samples from the posterior distribution of each EoS.
 Marker centers indicate median values, and error bars 90\% credible intervals.
 Left panel: Au+Au at $\sqrt{s_{\NN}}=200$ GeV compared to STAR data \cite{Adams:2004bi}.
 Middle panel: Pb+Pb at $\sqrt{s_{\NN}}=2.76$ TeV compared to ALICE data \cite{Adam:2016izf}.
 Right panel: Pb+Pb at $\sqrt{s_{\NN}}=5.02$ TeV compared to ALICE data \cite{Adam:2016izf}.}
 \label{fig:v2_posterior}
\end{figure*}

Finally, as an overall quality check, we show how well the favored
parameter combinations reproduce the experimental data. This is done by drawing 1000
samples from each posterior distribution and using the Gaussian
process emulator to predict the simulation output for these
values. The results for charged and identified particle
multiplicities, identified particle $\langle p_T \rangle$, and the
elliptic flow $v_2\{4\}$ are shown in Figs.~\ref{fig:nch_posterior},
\ref{fig:nid_posterior},~\ref{fig:meanptid_posterior},
and~\ref{fig:v2_posterior}, respectively.

The overall agreement with the data is quite good for all observables,
and the analysis is able to find equally good data fits for all four
EoSs.  As normal for thermal models, the charged
particle multiplicities tend to be underestimated due to the tension
between pion multiplicity on one hand, and kaon and proton
multiplicities on the other hand. As the analysis makes a compromise
between too few pions and too many kaons and protons, the overall
charged particle multiplicity (which is dominated by pions) will
remain below the data. Also the mean transverse momentum of pions is
slightly too large, which may prove difficult to alleviate
without the introduction of bulk viscosity~\cite{Ryu:2015vwa} and/or
improved treatment of resonances during the hadronic
phase~\cite{Huovinen:2016xxq}.

\section{Summary}
\label{sec:summary}

In this work, we have introduced three new parametri\-zations of the
equation of state based on the contemporary lattice data:
\begin{itemize}
\item $s87h_{04}$ connects the HRG based on the PDG 2004
  particle list to parametrized lattice data obtained using the HISQ
  discretization scheme.
\item $s88h_{18}$ is based on the HRG containing all strange and non-strange
  hadrons and resonances in the PDG 2018 summary tables, and
  the same HISQ lattice data as $s87h_{04}$.
\item $s83s_{18}$ is constructed using the PDG 2018
  resonances, and the continuum extrapolated lattice data obtained
  using the stout discretization.
\end{itemize}

We used these new parametrizations and the older $s95p$
parametrization to examine how sensitive the shear viscosity over
entropy density ratio $\eta/s$ is to the equation of state. We assumed
a piecewise linear parametrization for $(\eta/s)(T)$, and determined
the probability distributions of the best-fit parameter values within
the EKRT framework using a Bayesian statistics approach.

Using charged and identified particle multiplicities, identified
particle mean transverse momenta, and elliptic flow at three different
collision energies as calibration data, we were able to constrain the
value of $\eta/s$ to be between 0.08 and 0.23 with 90\% credibility in
the temperature range $150 \lton T/\mathrm{MeV} \lton 220$ when all
EoS parametrizations are taken into account. When we constrain the
EoSs to the most contemporary parametrizations $s83s_{18}$ and
$s88h_{18}$, we obtain $0.12 < \eta/s < 0.23$ in the above mentioned
temperature range. As the differences between the EoSs are well
covered by the 90\% credible intervals, the earlier results obtained
using the $s95p$ parametrization remain valid. The weak sensitivity to
the EoS is consistent with the old ideal fluid results for flow and
EoS: Based on flow alone, it is difficult to distinguish an EoS with a
smooth crossover from an EoS without phase transition~\cite{Huovinen:2005gy}.
Thus when the differences between EoSs are just details in the
crossover, the differences in flow, which should be compensated by
different shear viscosity, are small, and consequently differences in
the extracted $\eta/s$ are small.

The overall agreement with the data is quite good, and similar to
Refs.~\cite{Niemi:2015qia,Niemi:2015voa}, where event-by-event
fluctuations were included to the framework of EKRT initial conditions
and fluid dynamics, albeit without the Bayesian analysis. The good
agreement achieved here is partly due to the EKRT initial conditions.
In particular the centrality and $\sqrt{s_\NN}$ dependence of hadron
multiplicities follow mainly from the QCD dynamics of the EKRT model.
A noticeable difference to the earlier event-by-event analysis is that
here we used identified hadron multiplicities as constraint, which led
to the chemical freeze-out temperature $\Tchem \approx 154$ MeV, and a
slight overshoot of the pion average $p_T$ compared to the data. In
the earlier analysis $\Tchem \approx 175$ MeV was used to reproduce the
average $p_T$ data, which in turn led to too large proton multiplicity.
It is possible to solve this tension by introducing bulk
viscosity~\cite{Ryu:2015vwa}, but that is left for a future work.
We emphasize that compared to the (in principle) more detailed hydro +
cascade models our hydro + partial chemical equilibrium approach has
two major advantages: It allows us to parametrize $(\eta/s)(T)$ so that
it is continuous in the whole temperature range, and at the same time
it gives us a possibility to constrain the viscosity also in the
hadronic phase.

Inclusion of event-by-event fluctuations to the analysis
would provide access to several new flow observables such as higher
flow harmonics $v_n$, and flow correlations, which may give tighter
constraints in broader temperature interval on $(\eta/s)(T)$. However,
within the current uncertainties of the fitting procedure, we cannot
exclude the possibility that the effect of the EoS remains negligible
even when $\eta/s$ at $T>220$ MeV becomes better under control.

Since the sensitivity of flow to shear viscosity at high temperatures
is low, observables based on high $p_T$ particles may be useful to
constrain, not only the pre-equilibrium
dynamics~\cite{Noronha-Hostler:2016eow,Andres:2019eus,Zigic:2019sth},
but also the properties of the fluid when it is hottest.

\section*{Acknowledgments}

We thank V.~Mykhaylova for sharing her quasi-particle results with us.
JA and PH were supported by the European Research Council, Grant
No.~ERC-2016-COG:725741, PH was also supported by National Science
Center, Poland, under grant Polonez DEC-2015/19/P/ST2/03333 receiving
funding from the European Union's Horizon 2020 research and innovation
program under the Marie Sk\l odowska-Curie grant agreement No.~665778;
KJE and HN were supported by the Academy of Finland, Project
No.~297058; and PP was supported by U.S.~Department of Energy under
Contract No.~DE-SC0012704. We acknowledge the CSC -- IT Center for Science
in Espoo, Finland, for the allocation of the computational resources.

\appendix

\section{EoS parametrization}
  \label{appx:para}

\begin{table*}
  \centering
  \caption{The values of parameters for different fits of the trace anomaly.}
  \begin{tabular}{l|cccccccccc}
    \hline
    \hline
           & $d_0$      & $d_1$(GeV$^2$) & $d_2$(GeV$^4$)   & $d_3$(GeV$^{n_3}$)     & $d_4$(GeV$^{n_4}$) & $d_5$(GeV$^{n_5}$)&$n_3$&$n_4$&$n_5$&$T_0$(MeV) \\
    \hline
    $s83s_{18}$ & 5.688$\x 10^{-3}$& $0.3104$ & -6.217$\x 10^{-3}$& -6.680$\x 10^{-32}$   & 1.071$\x 10^{-32}$ & --              & 41  & 42  & -- & $166$ \\
    $s87h_{04}$ & 5.669$\x 10^{-2}$& $0.2974$ & -4.184$\x 10^{-3}$& -5.146$\x 10^{-8}\,\,$& 1.420$\x 10^{-33}$ & --              & 10  & 42  & -- & $172$ \\
    $s88h_{18}$ & 4.509$\x 10^{-2}$& $0.3082$ & -5.136$\x 10^{-3}$& -1.150$\x 10^{-10}$   & 2.076$\x 10^{-32}$ & -3.021$\x 10^{-33}$& 13 & 41 & 42 & $155$ \\
    $s95p$ & --              & $0.2660$ &\:2.403$\x 10^{-3}$& -2.809$\x 10^{-7}\,\,$& 6.073$\x 10^{-23}$ & --              &  10 & 30  & -- & $183.8$ \\
    \hline
    \hline
  \end{tabular}
  \label{tab:latticeparam}
\end{table*}

\begin{table*}
  \centering
  \caption{The values of parameters for different fits of the HRG trace anomaly.}
  \begin{tabular}{l|ccccccccc}\hline\hline
           & $a_1$(GeV$^{-m_1}$) & $a_2$(GeV$^{-m_2}$) & $a_3$(GeV$^{-m_3}$) & $a_4$(GeV$^{-m_4}$) & $m_1$ & $m_2$ & $m_3$ & $m_4$ & $\Th$(MeV) \\ \hline
    $s83s_{18}$ & $0.1850$          &  1.985$\x 10^4$    & 1.278$\x 10^5$     & -1.669$\x 10^7$   & 0     & 5    & 7     & 10     & 170 \\
    $s87h_{04}$ & $4.654\;\;$       & -$879\quad\;\,$    & $8081\quad\;\,$    & -7.039$\x 10^6$  & 1     & 3    & 4     & 10    & 190 \\
    $s88h_{18}$ & $0.1844$          &  2.043$\x 10^4$    & 8.550$\x 10^5$     & -2.434$\x 10^7$  & 0     & 5    & 8     & 10    & 169 \\
    $s95p$ & $4.654\;\;$       & -$879\quad\;\,$    & $8081\quad\;\,$    & -7.039$\x 10^6$  & 1     & 3    & 4     & 10    & 190 \\ \hline\hline
  \end{tabular}
  \label{tab:hrgparam}
\end{table*}

At high temperature the trace anomaly can be well parametrized by the
inverse polynomial form. Therefore we will use the following Ansatz
for the high temperature region:
\begin{equation}
 \frac{\epsilon -3p}{T^4} = d_0 + \frac{d_1}{T^2} + \frac{d_2}{T^4} + \frac{d_3}{T^{n_3}} + \frac{d_4}{T^{n_4}} + \frac{d_5}{T^{n_5}}. 
  \label{eq:e-3p_high}
\end{equation}
This form does not have the right asymptotic behavior in the high
temperature region, where we expect
$(\epsilon-3p)/T^4 \sim g^4(T) \sim 1/\ln^2(T/\Lambda_{QCD})$, but it works well in
the temperature range of interest. Furthermore, it is flexible enough
to match to the HRG result in the low temperature region. We
match this Ansatz to the HRG model at temperature $T_0$ by requiring that the
trace anomaly, and its first and second derivatives are
continuous. This requirement provides constraints for three
parameters, $d_0, d_1,$ and $d_2$, and leaves the remaining seven,
$d_3, d_4, d_5, n_3, n_4, n_5,$ and $T_0$ to be fixed by minimizing a
$\chi^2$ fit to the data. Fitting the powers $n_3$--$n_5$ would be a
highly non-linear problem, but we simplify the problem by requiring
that the powers are integers, and using brute force: We make a fit
with all the integer values $5\leq n_3 \leq 40$, $n_3 < n_4 \leq 41$,
and $n_4 < n_5 \leq 42$, and choose the values $n_3, n_4$, and $n_5$
which lead to the smallest $\chi^2$. When the powers and $T_0$ are
kept fixed, minimizing $\chi^2$ requires only a simple matrix
inversion. Thus to fix $T_0$ we are able to cast $\chi^2$ as a
function of only a single parameter, $T_0$. We require that
$155 \leq T_0/\mathrm{MeV} \leq 190$, and search for the
value of $T_0$ which minimizes $\chi^2$.

To obtain the continuum limit in the lattice calculations of the trace
anomaly, one has to perform interpolation in the temperature, and then
perform continuum extrapolations (see e.g.~\cite{Borsanyi:2013bia}).
This procedure can introduce additional uncertainties when providing
parametrization of the lattice results. As mentioned in the main text,
the lattice spacing ($N_t$) dependence of the lattice results
is small in the case of the HISQ discretization scheme
for $N_t\geq 8$. In fact, for $T>230$ MeV and $T<170$ MeV there is no
statistically significant $N_t$ dependence, so in these temperature
ranges we can use the HISQ lattice results with $N_t=8,10$ and
$12$. In the peak region, $170 < T/\mathrm{MeV} < 230$, the $N_t=8$
HISQ results are slightly higher than the $N_t=10$ and $N_t=12$
results, and therefore have been omitted from the fits. At
temperatures above $800$ MeV only lattice results with $N_t = 6$ and 4
are available~\cite{Bazavov:2014pvz,Bazavov:2017dsy}. To take the
larger discretization errors of the $N_t = 6$ and 4 results into
account, we follow Ref.~\cite{Bazavov:2017dsy}, scale them by
factors 1.4 and 1.2, and include systematic errors of 40\% and 20\%,
respectively. Contrary to the HISQ action results, we employ the
continuum extrapolated stout action results~\cite{Borsanyi:2013bia,
  Borsanyi:2010cj} for simplicity. The resulting parameters are shown
in Table~\ref{tab:latticeparam}. We find that only the parametrization
$s88h_{18}$ requires the use of all six terms in Eq.~\ref{eq:e-3p_high}.
In the cases of $s83s_{18}$ and $s87h_{04}$ we are able to obtain
equally good fits with only five terms, and thus set $d_5$ to zero by
hand.

For the sake of completeness, we also parametrize the HRG part of the
trace anomaly as
\begin{equation}
  \frac{\epsilon-3p}{T^4} = a_1T^{m_1} + a_2T^{m_2} + a_3T^{m_3} + a_4T^{m_4}.
  \label{eq:e-3p_low}
\end{equation}
To carry out the fit we evaluate HRG trace anomaly in temperature
interval $70 < T/\mathrm{MeV} < \Th$, where $\Th$ depends on the
parametrization, with 1 MeV steps assuming that each point has equal
``error''. The limits have entirely utilitarian origin: in
hydrodynamical applications the system decouples well above 70 MeV
temperature and only a rough approximation of the EoS,
$p=p(\epsilon)$, is needed at lower temperatures. On the other hand we
expect to switch to the lattice parametrization below $\Th$, and the
HRG EoS above that temperature is not needed either. We fix the powers
in Eq.~(\ref{eq:e-3p_low}) again using brute force. We require them to
be integers, go through all the combinations $0\leq
l_1<l_2<l_3<l_4\leq 10$, fit the parameters $a_1$, $a_2$, $a_3$, $a_4$
to the HRG trace anomaly evaluated with 1 MeV intervals, and choose
the values $l_1,~l_2,~l_3$ and $l_4$ which minimize the $\chi^2$. We
end up with parameters shown in Table~\ref{tab:hrgparam}.  To obtain
the EoS, one also needs the pressure at the lower limit of the
integration (see Eq.(\ref{eq:P-integral})) $T_\mathrm{low} = 0.07$
GeV: $p(T_\mathrm{low})/T^4_\mathrm{low}=0.1661$.  Our EoSs are
available in a tabulated form at arXiv as ancillary files for this
paper, and at Ref.~\cite{osf}. These tables also include the option of
a chemically frozen hadronic stage, and a list of resonances
included in the hadronic stage with their properties and decay channels.

\section{Predicting model output with Gaussian processes}
  \label{appx:GP}

Let us assume that we do not know exactly what the model's output $y$
for a particular input parameter $\vec{x}$ is, but we know its most
probable value $\mu(\vec{x})$. We postulate that the probability
distribution for the output value $P(y)$ is a normal distribution with
mean $\mu(\vec{x})$ and so far unknown width $\sigma$. Thus the
probability distribution for a set $Y_a$ of $N$ model outputs for
observable $a$, corresponding to a set $X$ of $N$ points in the
parameter space, is a multivariate normal distribution:
\begin{equation}
\mathcal{G}:X \rightarrow Y_a \sim \mathcal{N}(\boldsymbol{\mu},\boldsymbol{C})
\end{equation}
where $\boldsymbol{\mu}=\mu(X)=\{\mu(\vec{x}_1),\ldots,\mu(\vec{x}_N)\}$ is the mean of the distribution,
and $\boldsymbol{C}$ is the covariance matrix defined by the covariance function $c(\vec{x},\vec{x}')$:
\begin{equation}
 \boldsymbol{C}=\mathcal{C}_{X,X}=
 \begin{pmatrix}
 c(\vec{x}_1,\vec{x}_1) & \dots & c(\vec{x}_1,\vec{x}_N) \\
 \vdots & \ddots & \vdots \\
 c(\vec{x}_N,\vec{x}_1) & \dots & c(\vec{x}_N,\vec{x}_N) \\
 \end{pmatrix}.
\end{equation}

As we are only interested in interpolating within the training data,
we may set $\mu(X) \equiv 0$, and construct the covariance function
$c(\vec{x},\vec{x}')$ in such a way that the probability distribution
is narrow at the training points nevertheless. This way we minimize
our {\em a priori} assumptions about the model behavior in regions of
parameter space not covered by the training data\footnote{Note that we
  use Gaussian process to estimate the model output of the principal
  components, not the actual observables, see
  Appendix~\ref{appx:PCA}.}. Our chosen covariance function is a
radial-basis function (RBF) with a noise term
\begin{equation}
c(\vec{x},\vec{x}')=\theta_0\exp\left(-\sum\limits_{i=1}^{n}\frac{(x_i-x'_i)^2}{2\theta_i^2}\right)+\theta_{\text{noise}}\delta_{\vec{x}\vec{x}'}
\end{equation}
The hyperparameters
$\vec{\theta}=(\theta_0,\theta_1,\ldots,\theta_n,\theta_{\text{noise}})$,
where $n$ is the dimension of the input parameter space, are not known
a priori and must be estimated from training data, consisting of simulation output $U$ computed at training points $T$,
by maximizing the log-likelihood (see
Chapter 5 of \cite{Rasmussen:2006})
\begin{equation}
\label{eq:theta_likelihood}
\begin{split}
\log P(U|T,\vec{\theta})=&-\frac{1}{2}U^T\boldsymbol{C}^{-1}(T,\vec{\theta})U\\
&-\frac{1}{2}\log|\boldsymbol{C}(T,\vec{\theta})|\\
&-\frac{N}{2}\log(2\pi).
\end{split}
\end{equation}

Emulator prediction for the model output $y_0$ at a point $\vec{x_0}$
can then be determined by writing a joint probability distribution for the output
at various points in parameter space:
\begin{equation}
 \begin{pmatrix}
 y_0 \\
 U
 \end{pmatrix}
 \sim
 \mathcal{N}\left(
  \begin{pmatrix}
  0\\
  \vec{0}
 \end{pmatrix}
 ,
  \begin{pmatrix}
  \mathcal{C}_{0,0} & \mathcal{C}_{0,T}\\
  \mathcal{C}_{T,0} & \mathcal{C}_{T,T}
 \end{pmatrix}
 \right)
\end{equation}
from which we can derive the conditional predictive mean
$y^{\text{GP}}(\vec{x_0})$ and associated variance
$\sigma^{\text{GP}}(\vec{x_0})^2$ as (see e.g.\ Appendix A.2 of~\cite{Rasmussen:2006})
\begin{equation}
\label{eq:gp_estimate}
\begin{split}
 y^{\text{GP}}(\vec{x_0}) &= \mathcal{C}_{0,T}\mathcal{C}_{T,T}^{-1}U,\\
 \sigma^{\text{GP}}(\vec{x_0})^2 &= \mathcal{C}_{0,0}-\mathcal{C}_{0,T}\mathcal{C}_{T,T}^{-1}\mathcal{C}_{T,0}.
\end{split}
\end{equation}

Note that we use the training data $U$ twice: First in
Eq.~\eqref{eq:theta_likelihood} to determine the hyperparameters
$\vec{\theta}$ of the covariance function $c(\vec{x},\vec{x}')$ and
then in Eq.~\eqref{eq:gp_estimate} as a condition for the GP
prediction.

\section{Principal component analysis}
\label{appx:PCA}

We reduce the number of Gaussian processes needed for model emulation
with principal component analysis (PCA), which transforms the data in
the directions of maximal variance.

We represent the model output with a $N$ x $m$ matrix $Y$,
where $N$ is the number of simulation points and $m$ the number of observables.
In preparation for the PCA, the data columns are normalized with the corresponding experimental values to obtain dimensionless quantities,
and centered by subtracting the mean of each observable from the elements of each column; we denote this scaled and shifted data matrix by $\hat{Y}$.

We then want to find an eigenvalue decomposition of the covariance matrix $\hat{Y}^T\hat{Y}$:
\begin{equation}
\hat{Y}^{T}\hat{Y}=V \Lambda V^{T},
\end{equation}
where $\Lambda$ is the diagonal matrix containing the eigenvalues $\lambda_1,...,\lambda_m$
and $V$ is an orthogonal matrix containing the eigenvectors of the covariance matrix.

The eigenvalue decomposition is found by factorizing $\hat{Y}$ via the singular value decomposition:
\begin{equation}
\hat{Y} = U S V^{T},
\end{equation}
where $S$ is a diagonal matrix containing the singular values (square roots of the eigenvalues of $\hat{Y}^{T}\hat{Y}$)
and $V$ contains the right-singular vectors of $\hat{Y}$ (eigenvectors of $\hat{Y}^{T}\hat{Y}$);
these are the principal components (PCs).
Matrix $U$ contains the left-singular vectors of $\hat{Y}$, which are eigenvectors of $\hat{Y}\hat{Y}^T$.

The eigenvalues are proportional to the total variance of the data.
Since $\lambda_1 \geq \lambda_2 \geq ... \geq \lambda_m$,
the fraction of the total variance explained by the $k$th principal component,
$\lambda_k / (\sum\limits_{j=1}^m \lambda_j)$,
becomes negligible starting from some index $k < m$.
This allows us to define a lower-rank approximation of the original transformed data matrix $Z=\hat{Y}V$ as $Z_k=\hat{Y}V_k$,
where $V_k$ contains the first $k$ columns of $V$.

The transformation of a vector $\vec{y}$ from  the space of observables
to a vector $\vec{z}$ in the reduced-dimension principal component space is thus defined as
\begin{equation}
\vec{z} = \vec{y}\,V_k,
\end{equation}
while for matrices (such as the covariance matrix in the likelihood function \eqref{eq:likelihood}) the transformation is
\begin{equation}
\Sigma_z = V_k^T \, \Sigma_y\,V_k.
\end{equation}

To compare an emulator prediction $\vec{z}^{\,\text{GP}}$ against physical observables, we use the inverse transformation
\begin{equation}
\vec{y}^{\,\text{GP}} = \vec{z}^{\,\text{GP}} \, V_k^T.
\end{equation}

\section{Correlations between the model parameters}
\label{appx:posterior}

\begin{figure*}[t]
 \centering
 \includegraphics[width=18cm]{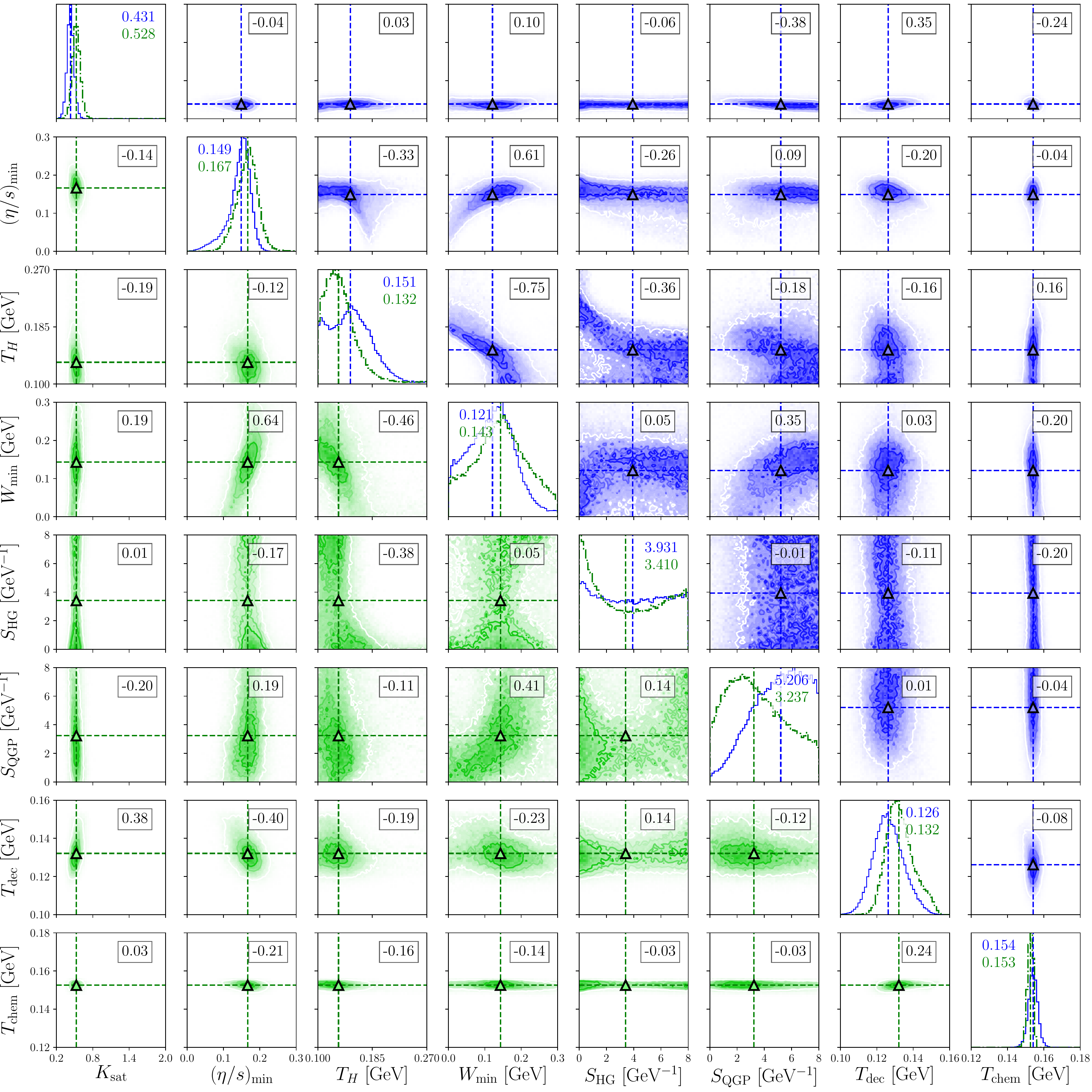}
 \caption{Posterior probability distribution for the $s88h_{18}$ (lower
   triangle, green color) and $s95p$ (upper triangle, blue color)
   EoSs. Diagonal panels: Marginalized 1-D distributions for each
   parameter. Solid blue line: $s95p$. Dash-dotted green line:
   $s88h_{18}$. Dashed lines and numbers indicate median value, with upper
   number corresponding to $s95p$ and lower number to $s88h_{18}$.
   Off-diagonal panels: 2-D projections of the posterior
   distributions. Dashed lines indicate median values for each
   parameter, while the framed numbers refer to Spearman rank
   correlation coefficients for each parameter pair.}
 \label{fig:posterior_s88s_s95p}
\end{figure*}

Figure \ref{fig:posterior_s88s_s95p} provides a more detailed view of
the 8-dimensional posterior probability distribution, using the
analysis results for the $s88h_{18}$ and $s95p$ EoSs as an
example. The diagonal panels show the marginalized one-dimensional
distributions for each parameter, which were summarized in
Figs.~\ref{fig:posterior_ksat}--\ref{fig:posterior_tmin} in
Section~\ref{sec:results}. The off-diagonal panels illustrate the
correlations between each parameter pair $(X,Y)$. The correlation
strength is quantified with the Spearman rank correlation coefficient
\cite{Spearman:1904}, which is the Pearson correlation coefficient
between the rank values $r_X$ and $r_Y$:
\begin{equation}
 \rho=\frac{C(r_X,r_Y)}{\sigma(r_X)\sigma(r_Y)},
\end{equation}
where $C$ refers to covariance and $\sigma$ to standard deviation. This
relaxes the assumption of a linear relationship, present in the Pearson
correlation coefficient, and is instead a measure of the monotonic
relationship between the two parameters.

\end{document}